\documentclass[reprint,amsmath,amssymb,aps,prapplied,floatfix]{revtex4-2}
\usepackage[english]{babel}
\usepackage{graphicx}
\usepackage{dcolumn}
\usepackage{bm}
\usepackage{ragged2e}
\usepackage{enumitem}
\usepackage{booktabs}
\usepackage{multirow}
\usepackage{array}
\usepackage{tabularx}
\newcolumntype{Y}{>{\centering\arraybackslash}X}
\usepackage[section]{placeins}
\usepackage{adjustbox}
\usepackage{comment}
\usepackage[normalem]{ulem}
\makeatletter
\newcommand*{\rom}[1]{\expandafter\@slowromancap\romannumeral #1@}
\makeatother

\usepackage[hidelinks,colorlinks=true,linkcolor=blue,citecolor=blue,urlcolor=blue]{hyperref}
\usepackage{bookmark}
\usepackage{cleveref}

\begin{document}
\preprint{{APS}/123-QED}
\title{Study of quark and gluon jet identification in photoproduction at EIC}
\author {Siddharth Narayan Singh}
\email{siddharthsingh@brandeis.edu}
\affiliation{Brandeis University, Waltham,~US}
\author {R. Aggarwal}
\email{ritu.aggarwal1@gmail.com}
\affiliation{{USAR} and {USBAS}, Guru Gobind Singh Indraprastha University, East Delhi Campus, 110092, India}
\author {M. Kaur}
\email{manjit@pu.ac.in}
\affiliation{Department of Physics, Panjab University, Chandigarh 160014, India\\Department of Physics, Amity University, Punjab, Mohali 140306, India}
\date{\today}
\begin{abstract}
\centering
\textbf{Abstract}
\begin{justify}
This paper presents a systematic study of the substructure of dijets produced in direct and resolved photoproduction events and subsequent
quark and gluon jet identification in dijets, considering the kinematic reach of the upcoming Electron-Ion Collider (EIC) experiment. We investigate the substructure of jets produced in dijet photoproduction events with center-of-mass energies $\sqrt{s} = 64 \text{--}140$ GeV at the EIC.~Events are simulated using event generator PYTHIA and contributions from direct and resolved photoproduction subprocesses are analyzed at different energies.~Jets are reconstructed using longitudinally invariant $k_T$ algorithm within the FastJet framework.~Substructure of quark and gluon initiated jets in the selected dijet photoproduction sample is studied in detail by using jet-shape variables.~Predictions for subjet multiplicities and integrated jet shapes are presented in the gluon-initiated and quark-initiated jets in the dijet photoproduction events at the EIC energies.~These results demonstrate the feasibility of distinguishing quark and gluon jets in the photoproduction events at the EIC and provide a baseline for further QCD studies.
\end{justify}
\end{abstract}
\maketitle

\section{Introduction}
The physics goals of the upcoming EIC collider experiment include the unparalleled measurements of hadron and nuclei structures, along with a better understanding of particle production dynamics.  The lepton-hadron scattering data, as published by the H1 and ZEUS experiments~\cite{Abramowicz2015} form the basis for understanding nucleon structure. The knowledge about the nucleon structures is summarised in the form of parton distribution functions (PDFs). The gluon contribution to PDFs enters only at next-to-leading order~(NLO) in deep inelastic scattering~(DIS); it is therefore important to study channels where the gluon from the parton sea participates in the interaction, such as dijet production in DIS or photoproduction~\cite{page2020experimental}. A recent study has shown that dijet photoproduction at the EIC has the potential to further constrain nuclear PDFs~\cite{PhysRevD.97.114013}. 

The dijet photoproduction processes have been widely studied at HERA~\cite{butterworth1996multiparton,Derrick1996-aj,zeus2003measurements} as they provide stringent tests of perturbative quantum chromodynamics (QCD). The direct and resolved photon processes through their angular distributions are known to provide a handle for a better understanding of production mechanisms~\cite{Derrick1996-aj}. If the jets produced in these processes are tagged, they also act as a sensitive ''spin-analyser" for the exchanged partons~\cite{lee2023machine}. Tagging the underlying hard interaction channel gives a way to experimentally determine and improve the knowledge of parton-in-photon PDFs~\cite{chu2017photon}.

EIC will not only facilitate collisions of electrons with hadrons but will also provide a basis for comparing data from collisions with heavier nuclei.~Precise information on the gluon contribution in the nuclear PDFs will give a better understanding of many nuclear effects, such as nuclear shadowing~\cite{Armesto2006-ms}, saturation~\cite{PhysRevD.59.014017} and the creation of quark-gluon plasma in heavy-nuclei collisions~\cite{Gelis2010-qt}.  

This paper focuses on the study of the substructure of jets produced in the dijet photoproduction processes at the EIC simulated using PYTHIA~\cite{10.21468/SciPostPhysCodeb.8}. A detailed study of the relative contributions of direct and resolved processes at each of the three different EIC energies is presented and compared with the HERA energy of 318 GeV. Predictions of the jet substructure in the direct and resolved processes are presented at the EIC energies. The main highlight of this paper is a systematic study of the substructure of dijets produced in direct and resolved photoproduction events and subsequent quark and gluon jet identification in dijets, considering the kinematic reach of the upcoming EIC experiment.~The parton that initiates the showering is usually associated with the type of jet at the leading order \!(LO), though there is no single or unique way of classifying quark and gluon jets~\cite{konishi1979jet, tumasyan2021study}.~In the LO QCD, gluons have a higher probability of radiating soft gluons compared to quarks and hence it is intuitive to find differences in the quark and gluon jet substructures~\cite{Ellis:1991qcd, PDG_QCD, opal1993study,opal1995model, buys1996energy, aleph2000measurements, buskulic1996quark, barate1998studies,  gras2017systematics}.~At large transverse jet energies, where the fragmentation effects are small, the jet substructure is governed largely by the initiating partons~\cite{Dissertori_Knowles_Schmelling_2009,PhysRevD.61.074009,Chekanov_2008, Glasman_2001}. 
\sloppy 
Using the subprocess information from PYTHIA, the dijet photoproduction events are labeled as quark-quark~(QQ), quark-gluon~(QG) and gluon-gluon~(GG) events depending upon the partons associated with the jets in the final state. This is explained in detail in Section 3. The predictions for the jet substructure of the QQ and GG events are given at the EIC energies, where low transverse momentum jets are expected to populate the detector.~The transverse energy profiles of jets in QQ and GG events are studied through their differential and integrated jet-shape distributions.~The integrated jet shape can be defined as the average fraction of the jet's transverse energy $E_{T}$ contained inside a cone of radius $r$ centered on the jet axis. The present study emphasizes the utility of jet substructure analysis, specifically through the integrated jet shape variable $\Psi(r)$, to distinguish between gluon-initiated and quark-initiated jets by labeling them as thick and thin jets respectively. 

A comparison is made between the jet shapes obtained at EIC and with the corresponding results from HERA, the first and only $ep$ collider. This serves as a reference for future EIC jet measurements. The analysis focused on the jet substructure in the direct and resolved events and further using it to produce gluon-enriched and quark enriched samples has not been performed previously in the EIC framework.

The study of separating quark and gluon jets is important because it increases the viability and precision of numerous physics studies at the present and upcoming high-energy colliders~\cite{Gallicchio_2013, PhysRevLett.107.172001, PhysRevD.frye_2017, Whitmore_2019,Bhattacherjee_2017,Chakraborty2018-wi,PhysRevD.95.034001}. The studies involving the beyond standard model (BSM) physics where signals would be hidden in the form of quark jets in the standard model~(SM) background of gluon jets, such a classification would be important~\cite{PhysRevD.95.034001, PhysRevD.99.114012}.  Various methods including both physical observables and modern machine learning techniques~\cite{Komiske_2022, Komiske_2018, Bhattacherjee_2015, Jankowiak2011-ht, Kasieczka2017-bu, Cheng2018-fg, Gras2017-me, Davighi_2018, Metodiev_2018, PhysRevD.41.1726, Larkoski:2014gra} have been used for this purpose.~Traditional jet tagging methods, such as substructure differences as presented in this paper, are important robust theory-driven methods, whereas the machine learning methods are unrobust theory-driven methods often trained on the simulated data ~\cite{larkoski2020jet}. In the era of increasing adaptability towards the machine learning methods, the theory-driven methods are still important to improve the interpretability of the results and for controlling the biases in the simulated or experimental data used for training purposes~\cite{lee2023machine}.

\begin{figure}
    \centering
    \includegraphics[width=0.35\textwidth]{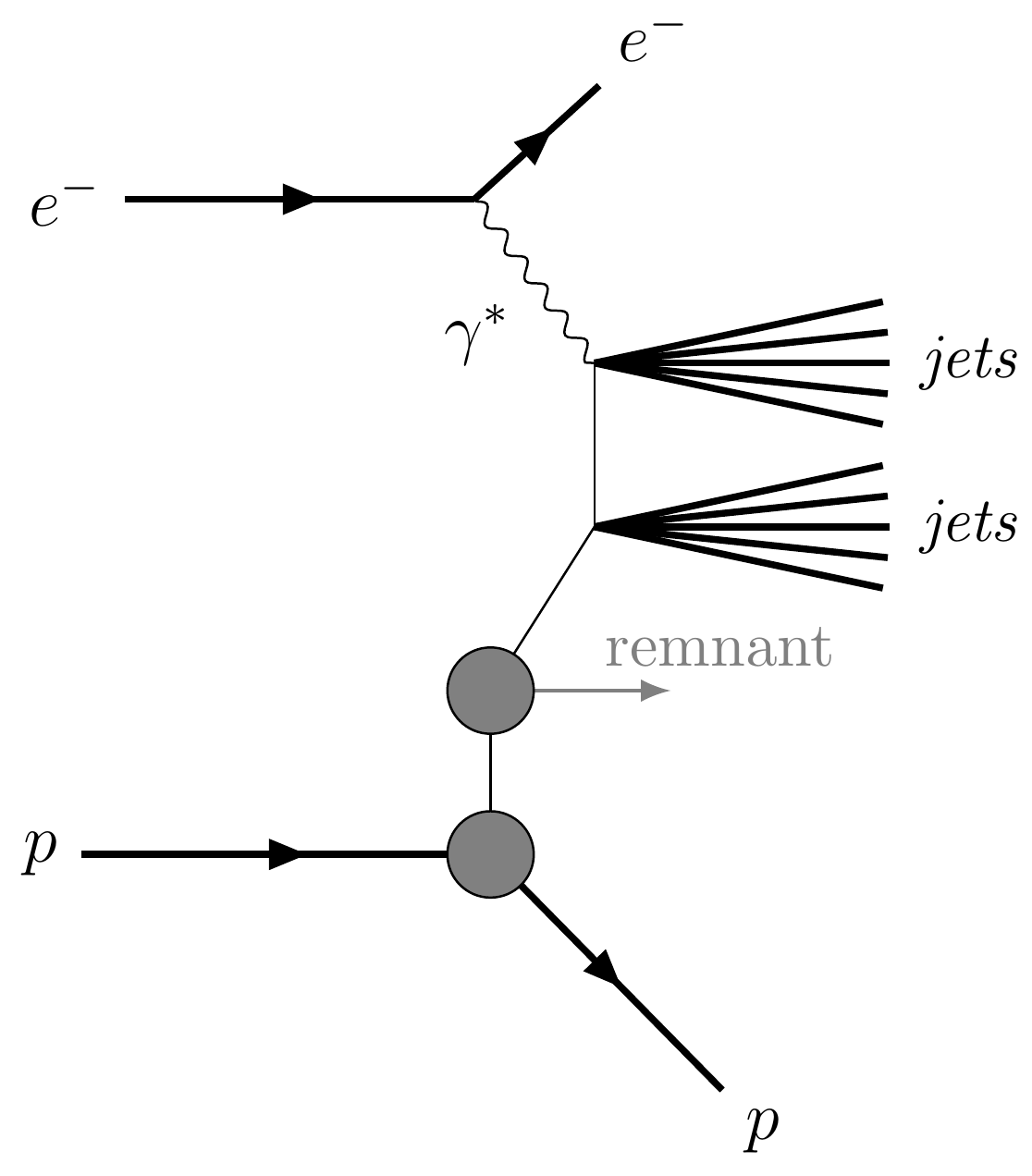} \\
    a. Direct Photoproduction Process \\
    \includegraphics[width=0.35\textwidth]{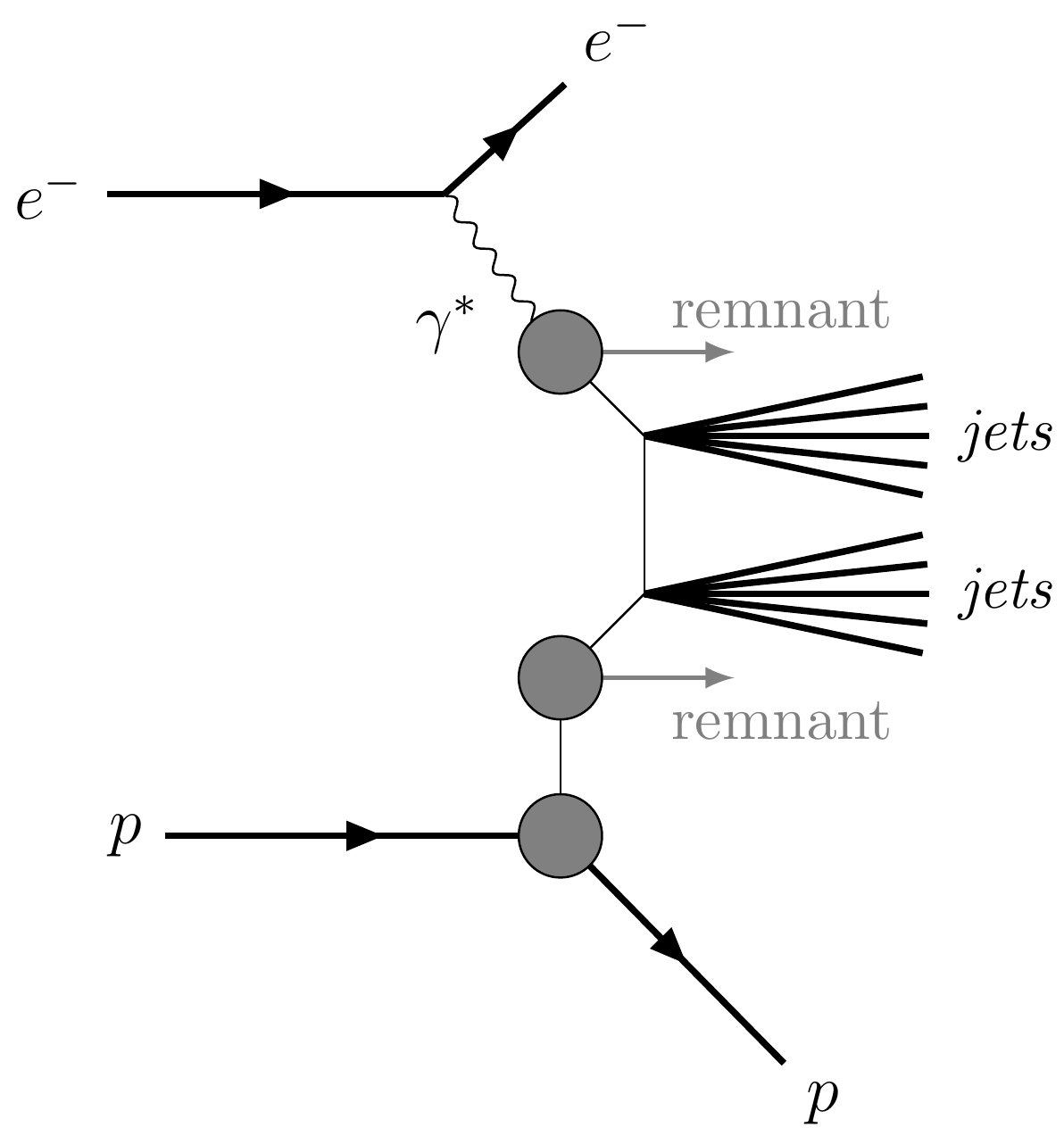} \\
    b. Resolved Photoproduction Process \\
    \caption{Direct~(a) and Resolved~(b) Photoproduction Processes}
    \label{fig:RandD}
\end{figure}

There  have been a lot of recent studies on the use of groomed jet substructure observables such as SoftDrop, $z_g$, $R_g$ and  Collinear Drop jet mass show some improvement in decreasing the non-perturbative disturbances, however they are prone to mistag when there is a large background, as in the case of ion collisions~\cite{mulligan2020identifying}.

Using the differences observed in integrated jet shapes between quark and gluon jet subsamples, selection cuts have been devised for the future EIC experiment that can be useful for producing quark-enriched and gluon-enriched jet samples.~Investigating the internal structure of jets provides insight into the transition from a parton produced in a hard process to the experimentally observable spray of hadrons~\cite{salam2010towards, klaus2018jet}.~ The Jet substructure~\cite{Marzani_2019, jet_subst, Frixione1997jet, chekanov2008three, PhysRevD.107.116002,zeus1998measurement} can be probed to understand the underlying physics of the collision.~The number of jet-like substructures resolved within jets by reapplying the jet algorithm at a smaller resolution scale $y_{\text{cut}}$ is known as subjet multiplicity. In this paper, a systematic comparison of subjet multiplicities in quark and gluon jets is presented using photoproduction processes at different EIC energies.

The results are useful for comparison with data and for constraining non-perturbative effects through improved event generator tuning.~At the EIC, jets are expected to serve as precise probes for initial data collection from lepton-nucleus collisions~\cite{Arratia_2020}, anticipated to yield cleaner results than those from hadron beam experiments.~This highlights the importance of the presented analysis, as well as to investigate the feasibility of jet classification and production of quark-enriched and gluon-enriched jet samples in photoproduction processes at various EIC energies.

This paper is structured as follows.~Section~2 covers event generation and the event sample selection procedure.~Section~3 introduces the jet algorithm used for jet reconstruction and examines jet shape observables, namely differential jet shape, integrated jet shape, and subjet multiplicity. The use of these observables to discriminate quark and gluon jets is described in Section~4. An outlook on the EIC experiment in view of this analysis is discussed in Section~5, followed by conclusion in Section~6.

\section{Event Generation and Selection}\label{EvtGen}
The Monte Carlo event generator PYTHIA v8.309~\cite{10.21468/SciPostPhysCodeb.8} is used to simulate $ep$ photoproduction events at low four-momentum transfers, with $10^{-5} < Q^{2} < 1.0\,\text{ GeV}^{2}$. For the simulations, we have used the default Monash tune~\cite{skands2014tuning} in PYTHIA, with NNPDF2.3 proton PDFs~\cite{deans2013progress} and CJKL photon PDFs~\cite{cornet2003new} and have allowed the Multi-parton Interactions.
\begin{table}[htbp!]
\caption{Beam energies of colliding particles, the corresponding center-of-mass, $\sqrt{s}$ energies and the number of events generated for simulating collisions at the EIC and HERA.}\label{table:COMval}
\small
\centering
\renewcommand{\arraystretch}{1.2}
\begin{tabular*}{\columnwidth}{@{\extracolsep{\fill}}ccc@{}}
\toprule
\textbf{Beam energies (GeV)} & \textbf{$\sqrt{s}$ (GeV)} & \textbf{Events} \\
\midrule
EIC (10, 100) & 63.2 & 998,511 \\
EIC (10, 275) & 104.9 & 999,444 \\
EIC (18, 275) & 141 & 999,633 \\
HERA (27.5, 820) & 300 & 999,881 \\
\bottomrule
\end{tabular*}
\end{table}

At leading order (LO), photoproduction events can be broadly classified into direct and resolved processes, as shown in Figs.~\ref{fig:RandD}(a) and (b), respectively. Both direct and resolved subprocesses are simulated. In the direct photon-parton interaction, the photon acts as a point-like particle, interacting either with a quark inside the proton via QCD Compton scattering ($\gamma q \rightarrow gq$), or with a gluon via boson-gluon fusion ($\gamma g \rightarrow q\bar{q}$). In the resolved process, the photon exhibits a partonic structure that interacts with partons from the hadron ($qq \rightarrow qq$, $q\bar{q} \rightarrow q\bar{q}$, $q\bar{q} \rightarrow gg$, $gg \rightarrow q\bar{q}$, $gg \rightarrow gg$, $gq \rightarrow gq$; charge conjugates and the heavy-flavor analogues with $c\bar{c}$ or $b\bar{b}$ in the final state included). The resolved processes included in the PYTHIA simulation have subprocess codes 111$-$116 and  121$-$124 for heavy flavors. The direct boson–gluon fusion processes have PYTHIA codes 271$-$273, whereas the QCD Compton process is included using 274 code.

The events are generated at three different EIC center-of-mass energies as listed in the Table~\ref{table:COMval}. For the validation of the study and the comparison of the results, events are also simulated for  $\sqrt{s}$=300 GeV corresponding to the published  HERA data~\cite{Chekanov2004-te}. A comparison of the direct and resolved subprocesses generated in the samples at various center-of-mass energies is given in Table~\ref{Table2SubProc}. The resolved processes are dominated by the presence of a quark jet and a gluon jet in the final state, whereas in the direct process two quark jets in the final state dominate. The percentage of direct events increases from about 26\% at $\sqrt{s}$=300 GeV to about 50\% at $\sqrt{s}$=64 GeV.

\begin{table}[hbtp!]
\centering
\caption{Photoproduction cross sections and subprocess contributions for different center-of-mass energies. Only the dominant subprocess contributions are shown for clarity.}
\label{Table2SubProc}
\begin{tabular}{@{}l@{\hspace{6em}}c@{}}
\toprule
\textbf{Subprocess} & \textbf{Contribution} \\
\midrule
\multicolumn{2}{c}{\textbf{EIC 64 GeV}} \\
\midrule
\textbf{Resolved} & \textbf{3.23 nb, 50.2 \%} \\
$qg \rightarrow qg$ & 27.09 \% \\
$qq \rightarrow qq$ & 16.47 \% \\
$gg \rightarrow gg$ & 5.42 \% \\
\textbf{Direct} & \textbf{3.21 nb, 49.8 \%} \\
$q\gamma \rightarrow qg$ & 26.73 \% \\
$\gamma g \rightarrow q\bar{q}$ & 14.29 \% \\
\midrule
\multicolumn{2}{c}{\textbf{EIC 105 GeV}} \\
\midrule
\textbf{Resolved} & \textbf{13.63 nb, 62.6 \%} \\
$qg \rightarrow qg$ & 35.26 \% \\
$qq \rightarrow qq$ & 15.81 \% \\
$gg \rightarrow gg$ & 10.33 \% \\
\textbf{Direct} & \textbf{8.16 nb, 37.4 \%} \\
$q\gamma \rightarrow qg$ & 16.24 \% \\
$\gamma g \rightarrow q\bar{q}$ & 12.91 \% \\
\midrule
\multicolumn{2}{c}{\textbf{EIC 141 GeV}} \\
\midrule
\textbf{Resolved} & \textbf{5.71 nb, 61.4 \%} \\
$qg \rightarrow qg$ & 34.66 \% \\
$qq \rightarrow qq$ & 16.62 \% \\
$gg \rightarrow gg$ & 8.92 \% \\
\textbf{Direct} & \textbf{3.59 nb, 38.6 \%} \\
$q\gamma \rightarrow qg$ & 16.80 \% \\
$\gamma g \rightarrow q\bar{q}$ & 12.74 \% \\
\midrule
\multicolumn{2}{c}{\textbf{HERA 300 GeV}} \\
\midrule
\textbf{Resolved} & \textbf{33.67 nb, 74.4 \%} \\
$qg \rightarrow qg$ & 42.36 \% \\
$gg \rightarrow gg$ & 17.37 \% \\
$qq \rightarrow qq$ & 13.48 \% \\
\textbf{Direct} & \textbf{11.61 nb, 25.6 \%} \\
$\gamma g \rightarrow q\bar{q}$ & 10.15 \% \\
$q\gamma \rightarrow qg$ & 8.10 \% \\
\bottomrule
\end{tabular}
\end{table}
To validate the analysis procedure, photoproduction events were simulated at HERA energies and compared with published HERA results~\cite{Chekanov2004-te}. The data were best described with $p_{T0}^{\text{ref}} = 3.0$~GeV/$c$, where $p_{T0}^{\text{ref}}$ is a key parameter in the PYTHIA tune, closely tied to choices such as colour flow description~\cite{10.21468/SciPostPhysCodeb.8}. This value lies between those optimal for $\gamma\gamma$ and $pp$ collisions, reflecting the photon's cleaner partonic structure compared to the proton, or indicating that more sophisticated energy scaling of $p_{T0}(\sqrt{s})$ may be required.

The parameter \texttt{PhaseSpace:pTHatMin} sets the minimum partonic transverse momentum in the hard $2\to2$ scattering process before parton showering and hadronization, serving as an infrared cutoff to avoid regions where perturbative QCD becomes unreliable.~The PYTHIA photoproduction examples use \texttt{PhaseSpace:pTHatMin} = 5.0~GeV as a standard choice for hard scattering studies~\cite{10.21468/SciPostPhysCodeb.8}.~The present analysis uses \texttt{PhaseSpace:pTHatMin} in the range 3-7~GeV to access softer processes relevant for jet shape studies.~For the HERA energy, \texttt{PhaseSpace:pTHatMin} = 7~GeV was used.~Both $p_{T0}^{\text{ref}}$ and \texttt{PhaseSpace:pTHatMin} should be tuned for future EIC studies when comparisons with real data become available.
\section{Jets}
In a particle collision, a jet is a collimated spray of hadrons produced when a high-energy parton fragments and hadronizes.~Jets are formed from these final-state hadrons and typically consist of a large number of particles traveling in roughly the same direction. Several jet algorithms have been developed over the years, each possessing its own strengths and weaknesses.~These include sequential clustering algorithms such as $k_T$, anti-$k_T$, and Cambridge/Aachen as well as traditional cone algorithms.~A detailed comparison of these different jet algorithms is provided in Refs.~\cite{chekanov2002jet}.

In this study, jets are reconstructed using the longitudinally invariant $k_T$ sequential recombination algorithm \cite{Catani:1993hr}, implemented with FastJet v3.4.0~\cite{Cacciari_2012}, with jet radius $R = 1.0$. The formulation of the longitudinally invariant $k_T$ jet algorithm is summarized as follows:
\begin{itemize}[leftmargin=*]
\item In the traditional $k_T$ algorithm, clustering is based on both transverse momentum and the distance in the pseudorapidity-azimuthal angle space. However, this definition makes the algorithm sensitive to boosts along the longitudinal direction. To overcome this limitation, the longitudinally invariant $k_T$ algorithm~\cite{Catani:1993hr} modifies the distance measure to ensure independence from the longitudinal direction. It specifically employs the following distance measures:
\begin{equation}
    d_{ij} = \min(p_{T,i}^{2}, p_{T,j}^{2}) \frac{\Delta R_{ij}^{2}}{R^{2}}
\end{equation}
\begin{equation}
    d_{iB} = p_{T,i}^{2}
\end{equation}
where $p_{T,i}$ and $p_{T,j}$ are the transverse momenta of particles $i$ and $j$, $\Delta R_{ij}^2 = (y_i - y_j)^2 + (\phi_i - \phi_j)^2$ is the squared distance in the rapidity-azimuthal angle plane, $R$ is the jet radius parameter, and $d_{iB}$ is the beam distance.
\item The algorithm iteratively finds the minimum distance $d_{\min}$ among all $d_{ij}$ and $d_{iB}$. If the minimum is $d_{ij}$, particles $i$ and $j$ are merged into a single cluster. If the minimum is $d_{iB}$, particle $i$ is declared a jet and removed from the list.
\end{itemize}
The analysis focuses on two-jet events, where the jet with the highest transverse energy $E_T$ is classified as the leading jet, and the jet with the second-highest $E_T$ is designated as the associated jet. The selection criteria require the leading jet to have $E_T > 10$~GeV and the associated jet to have $E_T > 7$~GeV.
\begin{table}[hbtp!]
\centering
\caption{Event counts and fraction information for photoproduction at different center-of-mass energies. Event data represents a sample of two-jet events from PYTHIA~8 Monte Carlo simulation output.}
\label{tab:event_cross_section_summary}
\renewcommand{\arraystretch}{1.1}
\begin{tabular*} {\columnwidth}{@{\extracolsep{\fill}}ccc@{}}
\toprule 
\textbf{Event Type} & \textbf{Count} & \textbf{Fraction (\%)} \\
\midrule
\multicolumn{3}{c}{\textbf{EIC 64 GeV}} \\
\midrule
QQ Events & 401,596 & 40.22 \\
GG Events & 58,928 & 5.90 \\
GQ Events & 537,987 & 53.88 \\
\midrule
\multicolumn{3}{c}{\textbf{EIC 105 GeV}} \\
\midrule
QQ Events & 378,245 & 37.84 \\
GG Events & 107,066 & 10.71 \\
GQ Events & 514,133 & 51.45 \\
\midrule
\multicolumn{3}{c}{\textbf{EIC 141 GeV}} \\
\midrule
QQ Events & 391,413 & 39.16 \\
GG Events & 93,259 & 9.33 \\
GQ Events & 514,961 & 51.51 \\
\midrule
\multicolumn{3}{c}{\textbf{HERA 300 GeV}} \\
\midrule
QQ Events & 318,964 & 31.90 \\
GG Events & 176,498 & 17.65 \\
GQ Events & 504,451 & 50.45 \\
\bottomrule
\end{tabular*}
\end{table}
\begin{figure}[ht]
\resizebox{4.5cm}{!}{\includegraphics{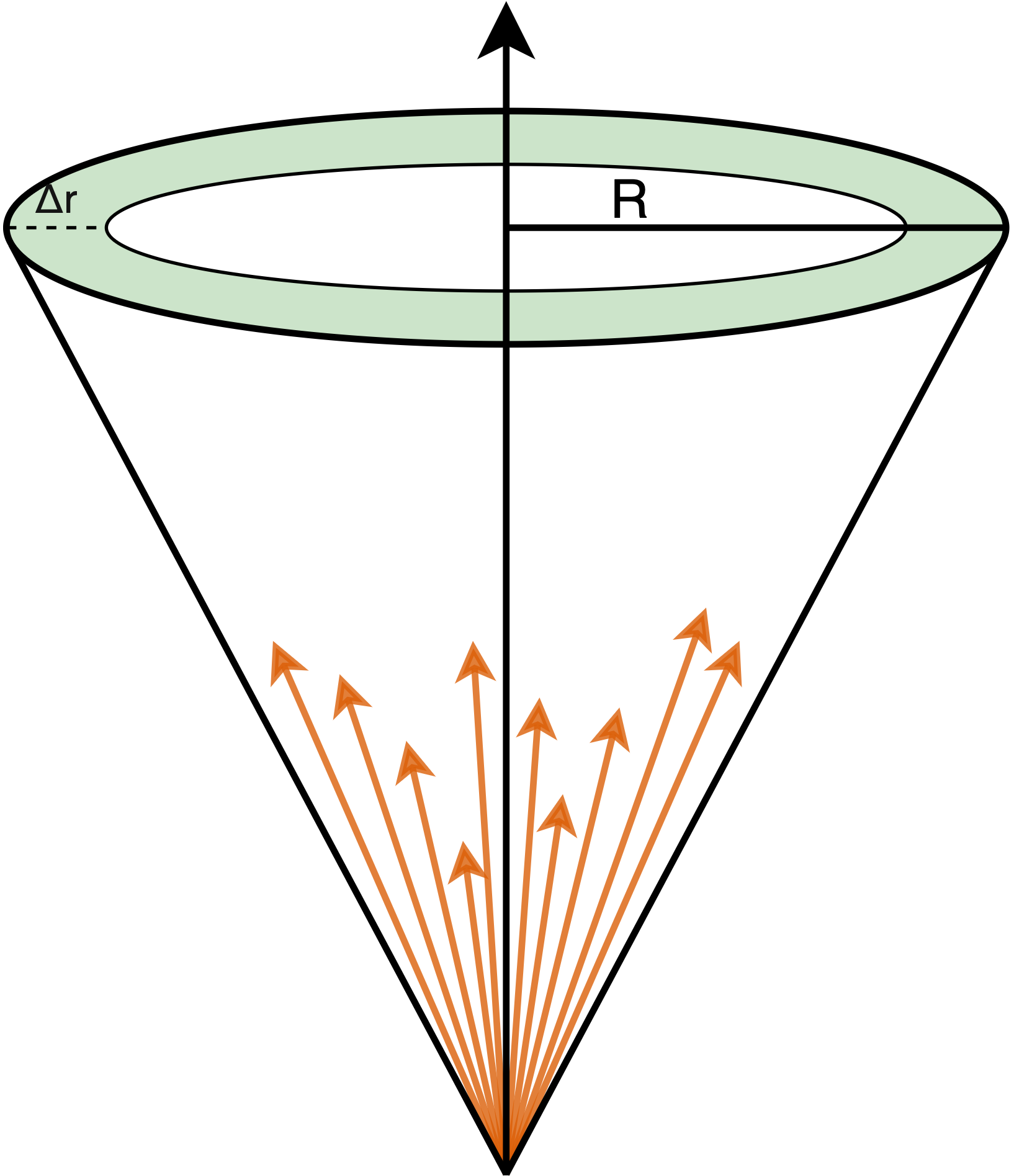}}
\caption{The differential jet shape is the average fraction of the jet’s transverse energy $E_{T,jet}$ contained inside an annulus of radius $r$, centred around the jet-axis.}
\label{fig:shapejet}
\end{figure}
\begin{figure*}[htbp!]
\centering
\includegraphics[width=0.45\textwidth]{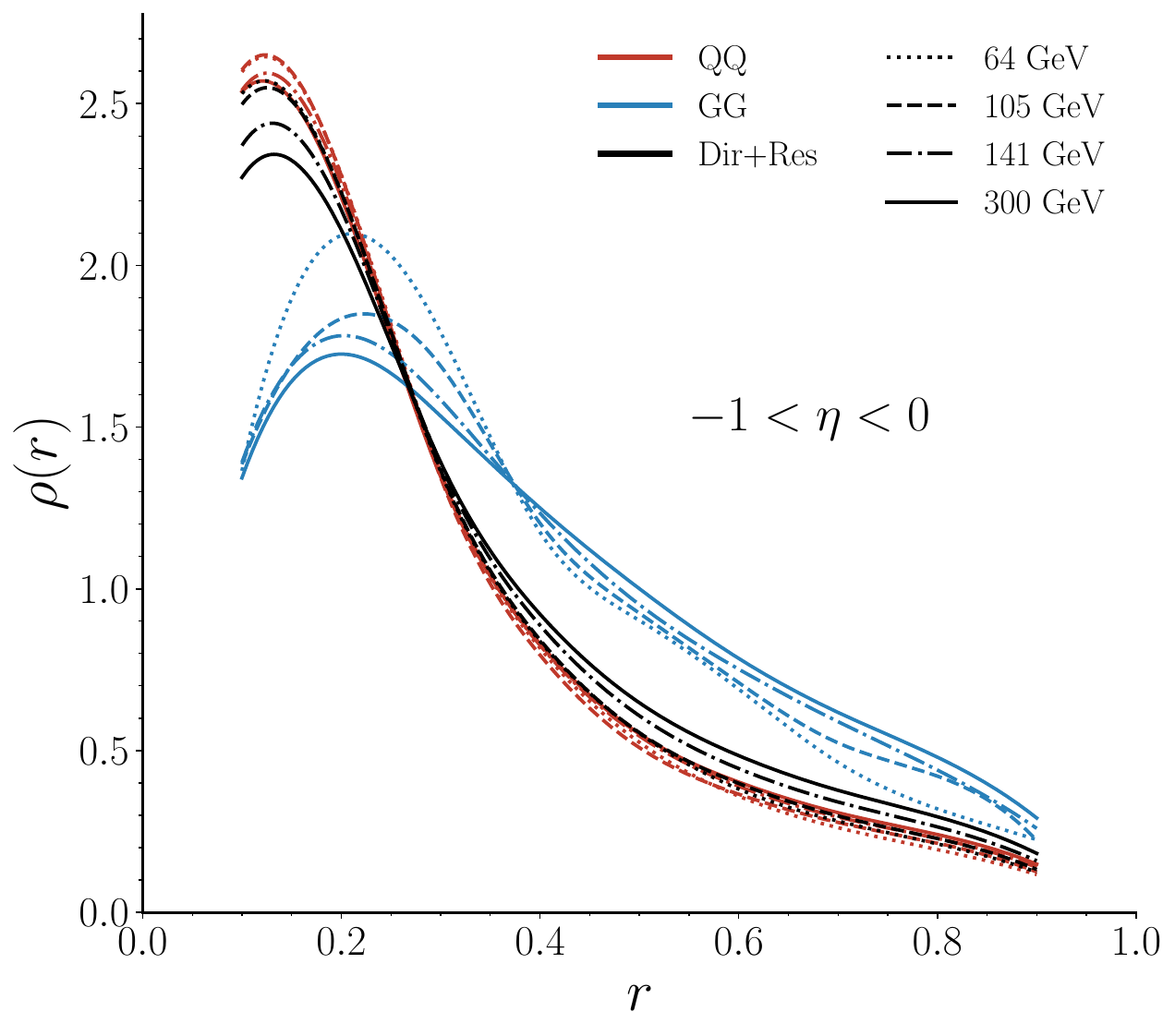}
\hfill
\includegraphics[width=0.45\textwidth]{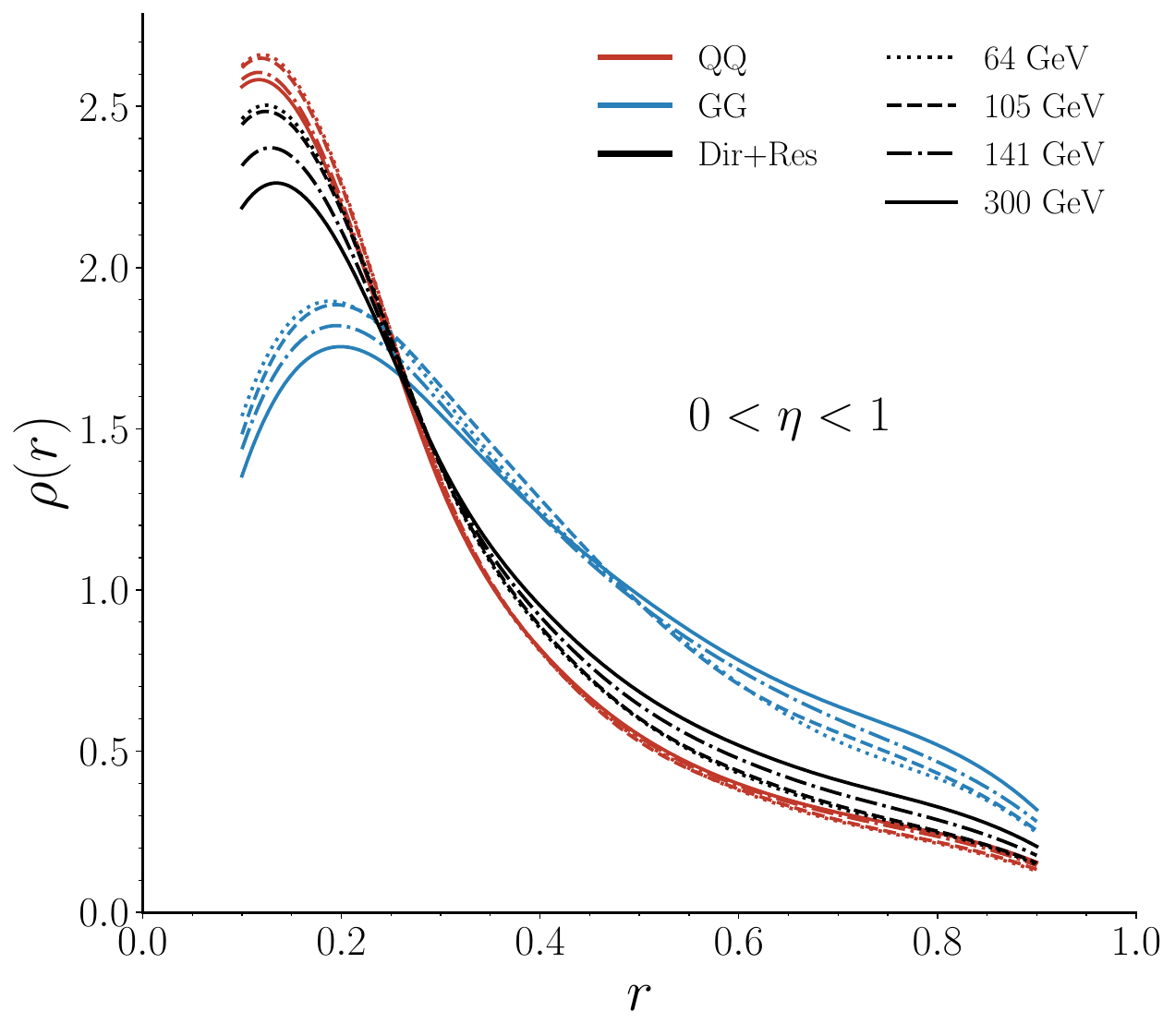}\\[6pt]
\includegraphics[width=0.45\textwidth]{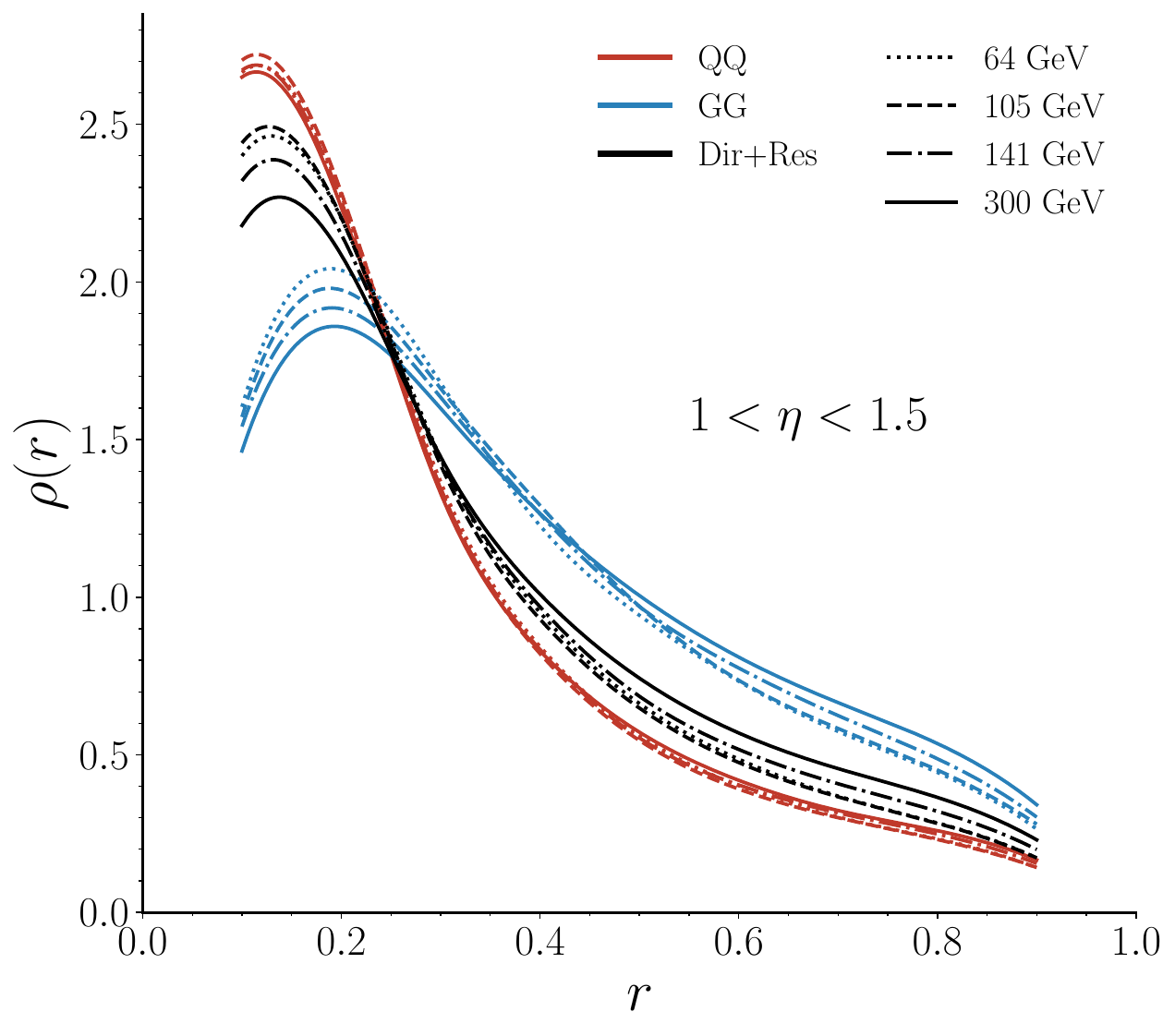}
\hfill
\includegraphics[width=0.45\textwidth]{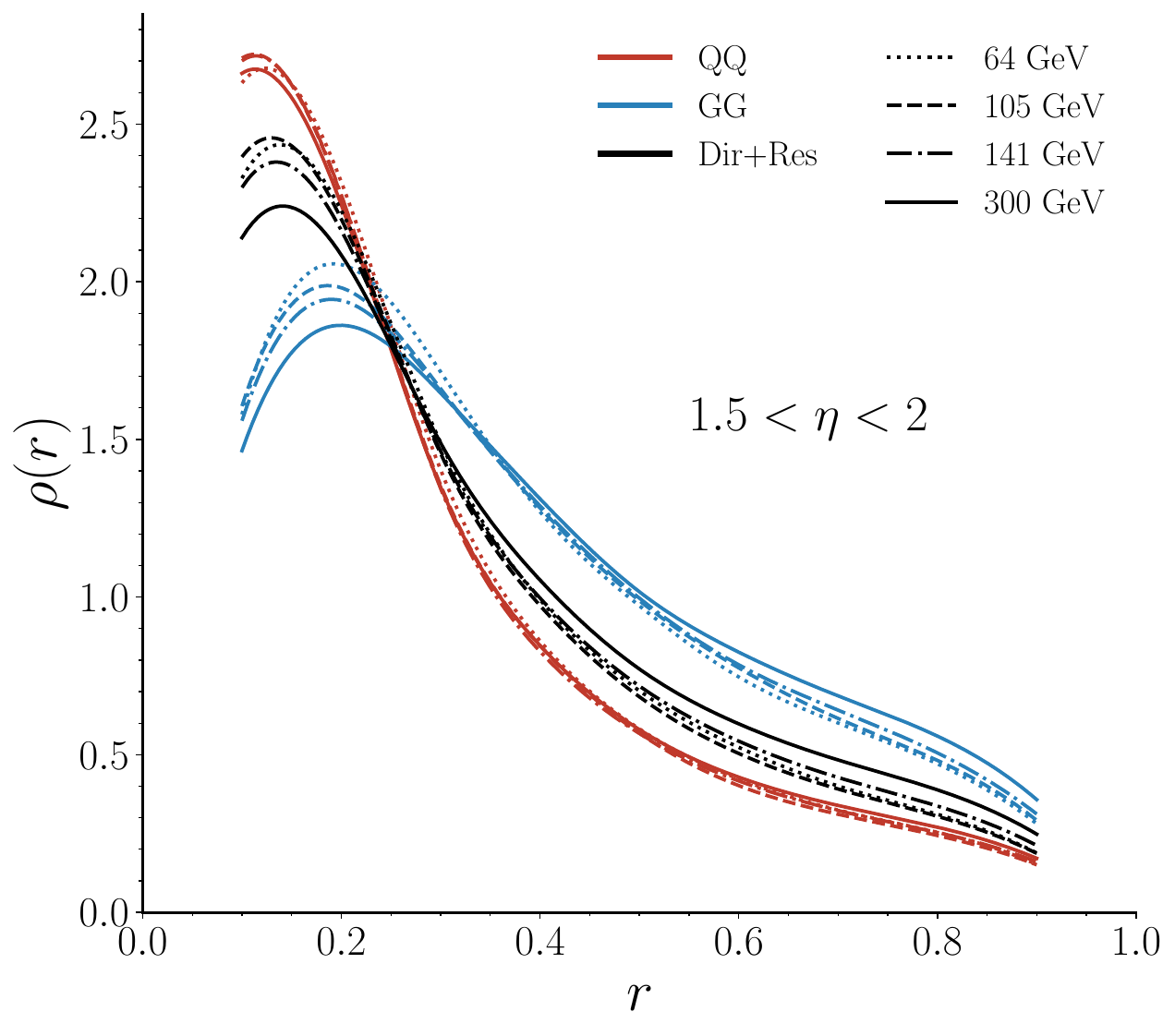}
\caption{Differential jet shape distribution $\rho(r)$ for the combined Direct+Resolved processes (black) as well as the GG (blue) 
and QQ (red) dijet photoproduction subsamples at different 
$ep$ energies in four pseudorapidity bins: 
$-1 < \eta < 0$ (top left), $0 < \eta < 1$ (top right), 
$1 < \eta < 1.5$ (bottom left), and $1.5 < \eta < 2$ (bottom right).}
\label{fig:diff_jet_shape_all}
\end{figure*}
\begin{figure*}[hbtp!]
\centering
\includegraphics[width=0.45\textwidth]{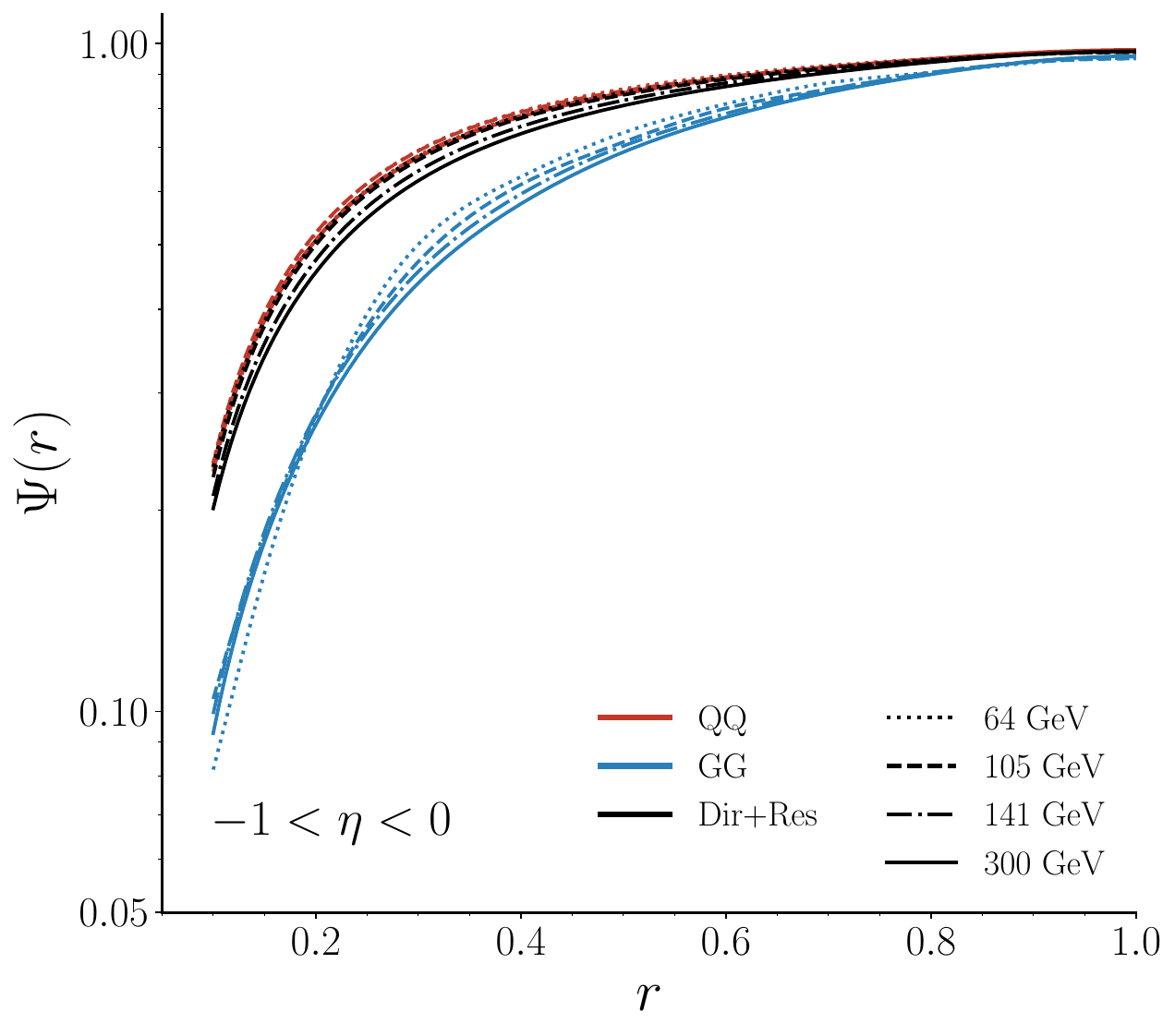}
\hfill
\includegraphics[width=0.45\textwidth]{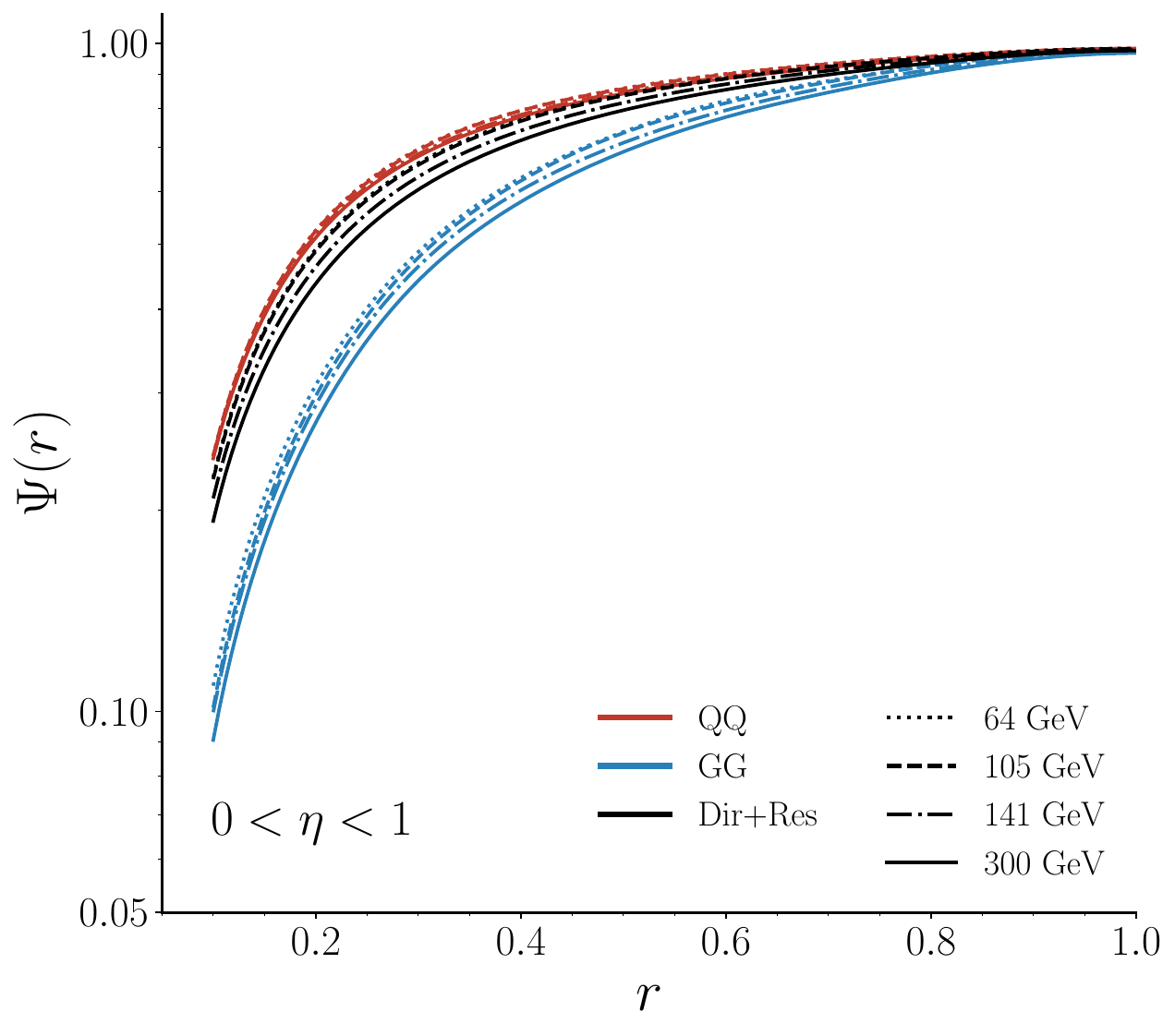}\\[6pt]
\includegraphics[width=0.45\textwidth]{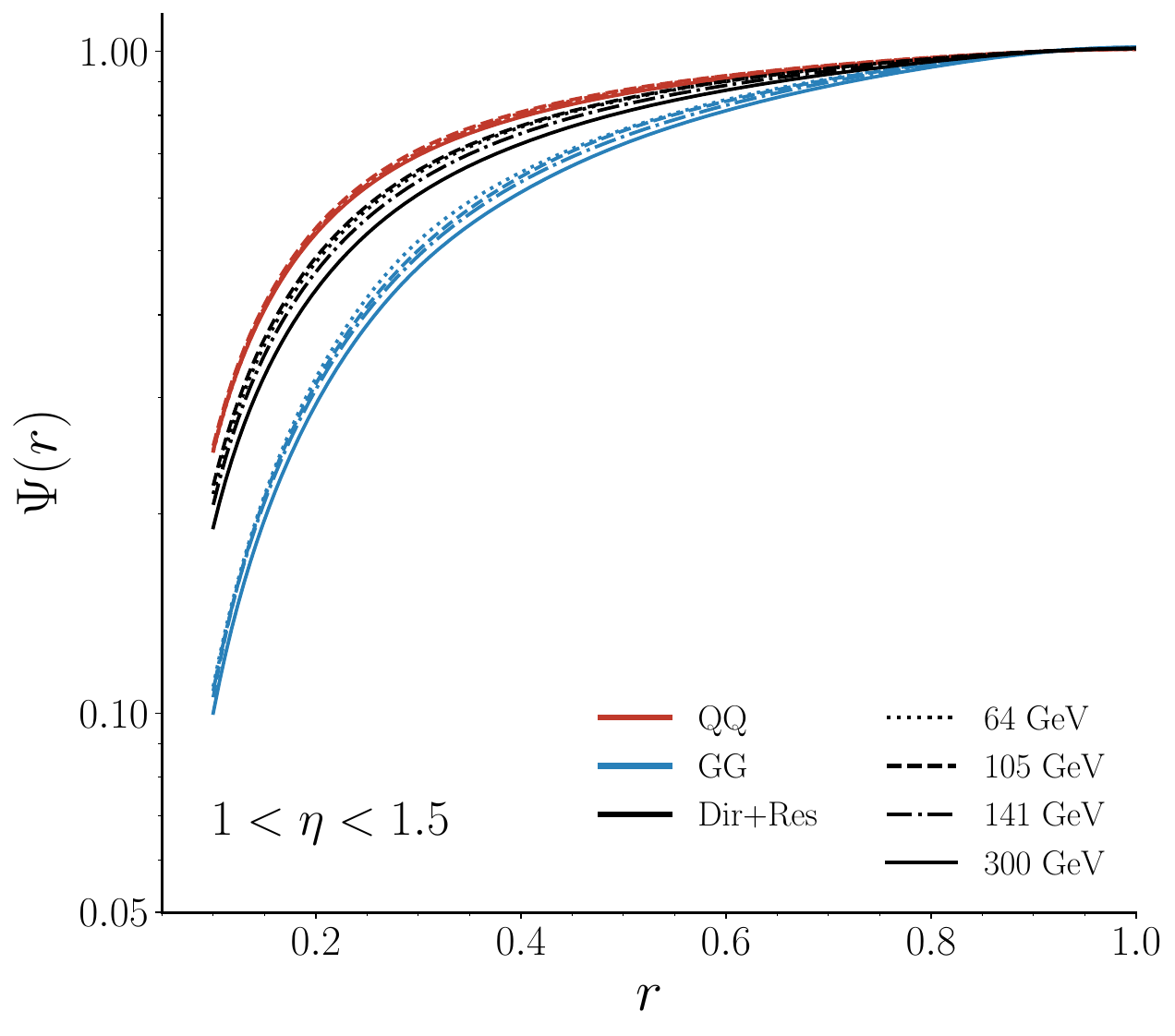}
\hfill
\includegraphics[width=0.45\textwidth]{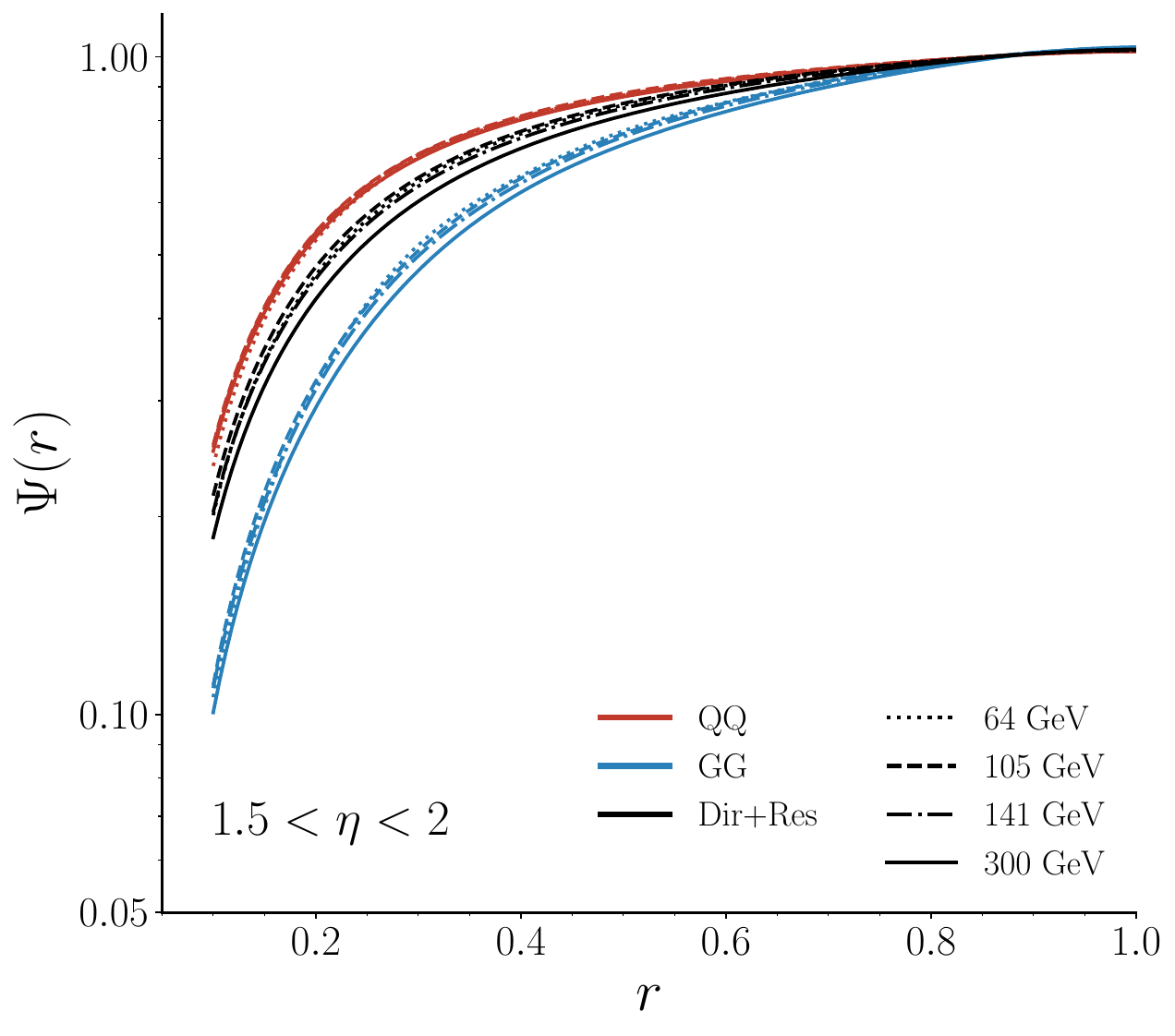}
\caption{Integrated jet shape distribution $\Psi(r)$ for the Direct and Resolved processes (black) as well as the GG (blue) 
and QQ (red) dijet photoproduction subsamples at different 
$ep$ energies in four pseudorapidity bins: 
$-1 < \eta < 0$ (top left), $0 < \eta < 1$ (top right), 
$1 < \eta < 1.5$ (bottom left), and $1.5 < \eta < 2$ (bottom right).}
\label{fig:int_jet_shape_all}
\end{figure*}
\begin{figure*}[hbtp!]
\includegraphics[width=0.45\textwidth]{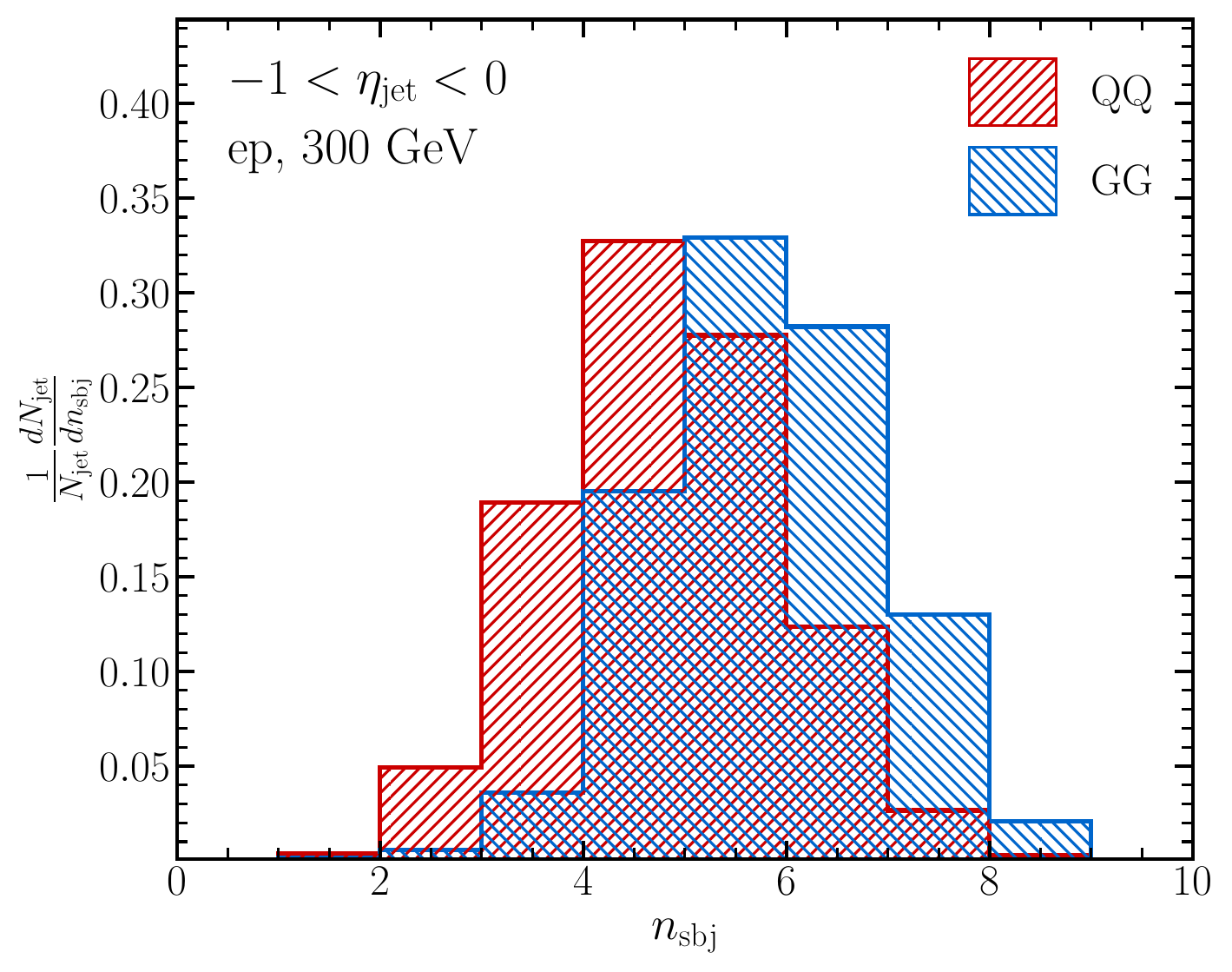}
\hfill
\includegraphics[width=0.45\textwidth]{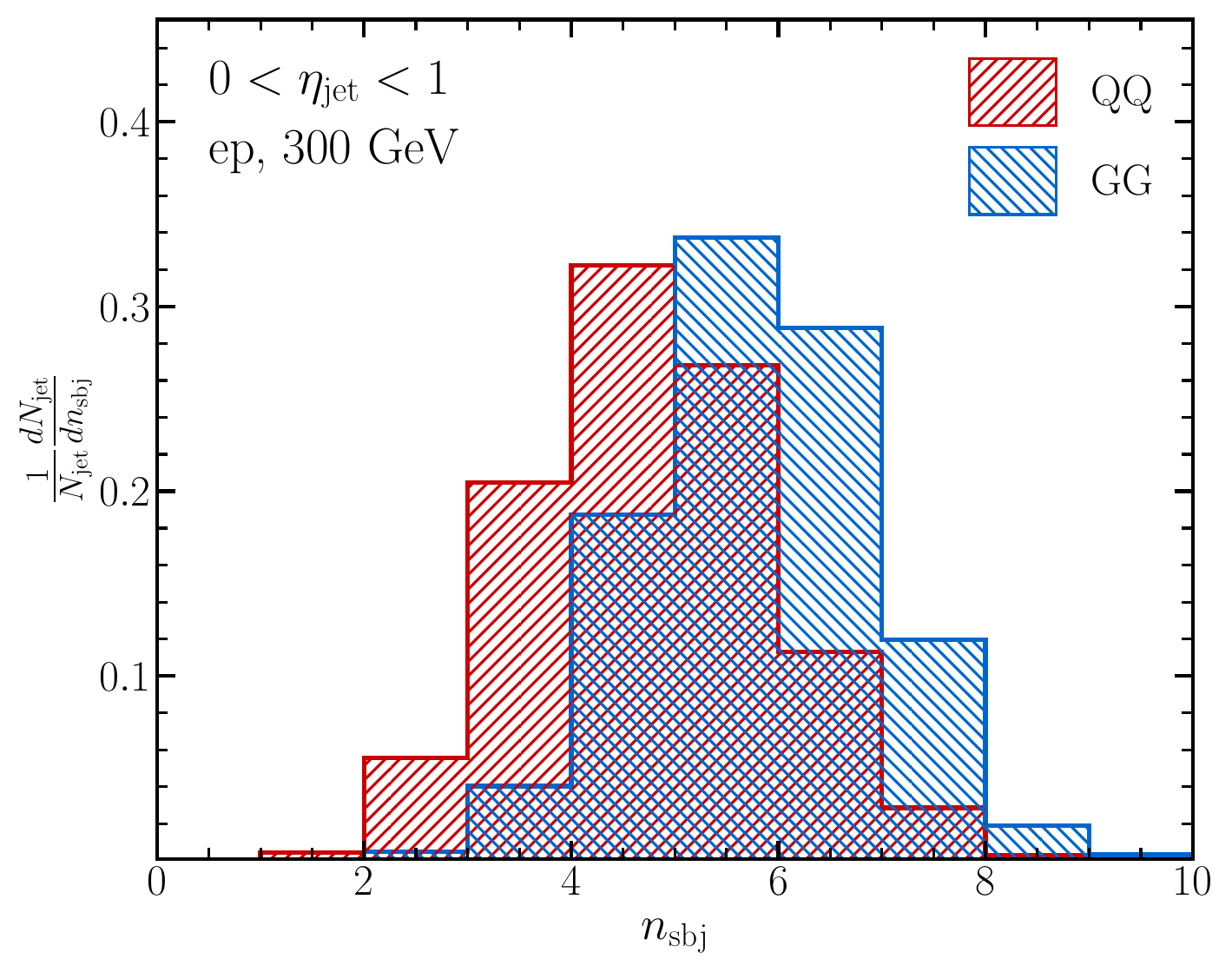}\\[6pt]
\includegraphics[width=0.45\textwidth]{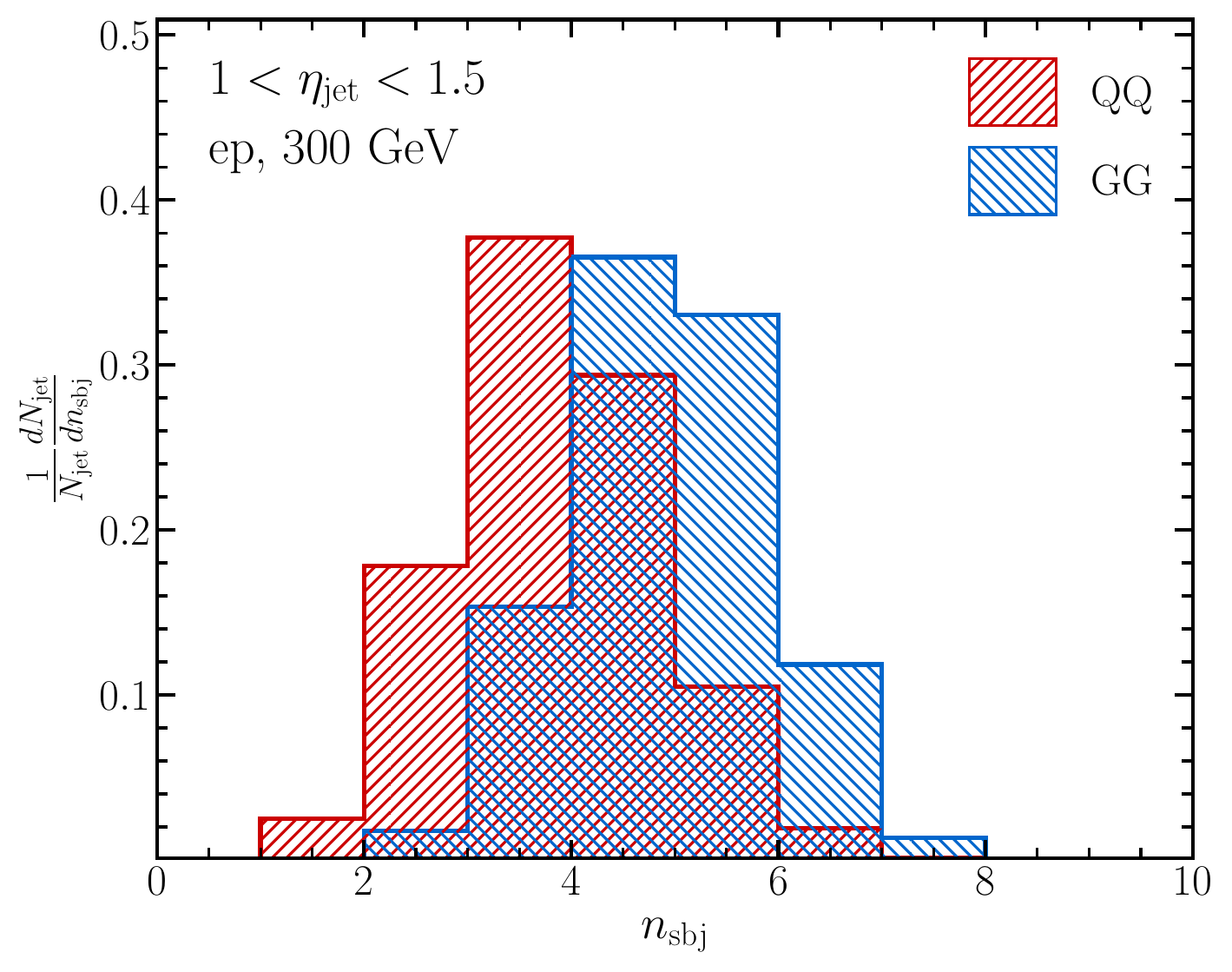}
\hfill
\includegraphics[width=0.45\textwidth]{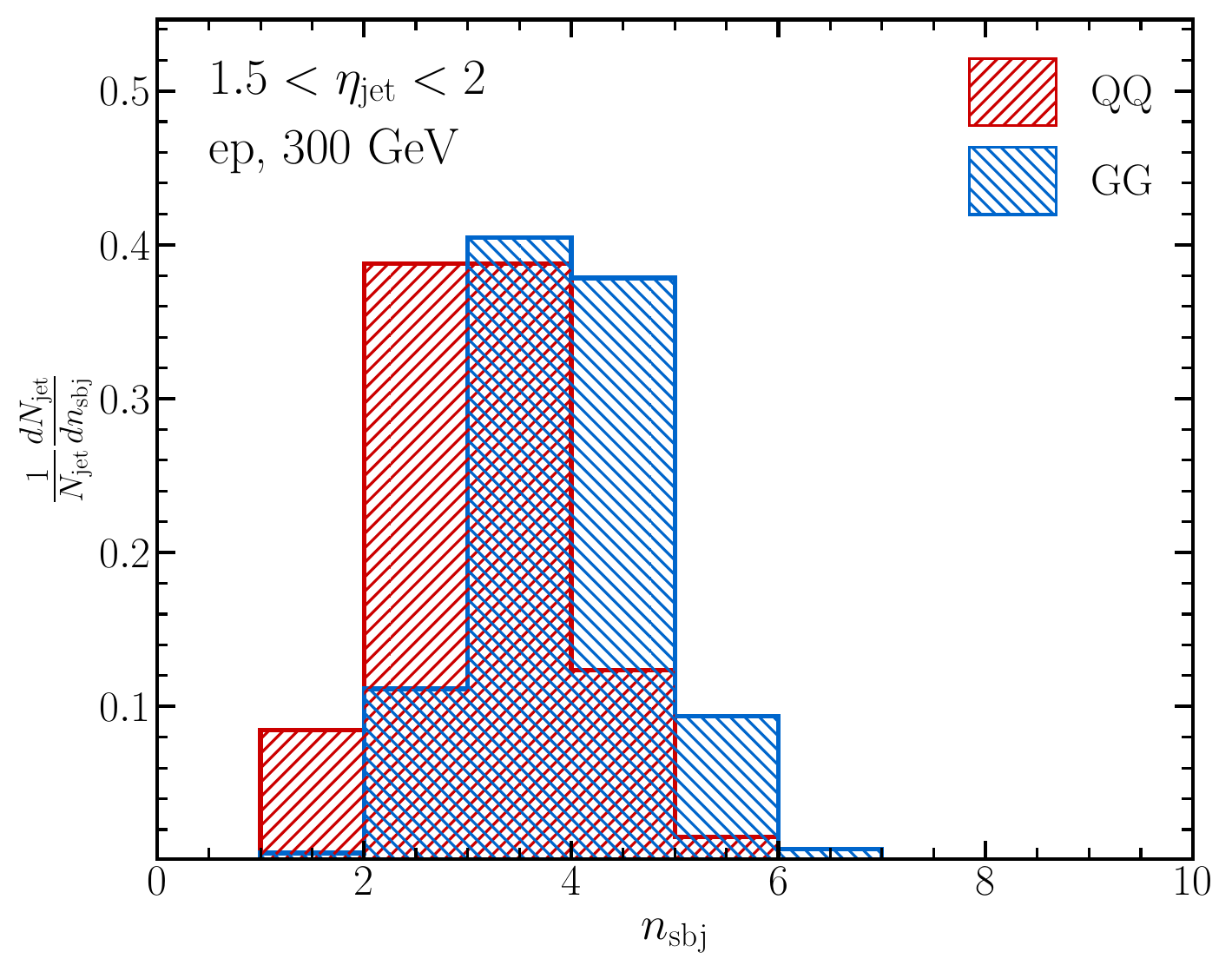}
\caption{Subjet multiplicities for quark and gluon initiated jets 
for HERA ($ep$, $\sqrt{s} = 300$ GeV) in four pseudorapidity bins: 
$-1 < \eta < 0$ (top left), $0 < \eta < 1$ (top right), 
$1 < \eta < 1.5$ (bottom left), and $1.5 < \eta < 2$ (bottom right).}
\label{fig:sbjmul_hera300}
\end{figure*}
\begin{figure*}[htbp!]
\centering
\includegraphics[width=0.45\textwidth]{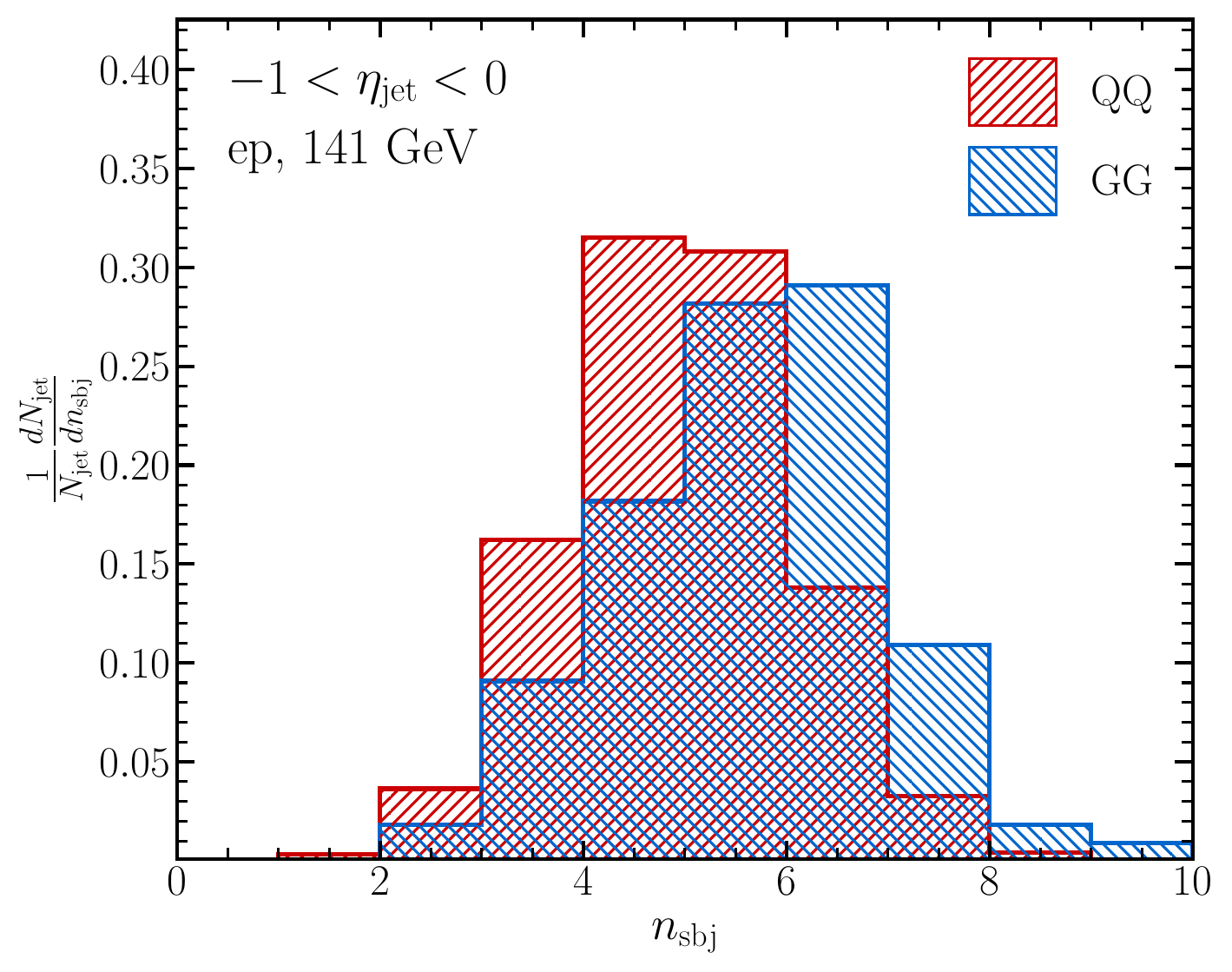}
\hfill
\includegraphics[width=0.45\textwidth]{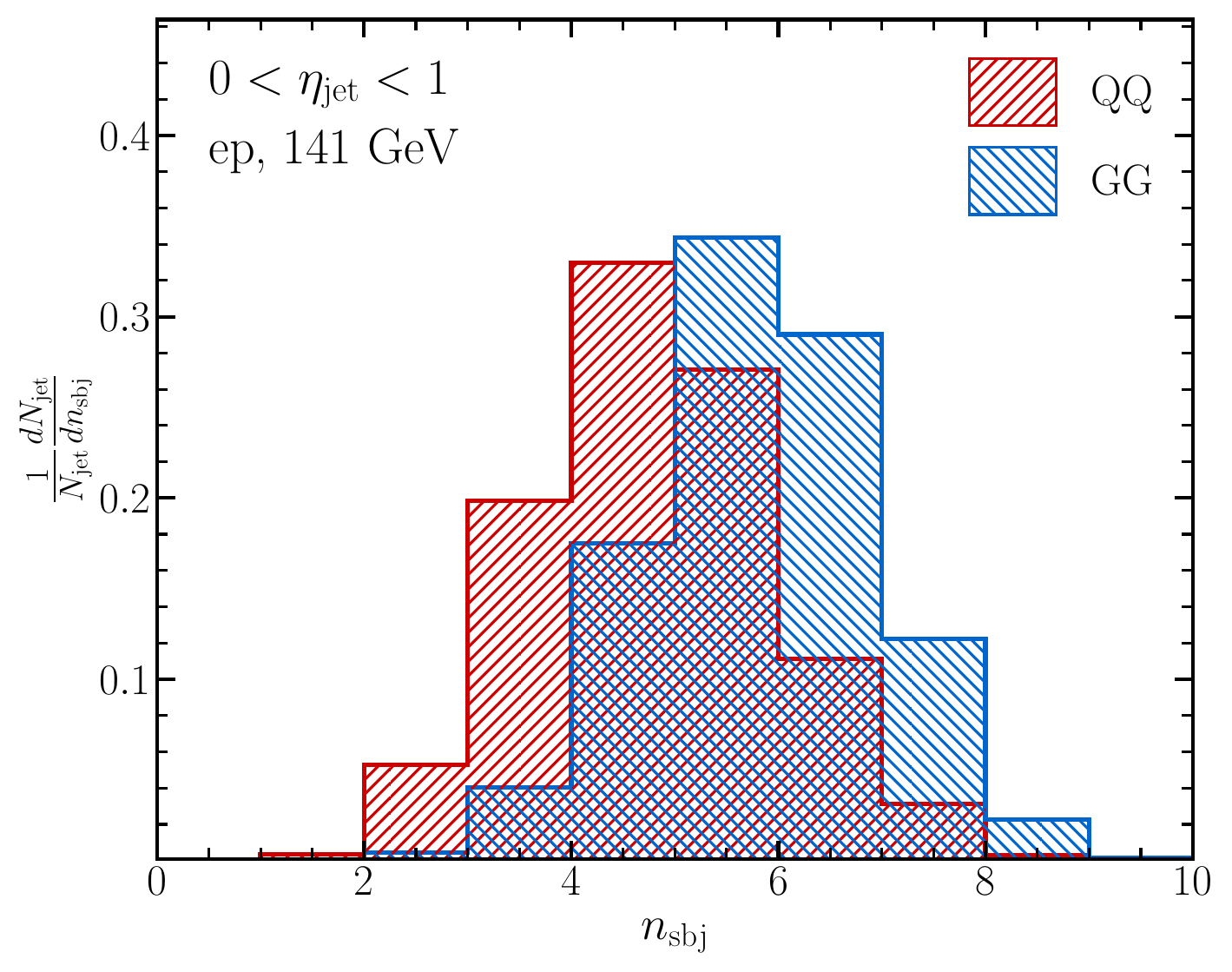}\\[6pt]
\includegraphics[width=0.45\textwidth]{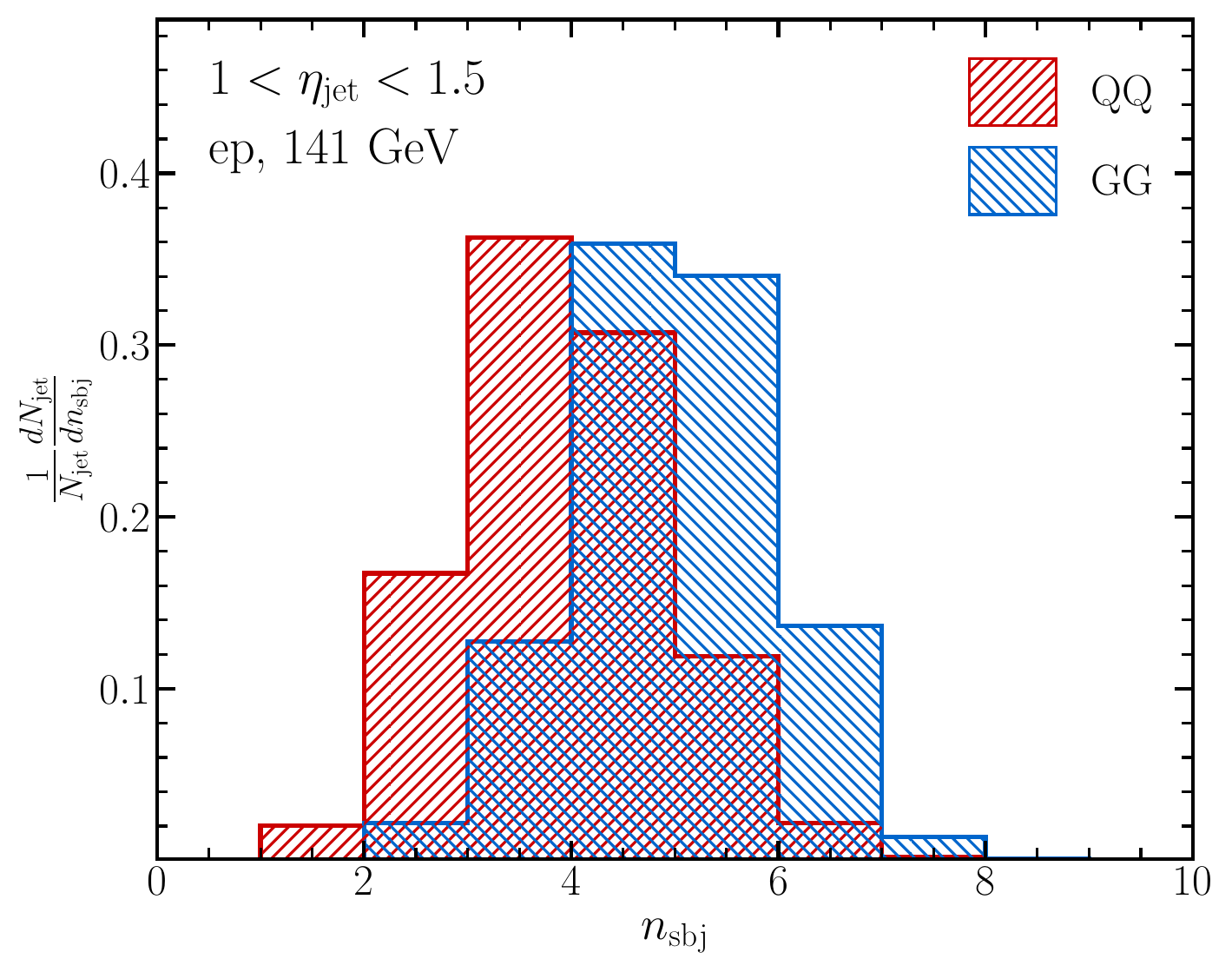}
\hfill
\includegraphics[width=0.45\textwidth]{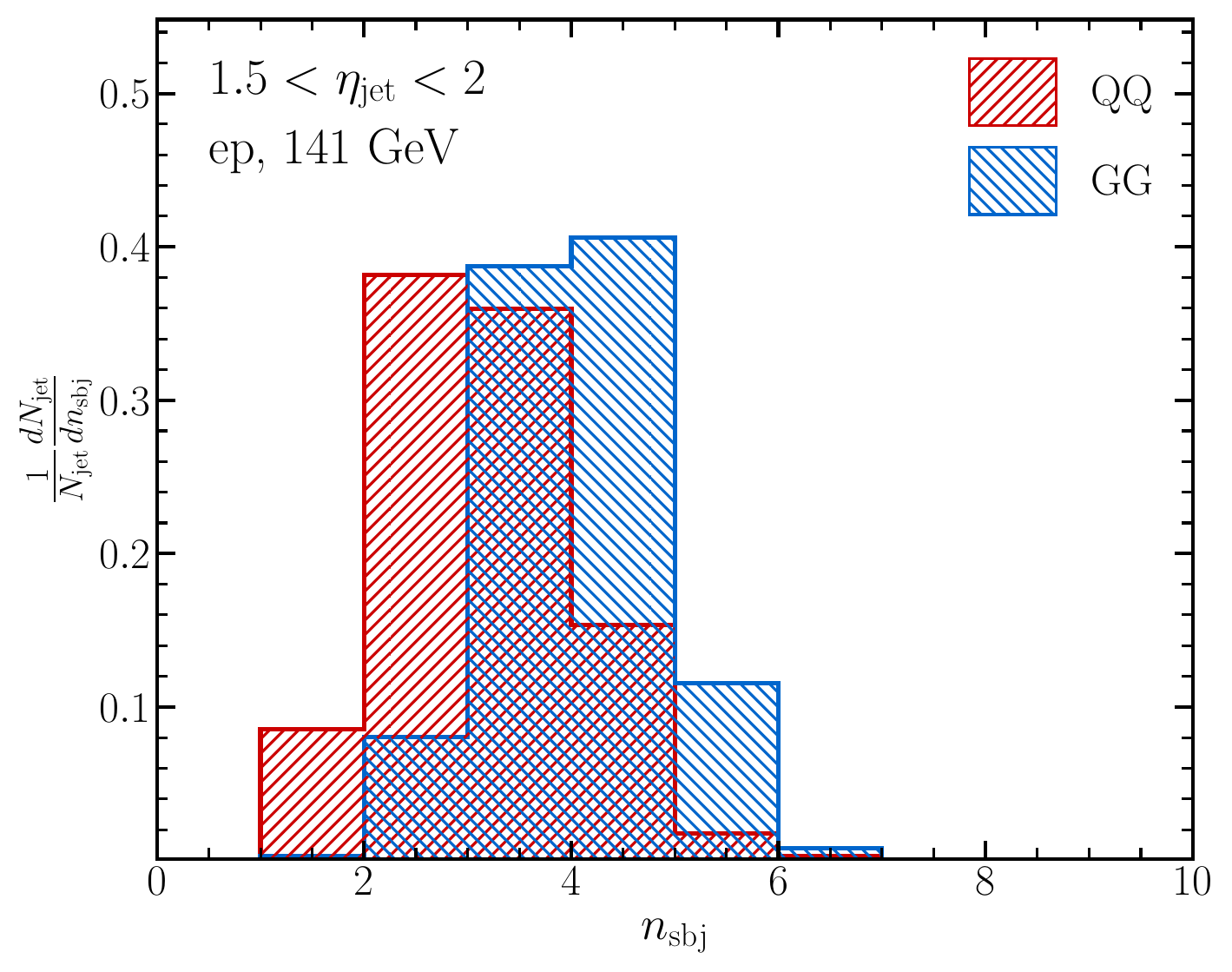}
\caption{Subjet multiplicities for quark and gluon initiated jets 
for EIC ($ep$, $\sqrt{s} = 141$ GeV) in four pseudorapidity bins: 
$-1 < \eta < 0$ (top left), $0 < \eta < 1$ (top right), 
$1 < \eta < 1.5$ (bottom left), and $1.5 < \eta < 2$ (bottom right).}
\label{fig:sbjmul_eic141}
\end{figure*}

Based on the parton-level information from PYTHIA, the two-jet final states in the generated samples are classified into three categories: QQ events (both jets initiated by quarks), GG events (both jets initiated by gluons), and QG events (one quark jet and one gluon jet). 

To be explicit, the $\mathrm{QQ}$ class is built from the resolved processes with PYTHIA subprocess codes 114 ($qq^{\prime} \to qq^{\prime}$), 116 ($q\bar{q} \to q^{\prime}\bar{q}^{\prime}$), 112 ($gg \to q\bar{q}$) and the heavy-flavor analogues 121--124 ($gg,\,q\bar{q} \to c\bar{c},\,b\bar{b}$), together with the direct boson--gluon fusion processes  271--273 ($\gamma g \to q\bar{q}, ,c\bar{c},\,b\bar{b}$). The $\mathrm{GG}$ class is built from the resolved processes 111 ($gg \to gg$) and 115 ($q\bar{q} \to gg$), and the $\mathrm{GQ}$ class from the resolved process 113 ($qg \to qg$) together with the direct QCD-Compton processes 274 ($\gamma q \to qg$).

\begin{figure}[htbp!]
    \centering
    \includegraphics[width=\columnwidth]{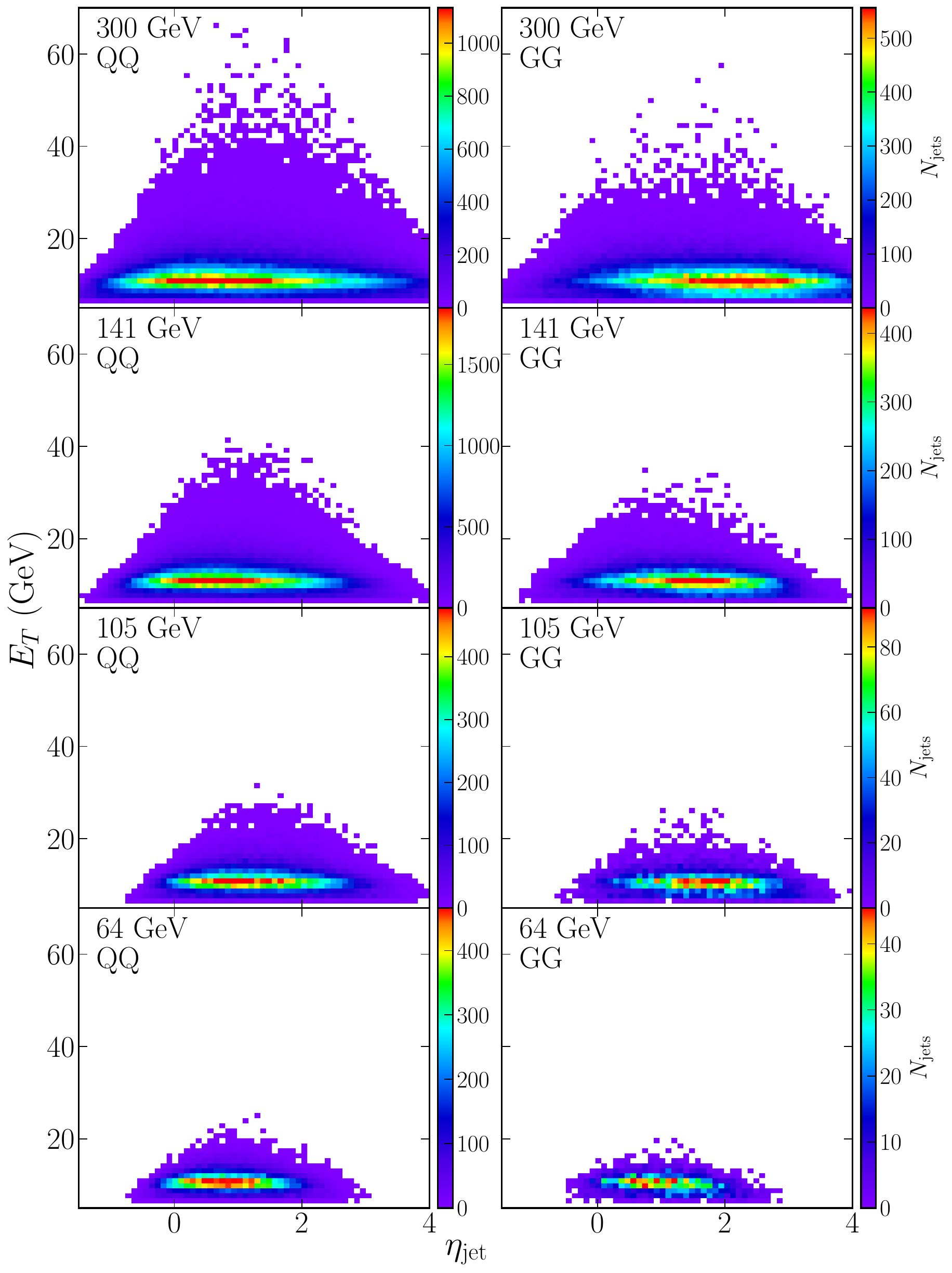} 
    \caption{The $E_{T}$ and $\eta$ distribution of jets in the generated GG and QQ dijet photoproduction subsamples.}
    \label{fig:et_eta}
\end{figure}

The relative fractions of these event types at various center-of-mass energies are summarized in Table~\ref{tab:event_cross_section_summary}. Figure~\ref{fig:et_eta} shows the transverse energy $E_T$ and pseudorapidity $\eta$ distributions for jets in the events required to have exactly two reconstructed jets, for the GG and QQ subsamples.~The properties of jets from these samples are studied in the following subsections, beginning with the energy distribution within jets and progressing to substructure through subjet analysis.

\subsection{Jet properties}
The internal structure of a jet can be characterized using several observables that probe different aspects of the energy and particle distributions within the jet~\cite{Marzani_2019}. Two primary categories of jet shape observables are commonly employed: differential jet shapes, which measure the transverse energy density in annular rings at fixed radial distances from the jet axis, and integrated jet shapes, which quantify the cumulative energy fraction within a cone of given radius. The differential jet shape provides the transverse energy density profile and makes the fundamental physics immediately visible, whereas the integrated jet shape gives the cumulative energy fraction contained within a cone of radius $r$. 
Complementary information is provided by subjet multiplicity, which counts the number of resolved substructures within a jet at a given resolution scale~\cite{PhysRevD.107.116002}.

\subsubsection{Differential Jet Shape}
The differential jet shape is defined as the mean proportion of a jet's transverse energy that is located within an annular region of radii $r\pm\triangle r/2$, concentric with the jet axis in the $\eta$-$\phi$ plane as shown in Figure~\ref{fig:shapejet}. 
\begin{equation}
\rho(r) = \frac{1}{N_{jets}}\frac{1}{\triangle r}\sum_{Jet}\frac{E_T(r-\triangle r/2 , r + \triangle r/2)}{E_T(0, r =1)}
\end{equation}
where, $N_{jets}$ is the number of jets in the sample and $E_T(r-\triangle r/2 , r + \triangle r/2)$ is the transverse energy within the given annulus of thickness, $\triangle r =0.1$. Here, \(r\) is a non-dimensional parameter that facilitates the comparison of jet profiles at different radial distances from the jet-axis.  The differential jet shape distributions $\rho(r)$ for the combined direct and resolved processes as well as GG and QQ photoproduction subsamples in $ep$ collisions across four pseudorapidity bins: $-1 < \eta < 0$, $0 < \eta < 1$, $1 < \eta < 1.5$, and $1.5 < \eta < 2$ are shown in Fig.~\ref{fig:diff_jet_shape_all}.~The smaller $\eta$ region is more populated by the quark-like jets as $\rho(r)$ distribution for combined direct and resolved processes lies closer to the QQ subsample.

It is observed that for quark jets, $\rho(r)$ peaks sharply near the jet axis, indicating that most of the jet's energy is concentrated in a narrow core. In contrast, gluon jets exhibit a flatter $\rho(r)$ distribution, with more energy deposited at larger radii, resulting in a broader jet profile.

This difference arises from the distinct colour charges of quarks and gluons in QCD. The probability of soft gluon emission is proportional to the colour factor: $C_A = 3$ for gluons and $C_F = 4/3$ for quarks. Since $C_A > C_F$, gluons radiate approximately twice as many soft gluons as quarks, leading to a wider spread of energy within gluon-initiated jets. Quark jets, with less radiation, retain a more collimated structure~\cite{Ellis:1991qcd, PDG_QCD}.

This characteristic difference in jet shapes provides a basis for distinguishing quark and gluon jets, as further explored through integrated jet shapes in the next subsection.

\subsubsection{Integrated Jet Shape}
The integrated jet shape $\Psi(r)$ quantifies the cumulative fraction of a jet's transverse energy contained within a cone of radius $r$ centered on the jet axis. Unlike the differential jet shape, which measures energy in an annular ring, the integrated jet shape provides a cumulative energy profile from the jet core outward. It is defined as:
\begin{equation}
\Psi(r) = \frac{1}{N_{\text{jets}}} \sum_{\text{jets}} \Psi_{\text{jet}}(r)
\end{equation}
\begin{equation}
\Psi_{\text{jet}}(r) = \frac{E_T(r)}{E_T(r=1)}
\end{equation}
where $E_T(r)$ is the transverse energy within radius $r$ from the jet axis, $E_T(r=1)$ is the total transverse energy within the jet cone of radius $R=1$, and $N_{\text{jets}}$ is the number of jets in the sample. By definition, $\Psi(r)$ increases monotonically from 0 at the jet axis to 1 at $r = R$. Figure~\ref{fig:int_jet_shape_all} shows $\Psi(r)$ for combined direct and resolved processed as well the the GG and QQ photoproduction subsamples across four pseudorapidity intervals: $-1 < \eta < 0$, $0 < \eta < 1$, $1 < \eta < 1.5$, and $1.5 < \eta < 2$.

The distributions show the variation in transverse energy fraction radially outward from the jet axis. Quark-initiated jets exhibit a narrow energy profile, with more than 50\% of the jet transverse energy concentrated within $r = 0.3$, as reflected by the rapidly saturating $\Psi(r)$ curves.~As expected, the gluon-initiated jets, by contrast, display a broader radial energy distribution that saturates more gradually.~At higher $\eta$, the integrated jet shape of combined direct and resolved processes moves towards the GG subsample shape showing that the region is more populated by the gluon-like jets.

This distinction is consistent across all center-of-mass energies and pseudorapidity intervals considered, indicating that the integrated jet shape profile is largely insensitive to the collision energy over the range studied. The integrated jet shape therefore serves as an effective observable for distinguishing quark- from gluon-initiated jets, as it directly probes differences in the underlying fragmentation dynamics. In the following section, we examine the subjet multiplicity, which provides complementary information on the internal structure of jets.

\subsection{Subjet Multiplicity}
Jet substructure can be further probed by clustering the constituents within a jet into smaller subgroups called subjets. This is achieved by reapplying the longitudinally invariant $k_T$ algorithm to the jet constituents at a finer resolution scale. Particles are successively merged until all pairwise distances satisfy the condition:
\begin{equation}
d_{ij} > d_{\text{cut}} = y_{\text{cut}} \cdot E_T^{2}
\end{equation}
where $y_{\text{cut}}$ is the resolution parameter, set to $5 \times 10^{-4}$ in this analysis. 

The remaining clusters at this stage are identified as subjets, and their number defines the subjet multiplicity $n_{\text{sbj}}$ for that jet. The mean subjet multiplicity $\langle n_{\text{sbj}} \rangle$ is then obtained by averaging over all jets in the sample. The resolution parameter $y_{\text{cut}}$ controls the granularity at which the internal structure of a jet is probed: a smaller value resolves finer substructure, yielding a higher number of subjets, while a larger value merges nearby constituents into broader clusters. For a given $y_{\text{cut}}$, the subjet multiplicity is sensitive to the nature of the initiating parton. Quark-initiated jets are typically more collimated, with a narrow distribution of hadrons near the jet axis, resulting in fewer resolved subjets. Gluon-initiated jets, by contrast, exhibit a broader and more diffuse internal structure due to their higher colour charge and the correspondingly enhanced rate of soft gluon radiation.

\begin{figure}[!htbp]
\resizebox{7cm}{!}{\includegraphics{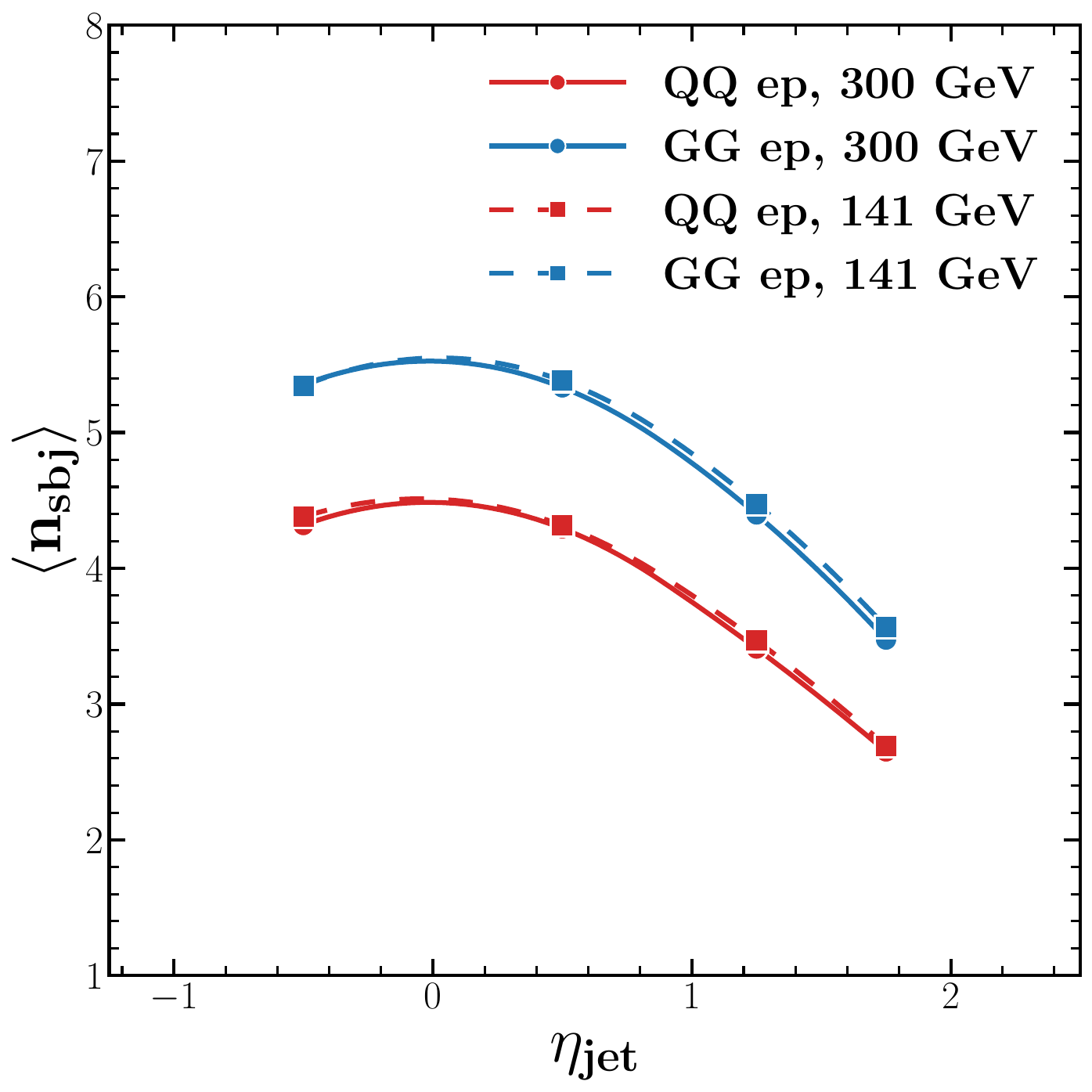}}
\caption{Mean subjet multiplicity $\langle n_{\mathrm{sbj}} \rangle$ as a function of jet pseudorapidity $\eta_{\mathrm{jet}}$ for quark (QQ) and gluon (GG) initiated dijets at $\sqrt{s} = 141$ GeV and $300$ GeV.}
\label{fig:meansubjetmultiplicity}
\end{figure}

The subjet multiplicity distributions from the GG and QQ photoproduction subsamples are shown in Figs.~\ref{fig:sbjmul_hera300} and~\ref{fig:sbjmul_eic141} for different pseudorapidity regions at HERA and the highest EIC center-of-mass energy of 141~GeV, respectively. The mean subjet multiplicity $\langle n_{\text{sbj}} \rangle$ for GG and QQ jets across different pseudorapidity intervals and center-of-mass energies is shown in Fig.~\ref{fig:meansubjetmultiplicity}.~To do a similar study at lower centre-of-mass energies at the EIC and also maintain good statistics, the $E_T$ cut on the dijets should be lowered. This is also evident from Fig.~\ref{fig:et_eta}.

As expected gluon jets consistently exhibit higher $\langle n_{\text{sbj}} \rangle$ than quark jets, as expected from their larger colour factor. To validate these simulation results, the following subsection presents a direct comparison of the integrated jet shapes with published ZEUS data~\cite{Chekanov2004-te}.

\subsection{Comparison with ZEUS data}
\begin{figure}[!htbp]
    \centering
    \includegraphics[width=\columnwidth]{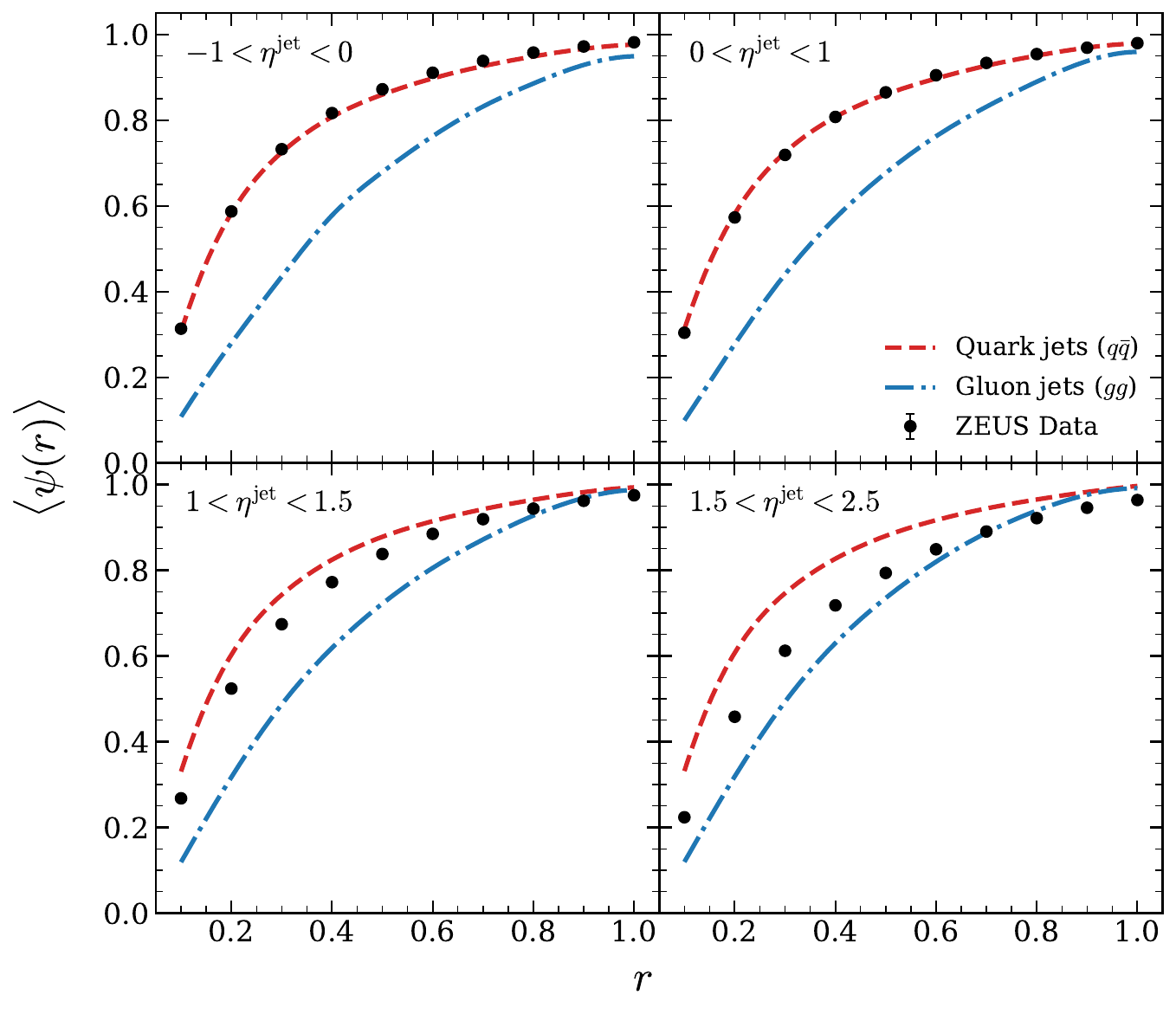}
    \caption{Integrated jet shape of quark and gluon jets with 
    $E_T > 17$~GeV in the simulated sample compared to published 
    ZEUS data at $\sqrt{s} = 300$~GeV~\cite{Chekanov2004-te}.}
    \label{fig:zeuscomparison1}
\end{figure}

The integrated jet shapes obtained from the simulated sample at $\sqrt{s} = 300$~GeV are compared with the published ZEUS photoproduction data~\cite{Chekanov2004-te} in Figure~\ref{fig:zeuscomparison1}. Since the ZEUS measurements were performed with a transverse energy threshold of $E_T^{\mathrm{jet}} > 17$~GeV, the same cut is applied to the simulated jets for this comparison.

In the backward pseudorapidity region ($-1 < \eta < 0$), the ZEUS data points lie close to the quark jet prediction, indicating that this region is predominantly populated by quark-initiated jets. As $\eta^{\mathrm{jet}}$ increases, the data move progressively from the quark toward the gluon jet prediction. In the most forward interval ($1.5 < \eta < 2.5$), the data fall closer to the gluon jet shape, consistent with an increasing fraction of gluon-initiated jets at higher $\eta^{\mathrm{jet}}$. 

This behaviour reflects the underlying parton dynamics: in the backward region, the direct process $\gamma g \to q\bar{q}$ produces predominantly quark jets, while in the forward region, resolved processes such as $q_\gamma g_p \to qg$ contribute an increasing proportion of gluon-initiated jets.

\begin{figure}[htbp!]
    \centering
    \includegraphics[width=\columnwidth]{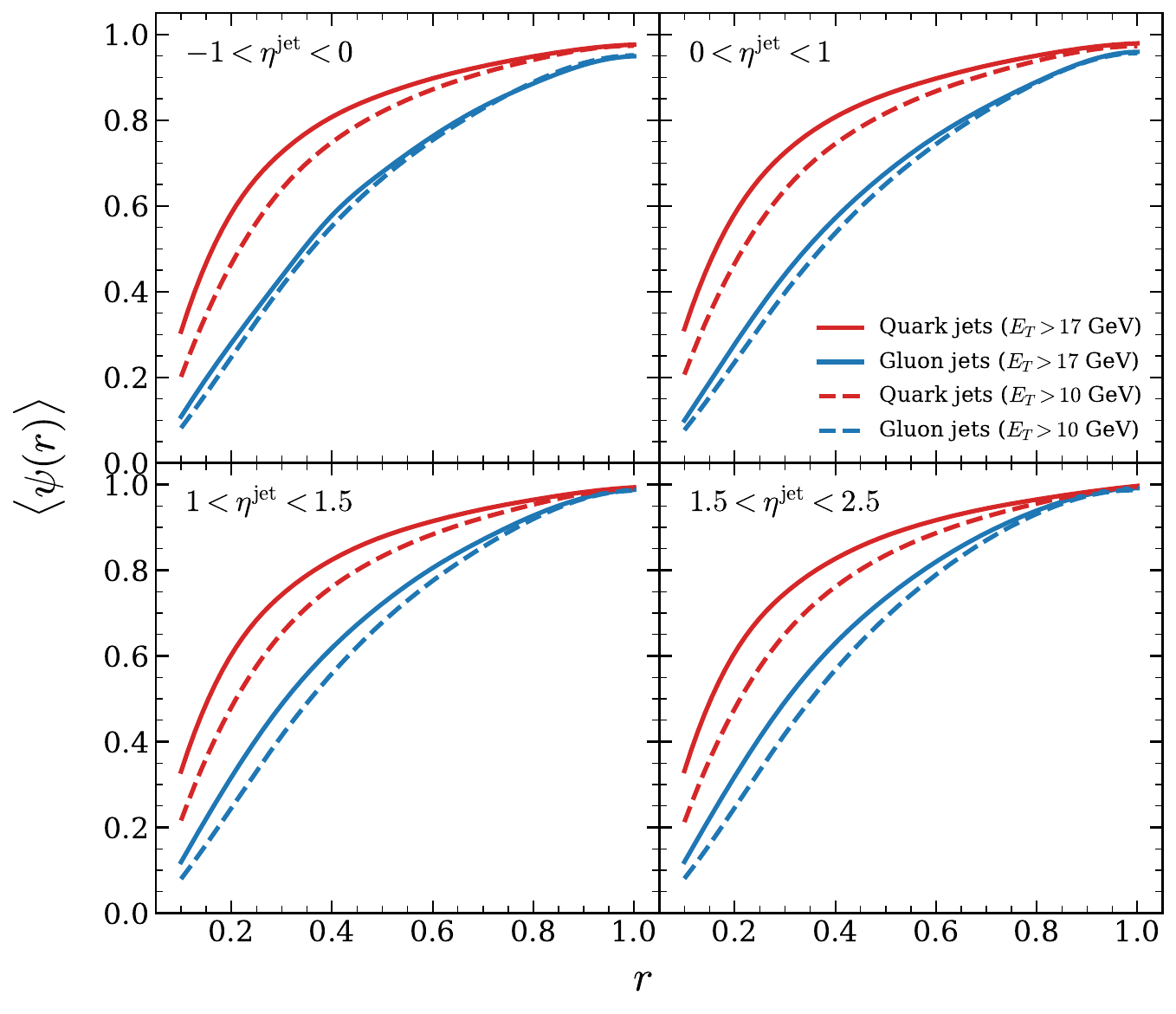}
    \caption{Integrated jet shape of the quark and gluon jets at 
    $E_T > 17$~GeV compared with jets at $E_T > 10$~GeV in the 
    simulated sample at $\sqrt{s} = 300$~GeV.}
    \label{fig:zeuscomparison2}
\end{figure}

Figure~\ref{fig:zeuscomparison2} compares the integrated jet shapes at two transverse energy thresholds: $E_T > 17$~GeV, corresponding to the ZEUS selection, and $E_T > 10$~GeV, used in the present analysis of jet substructure at the EIC energies. For both quark and gluon jets across all $\eta^{\mathrm{jet}}$ intervals, the integrated jet shape $\langle \psi(r) \rangle$ reaches a higher value at small $r$ for the higher $E_T$ threshold and saturates closer to unity at a smaller radius. To quantify the difference, the radius at which $\langle \psi(r) \rangle = 0.6$ is compared between the two selections. For quark jets in the backward region ($-1 < \eta^{\mathrm{jet}} < 0$), 60$\%$ of the jet transverse energy is contained within $r \approx 0.21$ at $E_T > 17$~GeV, compared to $r \approx 0.28$ at $E_T > 10$~GeV. For gluon jets in the same region, the corresponding radii are $r \approx 0.42$ and $r \approx 0.44$, respectively. In the most forward interval ($1.5 < \eta^{\mathrm{jet}} < 2.5$), the shift is more pronounced for gluon jets, with the 60$\%$ containment radius increasing from $r \approx 0.38$ to $r \approx 0.43$ when the threshold is lowered. Jets selected with the higher $E_T$ cut therefore exhibit narrower shapes, with a larger fraction of their transverse energy concentrated near the jet axis. This is consistent with the expectation from perturbative QCD that higher-$E_T$ jets are more collimated due to the reduced value of $\alpha_s$ and the correspondingly smaller phase space for wide-angle soft gluon radiation.

These results demonstrate that jet substructure observables remain well-defined and sensitive to the quark--gluon composition even at the lower transverse energies accessible in photoproduction at the EIC, making them viable probes for further QCD studies at these facilities.

\section{Quark and Gluon Jet Identification}
The differences in jet shapes and subjet multiplicities between quark and gluon jets, as presented in Section~3, suggest that these observables can be exploited for jet classification. In experimental data, however, the identity of the initiating parton is not directly accessible, and one must rely on observable jet properties to statistically distinguish quark-initiated from gluon-initiated jets. In this section, the integrated jet shape at $r = 0.3$ is used as a simple discriminator to produce quark-enriched and gluon-enriched jet samples.

\begin{figure*}[htbp!]
\centering
\includegraphics[width=0.45\textwidth]{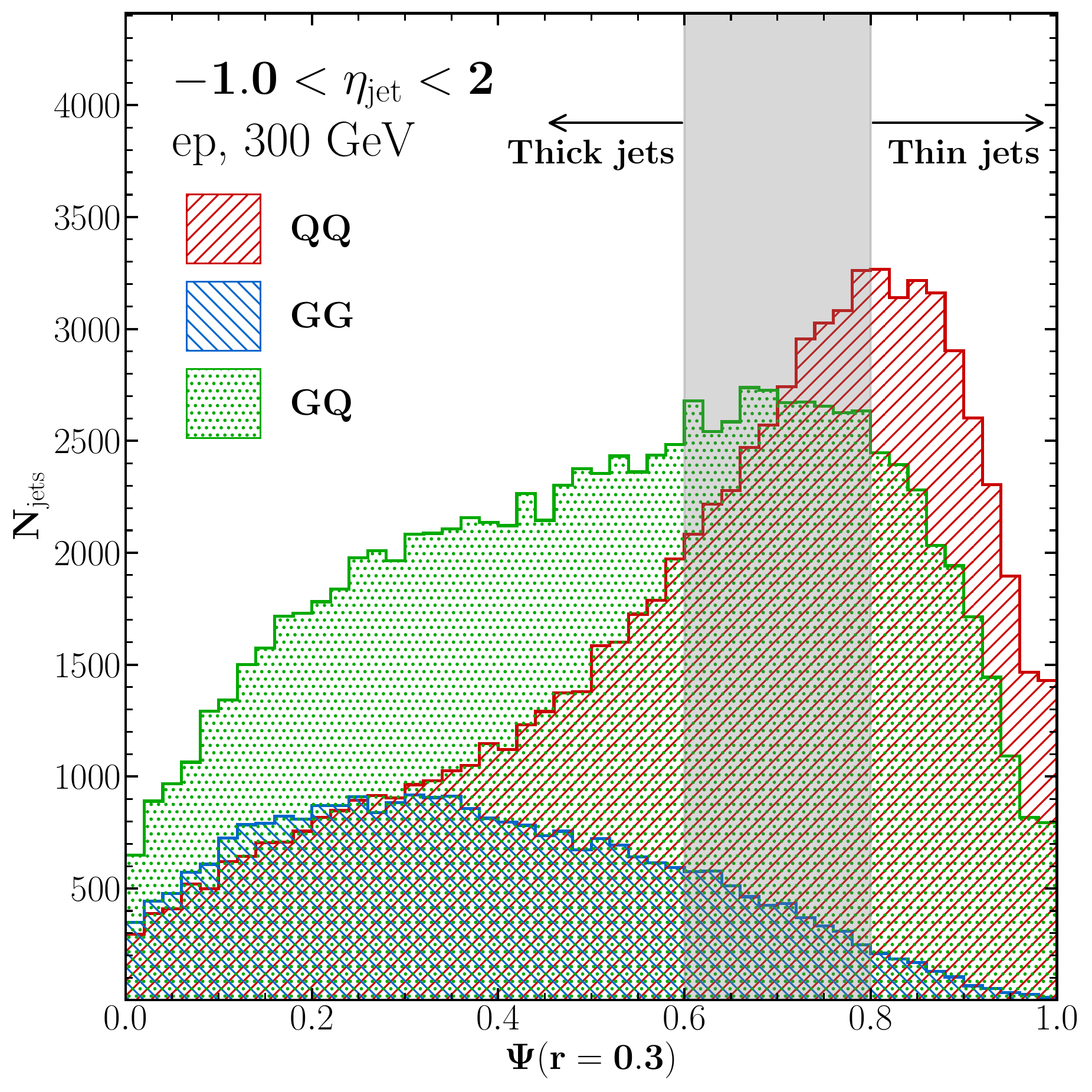}
\hfill
\includegraphics[width=0.45\textwidth]{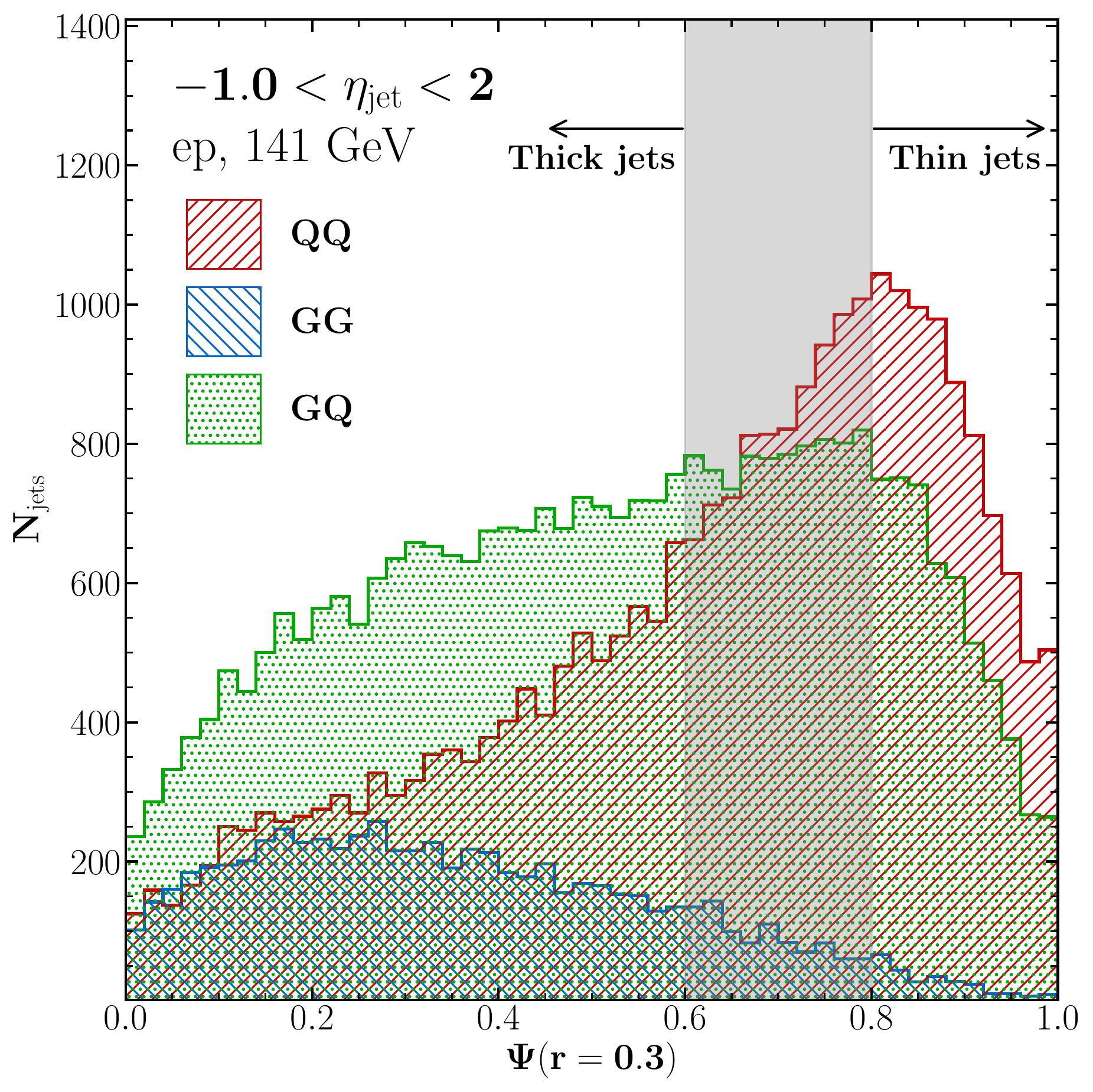}
\caption{Integrated jet shape $\Psi(r)$ at $r = 0.3$ for the 
QQ, QG, and GG dijet photoproduction subsamples at 
$\sqrt{s}$ = 300~GeV (left) and $\sqrt{s}$ = 141~GeV (right).}
\label{fig:intjetshapethickthin}
\end{figure*}

\begin{table}[htbp!]
\centering
\caption{Percentages of thin and thick jets identified from different dijet sub-samples at EIC and HERA energies.}
\label{tab:SubprocessComposition}
\renewcommand{\arraystretch}{1.6}
\begin{tabular*}{\columnwidth}{@{} c c c c c @{}}
\toprule
 & \textbf{EIC} & \textbf{EIC} & \textbf{EIC} & \textbf{HERA} \\
$\sqrt{s}$& $63.2$~GeV & $104.9$~GeV & $141$~GeV & $300$~GeV \\
\midrule
\textbf{Sample Type} & \multicolumn{4}{c}{\textbf{Thin Jets (\%)}} \\
\midrule
QQ & 64.04 & 62.57 & 61.37 & 53.52 \\
GG & 0.54  & 1.10  & 2.22  & 2.52 \\
GQ & 35.41 & 36.31 & 36.40 & 43.94 \\
\midrule
\textbf{Sample Type} & \multicolumn{4}{c}{\textbf{Thick Jets (\%)}} \\
\midrule
QQ & 39.35 & 34.78 & 26.16 & 26.44 \\
GG & 5.16  & 10.29 & 19.95 & 21.46 \\
GQ & 55.49 & 54.92 & 53.88 & 52.10 \\
\bottomrule
\end{tabular*}
\end{table}

The integrated jet shape $\Psi(r)$ at $r = 0.3$ for the GG, QG, and QQ photoproduction subsamples at the EIC energy of 141~GeV and HERA energy of 300~GeV is shown in Fig.~\ref{fig:intjetshapethickthin}. A higher value of $\Psi(0.3)$ for the quark jet sample indicates that the transverse energy is more collimated near the jet axis. The difference in $\Psi(0.3)$ values between the QQ and GG subsamples motivates the use of $\Psi(0.3)$ as a jet-type discriminator. A sample enriched in quark- or gluon-initiated jets can be prepared from the generated $ep$ photoproduction events using the following criterion. 

A jet is labeled as:
\begin{itemize}
    \item a ``thin'' jet, indicating a quark jet configuration, if its integrated jet shape $\Psi(r = 0.3) > 0.8$;
    \item a ``thick'' jet, indicating a gluon jet configuration, if $\Psi(r = 0.3) < 0.6$.
\end{itemize}
These jet shape cuts are motivated by studies performed at HERA~\cite{Glasman_2001,Chekanov2004-te}. 

Specifically, the sample with ``thick" jets predominantly involves subprocesses with gluon present as the exchanged parton, such as $gg \rightarrow gg$ and $qg \rightarrow  qg$ as shown in Table~\ref{Table2SubProc}. Conversely, the sample characterized by ``thin'' jets is largely dominated by subprocesses driven by quark as the exchanged parton as in the $\gamma g \rightarrow  q\overline{q}$ subprocess.

The quantitative assessment of thick and thin jet tagging in QQ, GG and QG dijet photo-production samples using the integrated jet shape is compiled in Table \ref{tab:SubprocessComposition}. At the lowest EIC energy of 63.2 GeV, $5.16\%$ jets of GG sample and  $55.49\%$ of GQ sample could be labeled as the thick jets indicative of gluon partonic behaviour, whereas  $39.35\%$ of jets from QQ sample are also incorrectly labeled as thick jets. The variation in these percentages at different cm energies are shown in Table \ref{tab:SubprocessComposition}. 

Conversely, the ``thin'' jets, which are linked with quark as the initiating partons in the final state, exhibit a contrasting distribution. For the lowest EIC energy of 63.2~GeV, $64.04\%$ jets from QQ sample and  $35.41\%$ jets of GQ sample were labeled as thin jets. The incorrect identification of jets from GG sample as thin jets was found to be only 0.54\%, which increases to about 2.5\% at the highest EIC energy of 141~GeV.

The ``thick'' jets at HERA consist of $26.44\%$ QQ, $21.46\%$ GG, and $52.10\%$ QG subsamples. In contrast, the ``thin'' jets are dominated by $53.52\%$ QQ, $2.52\%$ GG, and $43.94\%$ QG subsamples. These values are comparable to the published HERA results~\cite{Glasman_2001}, thereby validating the simulation and analysis procedures employed in this study.

\subsection{Purity Studies}
As shown in Table~\ref{tab:SubprocessComposition}, some false identifications occur during discrimination, and therefore the enriched samples are not 100\% pure. The purity of the gluon-enriched sample, obtained by selecting thick jets with $\Psi(0.3) < 0.6$, is quantified as:

\begin{equation}
\text{Purity}_{\text{gluon}} = \frac{N_{\text{jet}}^{GG} + 0.5 \times N_{\text{jet}}^{GQ}}{N_{\text{jet}}^{GG} + N_{\text{jet}}^{QQ} + 0.5 \times N_{\text{jet}}^{GQ}}
\end{equation}

where $N_{\text{jet}}^{GG}$, $N_{\text{jet}}^{GQ}$, and $N_{\text{jet}}^{QQ}$ are the number of jets from the GG, GQ, and QQ subsamples, respectively. Similarly, the purity of quark-enriched sample by selecting thin jets with $\Psi(0.3) > 0.8$ is given by:

\begin{equation}
\text{Purity}_{\text{quark}} = \frac{N_{\text{jet}}^{QQ} + 0.5 \times N_{\text{jet}}^{GQ}}{N_{\text{jet}}^{GG} + N_{\text{jet}}^{QQ} + 0.5 \times N_{\text{jet}}^{GQ}}
\end{equation}

\begin{figure}
\centering
\resizebox{8cm}{!}{\includegraphics{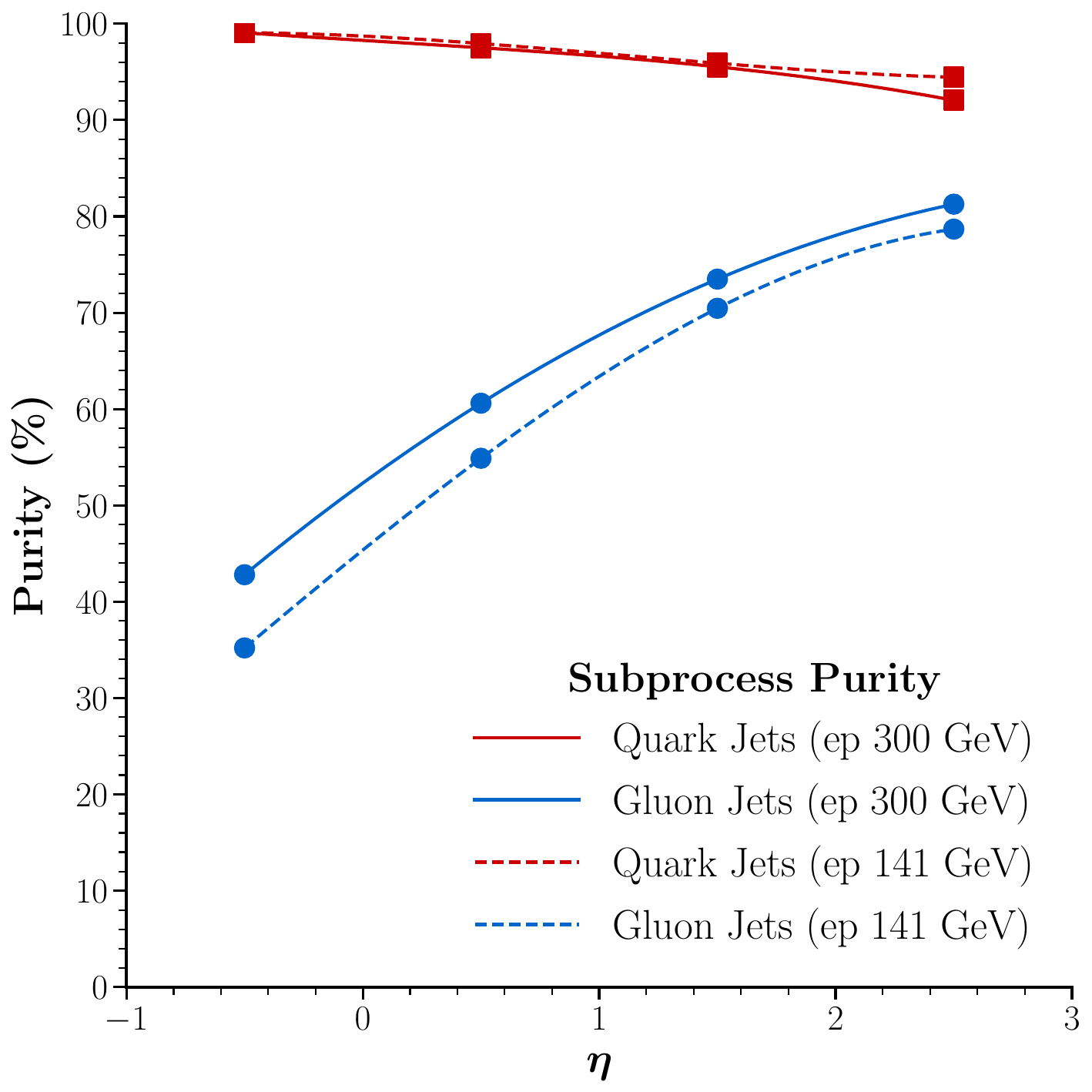}}
\caption{Purity of Quark and Gluon jets subsample selection.}
\label{fig:Eff_total}
\end{figure}

The purity of the gluon-enriched and quark-enriched samples produced by this discriminator is shown in Fig.~\ref{fig:Eff_total}. The quark-enriched sample maintains a purity above 90\% across all pseudorapidity intervals. In contrast, the gluon-enriched sample purity exhibits a strong $\eta$ dependence, increasing from approximately 35\% in the central region to about 80\% in the forward pseudorapidity intervals. This trend reflects the higher gluon jet population in the forward region ($\eta > 0$, corresponding to the proton beam direction), where resolved photoproduction processes dominate and gluon-initiated subprocesses such as $qg \rightarrow qg$ are enhanced, consistent with the expected QCD dynamics in $ep$ collisions. It should be noted that the present study is performed at the generator level; in a real detector environment with limited acceptance and finite resolution, the efficiency and purity of the selected samples would need to be re-evaluated using reconstructed data.

\section{Detector Outlook}
The results presented above are based on generator-level simulations without detector effects. In this section, we discuss how the proposed EIC design and its detector capabilities relate to the feasibility of the jet substructure measurements presented in this work. The EIC~\cite{Accardi:2012qut} will collide electrons with protons and nuclei, providing a unique tool for studying nuclear matter and the strong interaction.

The predictions of jet shapes presented here are given at the generator level, which are important and required to unfold the experimental results correctly. The jet shape, however, is also known to be sensitive to fragmentation and gluon radiation. A future scope of the study would be to use a different generator at the EIC energies, like HERWIG~\cite{bellm2016herwig}, which uses cluster model for fragmentation. At HERA energies, the uncertainties in the unfolding procedure were found to increase from 3\% to 5\% when hadron-level predictions were calculated with HERWIG rather than PYTHIA. Further, the experimental jet shapes unfolding can be done using either the predictions for the `direct and resolved processes' or exclusively from `quark and gluon jets', which adds to about  9\% uncertainty. The effect of using different proton and photon PDFs is another potential source of systematic uncertainty.

Three detector parts will be fundamentally important for jet substructure studies. Firstly, the tracker to measure the energy and direction of the charged particles, and then the electromagnetic and hadronic calorimeters to measure the momenta and positions of the electromagnetic and hadronic final-state particles. 

The yellow report design of EIC calorimetry consists of a high-granularity electromagnetic calorimeter and a hadronic calorimeter in the barrel and endcaps, covering nearly full azimuthal coverage of $-3.5 \le \eta \le 3.5$~\cite{khalek2022science}.  It also recommends a jet energy resolution of $\sigma_{E}/E \approx 50\%/\sqrt{E} \oplus 10\% $ for the forward hadron endcap region $1.5 \le \eta \le 3.5$ to enable precision substructure measurements. In the central barrel region ($|\eta| < 1.0 $), although the hadronic calorimeter has a degrading resolution ($\approx 75 -100\%/\sqrt{E}$ ). A future extension of the presented study could include detector smearing effects on the positions and energies of final-state particles, which could be used to address one of the major systematic uncertainties in jet substructure measurements due to the limited resolution of hadronic detectors. Another future scope of the study will be to examine the effects of jet grooming techniques on the jet substructure, and the performance of quark and gluon jet tagging in dijet photoproduction events.

The EIC strives to attain a luminosity in the range of $10^{33}$ to $10^{34}$~cm$^{-2}$s$^{-1}$ for $ep$ collisions and $10^{30} - 10^{32}$, $\text{cm}^{-2}$, $\text{s}^{-1}$ for \(eA\) collisions. Within the EIC framework, electron beam energies will span from 5 to 18~GeV, while proton beam energies will range from 41 to 275~GeV, yielding center-of-mass energies from approximately 29 to 140~GeV. The accessible $Q^2$ range in $ep$ collisions scales with the center-of-mass energy as $Q^2_{\text{max}} \sim s$. A minimum center-of-mass energy of approximately 20~GeV is required to access the perturbative regime ($Q^2 \gtrsim 10$~GeV$^2$) where quarks can be cleanly resolved, while center-of-mass energies of order 100~GeV are necessary to probe the high-$Q^2$ domain ($Q^2 \gtrsim 1000$~GeV$^2$) where electroweak effects become significant and precise determination of parton distributions at large Bjorken-$x$ is possible~\cite{Accardi:2012qut, khalek2022science}.

These energy ranges will enhance sensitivity to gluon distributions, and spin-polarized beams will significantly improve our understanding of the proton's spin structure. Recent studies have demonstrated that jet substructure analysis in photoproduction events at EIC energies is viable despite typically low transverse momenta~\cite{PhysRevD.97.114013}. 

The importance of jets as probes of gluon helicity contributions to the proton spin via dijet events in DIS has been highlighted~\cite{PhysRevD_101}. Furthermore, jet charge measurements have been proposed to classify jets by their originating partons, providing insight into nucleon flavor and spin dynamics~\cite{Kang_2023, Kang_2020}.

The variable center-of-mass energy, combined with the EIC's particle identification capabilities, will enable studies of hadronization processes within jets across a broad kinematic range in $x$ and $Q^2$. The phase space region with pseudorapidity $\eta > 0$ (the proton beam direction) is expected to be enriched in gluon jets, producing broader jet profiles. This forward region, accessible at the EIC, provides a promising testing ground for gluon-jet classification techniques in future data analyses.

\section{Discussion and Conclusion}
This paper presents a systematic study of the substructure of dijets produced in direct and resolved photoproduction events and subsequent quark and gluon jet identification in dijets, considering the kinematic reach of the upcoming EIC experiment.~In this study, we have investigated photoproduction processes in $ep$ collisions at the EIC energies ranging from $\sqrt{s} = 64$ to 141~GeV. The analysis focuses on the substructure of jets produced through direct and resolved photoproduction processes, simulated using the PYTHIA~8 event generator. Jets are reconstructed using the longitudinally invariant $k_T$ algorithm in FastJet~3.

In the present work, we have studied jets and their substructure in photoproduction events from electron-proton collisions across different EIC center-of-mass energies. We focused on three key observables: subjet multiplicity, differential jet shape, and integrated jet shape. This analysis allowed us to investigate the contributions of various direct and resolved photoproduction subprocesses as shown in Table~\ref{Table2SubProc}  to jet formation across both EIC and HERA energies, revealing the underlying jet substructure and their energy dependencies. Dijet events were selected for a detailed comparison of jet-shapes and subjet multiplicity in the labeled QQ and GG subsamples. The study demonstrated that as gluon-initiated jets exhibit a broader, more diffused structure compared to the collimated and narrow quark-initiated jets the jet-shape variables can be used to produce gluon-enriched and quark-enriched samples at the EIC.

The differential jet shape $\rho(r)$ of quark jet has a larger value for small r as compared to the gluon jets, showing its transverse energy is collimated near the jet-axis and then decreases steeply. A similar observation is made by studying the integrated jet shape $\Psi(r)$ for quark jets as compared to the gluon-initiated jets. With increasing radius, the $\Psi(r)$ for quark jet saturates quickly, showing more than 50\% energy is contained within a jet radius of 0.3. The gluon jets show a more gradual energy distribution over a larger radius. This variation can be used to identify whether the initiating parton is a gluon or a quark. 

The mean subjet multiplicity of the jets in QQ and GG is studied in different pseudorapidity intervals.~The predictions of the subjet multiplicities in the dijet photoproduction events with QQ and GG configuration  are presented at the EIC energy of 141 GeV and it is observed that the gluon jets have higher subjet multiplicity than the quark initiated jets in all pseudorapidity intervals.

We exploit the integrated jet shape variable to separate our samples into the quark and gluon jet enriched samples.  The jets are labeled as thick jets if their $\Psi(0.3)$ is less than 0.6 and as thin jets if their $\Psi(0.3)$ is greater than 0.8, respectively. The efficiency of this variable in selecting quark jet sample is calculated and found to be more than 90\%. The efficiency of this variable in producing gluon jet enriched sample is found to increase from 35\% to 80\% with pseudorapidity. The integrated jet shape variable $\Psi(r)$ proves to be an effective observable to differentiate between the gluon and quark jets. The study's findings are consistent with previous observations from HERA and provide a solid foundation for future experimental investigations at the EIC.

The classification of quark and gluon jets based on their substructure is crucial for enhancing the precision of scattered parton identification.~The analysis emphasizes the importance of jet substructure in probing the dynamics of parton fragmentation and hadronization processes.~By leveraging the integrated and differential jet shapes, one can effectively separate quark-dominated and gluon-dominated jets, facilitating more accurate measurements and interpretations of the data. 

The insights gained from this study are expected to significantly contribute to the physics program at the EIC, enabling detailed exploration of QCD phenomena and providing valuable benchmarks for tuning event generators. This, in turn, will enhance our capability to explore beyond the Standard Model physics, where new signals may be obscured by the Standard Model background. In conclusion, the detailed examination of jet substructures in photoproduction events at the EIC paves the way for more refined analyses and a deeper understanding of the strong force, parton dynamics, and the internal structure of nucleons. The aim of this paper is to highlight potentially intriguing jet-shape observables that could be utilized and further explored in detail as and when the EIC data become available.
\section*{Data Availability}
The data simulated at various energies to support the findings of this study are available from the corresponding author upon request.~The previously published data~\cite{Chekanov2004-te} are used for comparison in Figure~\ref{fig:zeuscomparison1}.
\clearpage
\bibliography{CC_ep.bib}

\begin{thebibliography}{70}%
\makeatletter
\providecommand \@ifxundefined [1]{%
 \@ifx{#1\undefined}
}%
\providecommand \@ifnum [1]{%
 \ifnum #1\expandafter \@firstoftwo
 \else \expandafter \@secondoftwo
 \fi
}%
\providecommand \@ifx [1]{%
 \ifx #1\expandafter \@firstoftwo
 \else \expandafter \@secondoftwo
 \fi
}%
\providecommand \natexlab [1]{#1}%
\providecommand \enquote  [1]{``#1''}%
\providecommand \bibnamefont  [1]{#1}%
\providecommand \bibfnamefont [1]{#1}%
\providecommand \citenamefont [1]{#1}%
\providecommand \href@noop [0]{\@secondoftwo}%
\providecommand \href [0]{\begingroup \@sanitize@url \@href}%
\providecommand \@href[1]{\@@startlink{#1}\@@href}%
\providecommand \@@href[1]{\endgroup#1\@@endlink}%
\providecommand \@sanitize@url [0]{\catcode `\\12\catcode `\$12\catcode `\&12\catcode `\#12\catcode `\^12\catcode `\_12\catcode `\%12\relax}%
\providecommand \@@startlink[1]{}%
\providecommand \@@endlink[0]{}%
\providecommand \url  [0]{\begingroup\@sanitize@url \@url }%
\providecommand \@url [1]{\endgroup\@href {#1}{\urlprefix }}%
\providecommand \urlprefix  [0]{URL }%
\providecommand \Eprint [0]{\href }%
\providecommand \doibase [0]{https://doi.org/}%
\providecommand \selectlanguage [0]{\@gobble}%
\providecommand \bibinfo  [0]{\@secondoftwo}%
\providecommand \bibfield  [0]{\@secondoftwo}%
\providecommand \translation [1]{[#1]}%
\providecommand \BibitemOpen [0]{}%
\providecommand \bibitemStop [0]{}%
\providecommand \bibitemNoStop [0]{.\EOS\space}%
\providecommand \EOS [0]{\spacefactor3000\relax}%
\providecommand \BibitemShut  [1]{\csname bibitem#1\endcsname}%
\let\auto@bib@innerbib\@empty
\bibitem [{\citenamefont {{ZEUS Collaboration}}\ \emph {et~al.}(2015)\citenamefont {{ZEUS Collaboration}}, \citenamefont {Abramowicz}, \citenamefont {Abt}, \citenamefont {Adamczyk} \emph {et~al.}}]{Abramowicz2015}%
  \BibitemOpen
  \bibfield  {author} {\bibinfo {author} {\bibnamefont {{ZEUS Collaboration}}}, \bibinfo {author} {\bibfnamefont {H.}~\bibnamefont {Abramowicz}}, \bibinfo {author} {\bibfnamefont {I.}~\bibnamefont {Abt}}, \bibinfo {author} {\bibfnamefont {L.}~\bibnamefont {Adamczyk}}, \emph {et~al.},\ }\bibfield  {title} {\bibinfo {title} {Combination of measurements of inclusive deep inelastic $e^{\pm} p$ scattering cross sections and {QCD} analysis of {HERA} data},\ }\bibfield  {journal} {\bibinfo  {journal} {Eur. Phys. J. C}\ }\textbf {\bibinfo {volume} {75}},\ \href {https://doi.org/10.1140/epjc/s10052-015-3710-4} {10.1140/epjc/s10052-015-3710-4} (\bibinfo {year} {2015})\BibitemShut {NoStop}%
\bibitem [{\citenamefont {Page}\ \emph {et~al.}(2020{\natexlab{a}})\citenamefont {Page}, \citenamefont {Chu},\ and\ \citenamefont {Aschenauer}}]{page2020experimental}%
  \BibitemOpen
  \bibfield  {author} {\bibinfo {author} {\bibfnamefont {B.~S.}\ \bibnamefont {Page}}, \bibinfo {author} {\bibfnamefont {X.}~\bibnamefont {Chu}},\ and\ \bibinfo {author} {\bibfnamefont {E.~C.}\ \bibnamefont {Aschenauer}},\ }\bibfield  {title} {\bibinfo {title} {Experimental aspects of jet physics at a future eic},\ }\href {https://doi.org/10.1103/PhysRevD.101.072003} {\bibfield  {journal} {\bibinfo  {journal} {Phys. Rev. D}\ }\textbf {\bibinfo {volume} {101}},\ \bibinfo {pages} {072003} (\bibinfo {year} {2020}{\natexlab{a}})}\BibitemShut {NoStop}%
\bibitem [{\citenamefont {Klasen}\ and\ \citenamefont {Kova\ifmmode~\check{r}\else \v{r}\fi{}\'{\i}k}(2018)}]{PhysRevD.97.114013}%
  \BibitemOpen
  \bibfield  {author} {\bibinfo {author} {\bibfnamefont {M.}~\bibnamefont {Klasen}}\ and\ \bibinfo {author} {\bibfnamefont {K.}~\bibnamefont {Kova\ifmmode~\check{r}\else \v{r}\fi{}\'{\i}k}},\ }\bibfield  {title} {\bibinfo {title} {Nuclear parton density functions from dijet photoproduction at the eic},\ }\href {https://doi.org/10.1103/PhysRevD.97.114013} {\bibfield  {journal} {\bibinfo  {journal} {Phys. Rev. D}\ }\textbf {\bibinfo {volume} {97}},\ \bibinfo {pages} {114013} (\bibinfo {year} {2018})}\BibitemShut {NoStop}%
\bibitem [{\citenamefont {Butterworth}\ \emph {et~al.}(1996)\citenamefont {Butterworth}, \citenamefont {Forshaw},\ and\ \citenamefont {Seymour}}]{butterworth1996multiparton}%
  \BibitemOpen
  \bibfield  {author} {\bibinfo {author} {\bibfnamefont {J.~M.}\ \bibnamefont {Butterworth}}, \bibinfo {author} {\bibfnamefont {J.~R.}\ \bibnamefont {Forshaw}},\ and\ \bibinfo {author} {\bibfnamefont {M.~H.}\ \bibnamefont {Seymour}},\ }\bibfield  {title} {\bibinfo {title} {Multiparton interactions in photoproduction at hera},\ }\href {https://doi.org/10.1007/s002880050286} {\bibfield  {journal} {\bibinfo  {journal} {Zeitschrift f{\"u}r Physik C: Particles and Fields}\ }\textbf {\bibinfo {volume} {72}},\ \bibinfo {pages} {637} (\bibinfo {year} {1996})}\BibitemShut {NoStop}%
\bibitem [{\citenamefont {{ZEUS Collaboration}}\ \emph {et~al.}(1996)\citenamefont {{ZEUS Collaboration}}, \citenamefont {Derrick}, \citenamefont {Krakauer}, \citenamefont {Magill}, \citenamefont {Mikunas} \emph {et~al.}}]{Derrick1996-aj}%
  \BibitemOpen
  \bibfield  {author} {\bibinfo {author} {\bibnamefont {{ZEUS Collaboration}}}, \bibinfo {author} {\bibfnamefont {M.}~\bibnamefont {Derrick}}, \bibinfo {author} {\bibfnamefont {D.}~\bibnamefont {Krakauer}}, \bibinfo {author} {\bibfnamefont {S.}~\bibnamefont {Magill}}, \bibinfo {author} {\bibfnamefont {D.}~\bibnamefont {Mikunas}}, \emph {et~al.},\ }\bibfield  {title} {\bibinfo {title} {Dijet angular distributions in direct and resolved photoproduction at {HERA}},\ }\href {https://doi.org/10.1016/0370-2693(96)00931-8} {\bibfield  {journal} {\bibinfo  {journal} {Phys. Lett. B}\ }\textbf {\bibinfo {volume} {384}},\ \bibinfo {pages} {401} (\bibinfo {year} {1996})}\BibitemShut {NoStop}%
\bibitem [{\citenamefont {collaboration}(2003)}]{zeus2003measurements}%
  \BibitemOpen
  \bibfield  {author} {\bibinfo {author} {\bibfnamefont {Z.}~\bibnamefont {collaboration}},\ }\bibfield  {title} {\bibinfo {title} {Measurements of inelastic j$/\psi$ and $\psi^{\prime}$ photoproduction at {HERA}},\ }\href {https://doi.org/10.1140/epjc/s2002-01130-2} {\bibfield  {journal} {\bibinfo  {journal} {The European Physical Journal C-Particles and Fields}\ }\textbf {\bibinfo {volume} {27}},\ \bibinfo {pages} {173} (\bibinfo {year} {2003})}\BibitemShut {NoStop}%
\bibitem [{\citenamefont {Lee}\ \emph {et~al.}(2023)\citenamefont {Lee}, \citenamefont {Mulligan}, \citenamefont {P{\l}osko{\'n}}, \citenamefont {Ringer},\ and\ \citenamefont {Yuan}}]{lee2023machine}%
  \BibitemOpen
  \bibfield  {author} {\bibinfo {author} {\bibfnamefont {K.}~\bibnamefont {Lee}}, \bibinfo {author} {\bibfnamefont {J.}~\bibnamefont {Mulligan}}, \bibinfo {author} {\bibfnamefont {M.}~\bibnamefont {P{\l}osko{\'n}}}, \bibinfo {author} {\bibfnamefont {F.}~\bibnamefont {Ringer}},\ and\ \bibinfo {author} {\bibfnamefont {F.}~\bibnamefont {Yuan}},\ }\bibfield  {title} {\bibinfo {title} {Machine learning-based jet and event classification at the electron-ion collider with applications to hadron structure and spin physics},\ }\href {https://doi.org/10.1007/JHEP03(2023)085} {\bibfield  {journal} {\bibinfo  {journal} {Journal of High Energy Physics}\ }\textbf {\bibinfo {volume} {2023}},\ \bibinfo {pages} {85} (\bibinfo {year} {2023})}\BibitemShut {NoStop}%
\bibitem [{\citenamefont {Chu}\ \emph {et~al.}(2017)\citenamefont {Chu}, \citenamefont {Aschenauer}, \citenamefont {Lee},\ and\ \citenamefont {Zheng}}]{chu2017photon}%
  \BibitemOpen
  \bibfield  {author} {\bibinfo {author} {\bibfnamefont {X.}~\bibnamefont {Chu}}, \bibinfo {author} {\bibfnamefont {E.-C.}\ \bibnamefont {Aschenauer}}, \bibinfo {author} {\bibfnamefont {J.-H.}\ \bibnamefont {Lee}},\ and\ \bibinfo {author} {\bibfnamefont {L.}~\bibnamefont {Zheng}},\ }\bibfield  {title} {\bibinfo {title} {Photon structure studied at an electron ion collider},\ }\href {https://doi.org/10.1103/PhysRevD.96.074035} {\bibfield  {journal} {\bibinfo  {journal} {Physical Review D}\ }\textbf {\bibinfo {volume} {96}},\ \bibinfo {pages} {074035} (\bibinfo {year} {2017})}\BibitemShut {NoStop}%
\bibitem [{\citenamefont {Armesto}(2006)}]{Armesto2006-ms}%
  \BibitemOpen
  \bibfield  {author} {\bibinfo {author} {\bibfnamefont {N.}~\bibnamefont {Armesto}},\ }\bibfield  {title} {\bibinfo {title} {Nuclear shadowing},\ }\href {https://doi.org/10.1088/0954-3899/32/11/R01} {\bibfield  {journal} {\bibinfo  {journal} {J. Phys. G Nucl. Part. Phys.}\ }\textbf {\bibinfo {volume} {32}},\ \bibinfo {pages} {R367} (\bibinfo {year} {2006})}\BibitemShut {NoStop}%
\bibitem [{\citenamefont {Golec-Biernat}\ and\ \citenamefont {W\"usthoff}(1998)}]{PhysRevD.59.014017}%
  \BibitemOpen
  \bibfield  {author} {\bibinfo {author} {\bibfnamefont {K.}~\bibnamefont {Golec-Biernat}}\ and\ \bibinfo {author} {\bibfnamefont {M.}~\bibnamefont {W\"usthoff}},\ }\bibfield  {title} {\bibinfo {title} {Saturation effects in deep inelastic scattering at low ${Q}^{2}$ and its implications on diffraction},\ }\href {https://doi.org/10.1103/PhysRevD.59.014017} {\bibfield  {journal} {\bibinfo  {journal} {Phys. Rev. D}\ }\textbf {\bibinfo {volume} {59}},\ \bibinfo {pages} {014017} (\bibinfo {year} {1998})}\BibitemShut {NoStop}%
\bibitem [{\citenamefont {Gelis}\ \emph {et~al.}(2010)\citenamefont {Gelis}, \citenamefont {Iancu}, \citenamefont {Jalilian-Marian},\ and\ \citenamefont {Venugopalan}}]{Gelis2010-qt}%
  \BibitemOpen
  \bibfield  {author} {\bibinfo {author} {\bibfnamefont {F.}~\bibnamefont {Gelis}}, \bibinfo {author} {\bibfnamefont {E.}~\bibnamefont {Iancu}}, \bibinfo {author} {\bibfnamefont {J.}~\bibnamefont {Jalilian-Marian}},\ and\ \bibinfo {author} {\bibfnamefont {R.}~\bibnamefont {Venugopalan}},\ }\bibfield  {title} {\bibinfo {title} {The color glass condensate},\ }\href {https://doi.org/10.1146/annurev.nucl.010909.083629} {\bibfield  {journal} {\bibinfo  {journal} {Annu. Rev. Nucl. Part. Sci.}\ }\textbf {\bibinfo {volume} {60}},\ \bibinfo {pages} {463} (\bibinfo {year} {2010})}\BibitemShut {NoStop}%
\bibitem [{\citenamefont {Bierlich}\ \emph {et~al.}(2022)\citenamefont {Bierlich} \emph {et~al.}}]{10.21468/SciPostPhysCodeb.8}%
  \BibitemOpen
  \bibfield  {author} {\bibinfo {author} {\bibfnamefont {C.}~\bibnamefont {Bierlich}} \emph {et~al.},\ }\bibfield  {title} {\bibinfo {title} {A comprehensive guide to the physics and usage of {PYTHIA} 8.3},\ }\href {10.21468/SciPostPhysCodeb.8} {\bibfield  {journal} {\bibinfo  {journal} {SciPost Physics Codebases}\ ,\ \bibinfo {pages} {8}} (\bibinfo {year} {2022})}\BibitemShut {NoStop}%
\bibitem [{\citenamefont {Konishi}\ \emph {et~al.}(1979)\citenamefont {Konishi}, \citenamefont {Ukawa},\ and\ \citenamefont {Veneziano}}]{konishi1979jet}%
  \BibitemOpen
  \bibfield  {author} {\bibinfo {author} {\bibfnamefont {K.}~\bibnamefont {Konishi}}, \bibinfo {author} {\bibfnamefont {A.}~\bibnamefont {Ukawa}},\ and\ \bibinfo {author} {\bibfnamefont {G.}~\bibnamefont {Veneziano}},\ }\bibfield  {title} {\bibinfo {title} {Jet calculus: A simple algorithm for resolving qcd jets},\ }\href {https://doi.org/https://doi.org/10.1016/0550-3213(79)90053-1} {\bibfield  {journal} {\bibinfo  {journal} {Nuclear Physics B}\ }\textbf {\bibinfo {volume} {157}},\ \bibinfo {pages} {45} (\bibinfo {year} {1979})}\BibitemShut {NoStop}%
\bibitem [{\citenamefont {{CMS Collaboration}}\ \emph {et~al.}(2022)\citenamefont {{CMS Collaboration}}, \citenamefont {Tumasyan}, \citenamefont {Adam}, \citenamefont {Andrejkovic} \emph {et~al.}}]{tumasyan2021study}%
  \BibitemOpen
  \bibfield  {author} {\bibinfo {author} {\bibnamefont {{CMS Collaboration}}}, \bibinfo {author} {\bibfnamefont {A.}~\bibnamefont {Tumasyan}}, \bibinfo {author} {\bibfnamefont {W.}~\bibnamefont {Adam}}, \bibinfo {author} {\bibfnamefont {J.~W.}\ \bibnamefont {Andrejkovic}}, \emph {et~al.},\ }\bibfield  {title} {\bibinfo {title} {Study of quark and gluon jet substructure in z+jet and dijet events from pp collisions},\ }\href {https://doi.org/10.1007/JHEP01(2022)188} {\bibfield  {journal} {\bibinfo  {journal} {Journal of High Energy Physics}\ }\textbf {\bibinfo {volume} {2022}},\ \bibinfo {pages} {188} (\bibinfo {year} {2022})}\BibitemShut {NoStop}%
\bibitem [{\citenamefont {Ellis}\ \emph {et~al.}(2011)\citenamefont {Ellis}, \citenamefont {Stirling},\ and\ \citenamefont {Webber}}]{Ellis:1991qcd}%
  \BibitemOpen
  \bibfield  {author} {\bibinfo {author} {\bibfnamefont {R.~K.}\ \bibnamefont {Ellis}}, \bibinfo {author} {\bibfnamefont {W.~J.}\ \bibnamefont {Stirling}},\ and\ \bibinfo {author} {\bibfnamefont {B.~R.}\ \bibnamefont {Webber}},\ }\href {https://doi.org/10.1017/CBO9780511628788} {\emph {\bibinfo {title} {{QCD and collider physics}}}},\ Vol.~\bibinfo {volume} {8}\ (\bibinfo  {publisher} {Cambridge University Press},\ \bibinfo {year} {2011})\BibitemShut {NoStop}%
\bibitem [{\citenamefont {Tanabashi}\ \emph {et~al.}(2018)\citenamefont {Tanabashi}, \citenamefont {Hagiwara}, \citenamefont {Hikasa} \emph {et~al.}}]{PDG_QCD}%
  \BibitemOpen
  \bibfield  {author} {\bibinfo {author} {\bibfnamefont {M.}~\bibnamefont {Tanabashi}}, \bibinfo {author} {\bibfnamefont {K.}~\bibnamefont {Hagiwara}}, \bibinfo {author} {\bibnamefont {Hikasa}}, \emph {et~al.} (\bibinfo {collaboration} {{Particle Data Group}}),\ }\bibfield  {title} {\bibinfo {title} {Review of particle physics},\ }\href {https://doi.org/10.1103/PhysRevD.98.030001} {\bibfield  {journal} {\bibinfo  {journal} {Phys. Rev. D}\ }\textbf {\bibinfo {volume} {98}},\ \bibinfo {pages} {030001} (\bibinfo {year} {2018})}\BibitemShut {NoStop}%
\bibitem [{\citenamefont {{OPAL collaboration}}\ \emph {et~al.}(1993)\citenamefont {{OPAL collaboration}}, \citenamefont {Acton}, \citenamefont {Alexander}, \citenamefont {Allison} \emph {et~al.}}]{opal1993study}%
  \BibitemOpen
  \bibfield  {author} {\bibinfo {author} {\bibnamefont {{OPAL collaboration}}}, \bibinfo {author} {\bibfnamefont {P.~D.}\ \bibnamefont {Acton}}, \bibinfo {author} {\bibfnamefont {G.}~\bibnamefont {Alexander}}, \bibinfo {author} {\bibfnamefont {J.}~\bibnamefont {Allison}}, \emph {et~al.},\ }\bibfield  {title} {\bibinfo {title} {A study of differences between quark and gluon jets using vertex tagging of quark jets},\ }\href {https://doi.org/10.1007/BF01557696} {\bibfield  {journal} {\bibinfo  {journal} {Zeitschrift f{\"u}r Physik C}\ }\textbf {\bibinfo {volume} {58}},\ \bibinfo {pages} {387} (\bibinfo {year} {1993})}\BibitemShut {NoStop}%
\bibitem [{\citenamefont {{OPAL collaboration}}\ \emph {et~al.}(1995)\citenamefont {{OPAL collaboration}}, \citenamefont {Akers}, \citenamefont {Alexander} \emph {et~al.}}]{opal1995model}%
  \BibitemOpen
  \bibfield  {author} {\bibinfo {author} {\bibnamefont {{OPAL collaboration}}}, \bibinfo {author} {\bibfnamefont {R.}~\bibnamefont {Akers}}, \bibinfo {author} {\bibfnamefont {G.}~\bibnamefont {Alexander}}, \emph {et~al.},\ }\bibfield  {title} {\bibinfo {title} {A model independent measurement of quark and gluon jet properties and differences},\ }\href {https://doi.org/10.1007/BF01566667} {\bibfield  {journal} {\bibinfo  {journal} {Zeitschrift f{\"u}r Physik C}\ }\textbf {\bibinfo {volume} {68}},\ \bibinfo {pages} {179} (\bibinfo {year} {1995})}\BibitemShut {NoStop}%
\bibitem [{\citenamefont {{DELPHI Collaboration}}(1996)}]{buys1996energy}%
  \BibitemOpen
  \bibfield  {author} {\bibinfo {author} {\bibnamefont {{DELPHI Collaboration}}},\ }\bibfield  {title} {\bibinfo {title} {Energy dependence of the differences between the quark and gluon jet fragmentation},\ }\href {https://doi.org/10.1007/s002880050095} {\bibfield  {journal} {\bibinfo  {journal} {Zeitschrift f{\"u}r Physik C}\ }\textbf {\bibinfo {volume} {70}},\ \bibinfo {pages} {179} (\bibinfo {year} {1996})}\BibitemShut {NoStop}%
\bibitem [{\citenamefont {{ALEPH collaboration}}\ \emph {et~al.}(2000)\citenamefont {{ALEPH collaboration}}, \citenamefont {Barate} \emph {et~al.}}]{aleph2000measurements}%
  \BibitemOpen
  \bibfield  {author} {\bibinfo {author} {\bibnamefont {{ALEPH collaboration}}}, \bibinfo {author} {\bibfnamefont {R.}~\bibnamefont {Barate}}, \emph {et~al.},\ }\bibfield  {title} {\bibinfo {title} {Measurements of the structure of quark and gluon jets in hadronic z decays},\ }\href {https://doi.org/10.1007/s100520000474} {\bibfield  {journal} {\bibinfo  {journal} {The European Physical Journal C}\ }\textbf {\bibinfo {volume} {17}},\ \bibinfo {pages} {1} (\bibinfo {year} {2000})}\BibitemShut {NoStop}%
\bibitem [{\citenamefont {{ALEPH collaboration}}\ \emph {et~al.}(1996)\citenamefont {{ALEPH collaboration}}, \citenamefont {Buskulic} \emph {et~al.}}]{buskulic1996quark}%
  \BibitemOpen
  \bibfield  {author} {\bibinfo {author} {\bibnamefont {{ALEPH collaboration}}}, \bibinfo {author} {\bibfnamefont {D.}~\bibnamefont {Buskulic}}, \emph {et~al.},\ }\bibfield  {title} {\bibinfo {title} {Quark and gluon jet properties in symmetric three-jet events},\ }\href {https://doi.org/10.1016/0370-2693(96)00930-6} {\bibfield  {journal} {\bibinfo  {journal} {Physics Letters B}\ }\textbf {\bibinfo {volume} {384}},\ \bibinfo {pages} {353} (\bibinfo {year} {1996})}\BibitemShut {NoStop}%
\bibitem [{\citenamefont {{ALEPH collaboration}}\ \emph {et~al.}(1998)\citenamefont {{ALEPH collaboration}}, \citenamefont {Barate}, \citenamefont {D.} \emph {et~al.}}]{barate1998studies}%
  \BibitemOpen
  \bibfield  {author} {\bibinfo {author} {\bibnamefont {{ALEPH collaboration}}}, \bibinfo {author} {\bibfnamefont {R.}~\bibnamefont {Barate}}, \bibinfo {author} {\bibfnamefont {B.}~\bibnamefont {D.}}, \emph {et~al.},\ }\bibfield  {title} {\bibinfo {title} {Studies of quantum chromodynamics with the aleph detector},\ }\href {https://doi.org/10.1016/S0370-1573(97)00045-8} {\bibfield  {journal} {\bibinfo  {journal} {Physics Reports}\ }\textbf {\bibinfo {volume} {294}},\ \bibinfo {pages} {1} (\bibinfo {year} {1998})}\BibitemShut {NoStop}%
\bibitem [{\citenamefont {Gras}\ \emph {et~al.}(2017{\natexlab{a}})\citenamefont {Gras}, \citenamefont {H\"o~che}, \citenamefont {Kar} \emph {et~al.}}]{gras2017systematics}%
  \BibitemOpen
  \bibfield  {author} {\bibinfo {author} {\bibfnamefont {P.}~\bibnamefont {Gras}}, \bibinfo {author} {\bibfnamefont {S.}~\bibnamefont {H\"o~che}}, \bibinfo {author} {\bibfnamefont {D.}~\bibnamefont {Kar}}, \emph {et~al.},\ }\bibfield  {title} {\bibinfo {title} {Systematics of quark/gluon tagging},\ }\href {https://doi.org/10.1007/JHEP07(2017)091} {\bibfield  {journal} {\bibinfo  {journal} {Journal of High Energy Physics}\ }\textbf {\bibinfo {volume} {2017}},\ \bibinfo {pages} {091} (\bibinfo {year} {2017}{\natexlab{a}})}\BibitemShut {NoStop}%
\bibitem [{\citenamefont {Dissertori}\ \emph {et~al.}(2009)\citenamefont {Dissertori}, \citenamefont {Knowles},\ and\ \citenamefont {Schmelling}}]{Dissertori_Knowles_Schmelling_2009}%
  \BibitemOpen
  \bibfield  {author} {\bibinfo {author} {\bibfnamefont {G.}~\bibnamefont {Dissertori}}, \bibinfo {author} {\bibfnamefont {I.~G.}\ \bibnamefont {Knowles}},\ and\ \bibinfo {author} {\bibfnamefont {M.}~\bibnamefont {Schmelling}},\ }\href {https://doi.org/10.1093/acprof:oso/9780199566419.001.0001} {\emph {\bibinfo {title} {Quantum Chromodynamics: High Energy Experiments and theory}}}\ (\bibinfo  {publisher} {Clarendon Press; Oxford University Press},\ \bibinfo {year} {2009})\BibitemShut {NoStop}%
\bibitem [{\citenamefont {Capella}\ \emph {et~al.}(2000)\citenamefont {Capella}, \citenamefont {Dremin}, \citenamefont {Gary}, \citenamefont {Nechitailo},\ and\ \citenamefont {Van}}]{PhysRevD.61.074009}%
  \BibitemOpen
  \bibfield  {author} {\bibinfo {author} {\bibfnamefont {A.}~\bibnamefont {Capella}}, \bibinfo {author} {\bibfnamefont {I.~M.}\ \bibnamefont {Dremin}}, \bibinfo {author} {\bibfnamefont {J.~W.}\ \bibnamefont {Gary}}, \bibinfo {author} {\bibfnamefont {V.~A.}\ \bibnamefont {Nechitailo}},\ and\ \bibinfo {author} {\bibfnamefont {J.~T.~T.}\ \bibnamefont {Van}},\ }\bibfield  {title} {\bibinfo {title} {Evolution of average multiplicities of quark and gluon jets},\ }\href {https://doi.org/10.1103/PhysRevD.61.074009} {\bibfield  {journal} {\bibinfo  {journal} {Phys. Rev. D}\ }\textbf {\bibinfo {volume} {61}},\ \bibinfo {pages} {074009} (\bibinfo {year} {2000})}\BibitemShut {NoStop}%
\bibitem [{\citenamefont {{ZEUS Collaboration}}\ \emph {et~al.}(2008)\citenamefont {{ZEUS Collaboration}}, \citenamefont {Chekanov}, \citenamefont {Derrick}, \citenamefont {Magill} \emph {et~al.}}]{Chekanov_2008}%
  \BibitemOpen
  \bibfield  {author} {\bibinfo {author} {\bibnamefont {{ZEUS Collaboration}}}, \bibinfo {author} {\bibfnamefont {S.}~\bibnamefont {Chekanov}}, \bibinfo {author} {\bibfnamefont {M.}~\bibnamefont {Derrick}}, \bibinfo {author} {\bibfnamefont {S.}~\bibnamefont {Magill}}, \emph {et~al.},\ }\bibfield  {title} {\bibinfo {title} {Diffractive photoproduction of dijets in ep collisions at {HERA}},\ }\href {https://doi.org/10.1140/epjc/s10052-008-0598-2} {\bibfield  {journal} {\bibinfo  {journal} {The European Physical Journal C}\ }\textbf {\bibinfo {volume} {55}},\ \bibinfo {pages} {177} (\bibinfo {year} {2008})}\BibitemShut {NoStop}%
\bibitem [{\citenamefont {Glasman}(2001)}]{Glasman_2001}%
  \BibitemOpen
  \bibfield  {author} {\bibinfo {author} {\bibfnamefont {C.}~\bibnamefont {Glasman}},\ }\bibfield  {title} {\bibinfo {title} {Jet substructure at {HERA}},\ }\href {https://doi.org/10.1063/1.1402839} {\bibfield  {journal} {\bibinfo  {journal} {AIP Conference Proceedings}\ }\textbf {\bibinfo {volume} {571}},\ \bibinfo {pages} {203} (\bibinfo {year} {2001})}\BibitemShut {NoStop}%
\bibitem [{\citenamefont {Gallicchio}\ and\ \citenamefont {Schwartz}(2013)}]{Gallicchio_2013}%
  \BibitemOpen
  \bibfield  {author} {\bibinfo {author} {\bibfnamefont {J.}~\bibnamefont {Gallicchio}}\ and\ \bibinfo {author} {\bibfnamefont {M.~D.}\ \bibnamefont {Schwartz}},\ }\bibfield  {title} {\bibinfo {title} {Quark and gluon jet substructure},\ }\href {https://doi.org/10.1007/JHEP04(2013)090} {\bibfield  {journal} {\bibinfo  {journal} {Journal of High Energy Physics}\ }\textbf {\bibinfo {volume} {2013}},\ \bibinfo {pages} {090} (\bibinfo {year} {2013})}\BibitemShut {NoStop}%
\bibitem [{\citenamefont {Gallicchio}\ and\ \citenamefont {Schwartz}(2011)}]{PhysRevLett.107.172001}%
  \BibitemOpen
  \bibfield  {author} {\bibinfo {author} {\bibfnamefont {J.}~\bibnamefont {Gallicchio}}\ and\ \bibinfo {author} {\bibfnamefont {M.~D.}\ \bibnamefont {Schwartz}},\ }\bibfield  {title} {\bibinfo {title} {Quark and gluon tagging at the lhc},\ }\href {https://doi.org/10.1103/PhysRevLett.107.172001} {\bibfield  {journal} {\bibinfo  {journal} {Phys. Rev. Lett.}\ }\textbf {\bibinfo {volume} {107}},\ \bibinfo {pages} {172001} (\bibinfo {year} {2011})}\BibitemShut {NoStop}%
\bibitem [{\citenamefont {Frye}\ \emph {et~al.}(2017)\citenamefont {Frye} \emph {et~al.}}]{PhysRevD.frye_2017}%
  \BibitemOpen
  \bibfield  {author} {\bibinfo {author} {\bibfnamefont {C.}~\bibnamefont {Frye}} \emph {et~al.},\ }\bibfield  {title} {\bibinfo {title} {Factorization for groomed jet substructure beyond the next-to-leading logarithm},\ }\href {https://doi.org/10.1007/JHEP07(2016)064} {\bibfield  {journal} {\bibinfo  {journal} {Physical Review D}\ }\textbf {\bibinfo {volume} {95}},\ \bibinfo {pages} {034001} (\bibinfo {year} {2017})}\BibitemShut {NoStop}%
\bibitem [{\citenamefont {Whitmore}(2019)}]{Whitmore_2019}%
  \BibitemOpen
  \bibfield  {author} {\bibinfo {author} {\bibfnamefont {B.}~\bibnamefont {Whitmore}},\ }\emph {\bibinfo {title} {Search for Beyond the Standard Model signals in a quark-gluon tagged dijet final state with the {ATLAS} detector}},\ \href {https://doi.org/10.17635/lancaster/thesis/738} {Ph.D. thesis},\ \bibinfo  {school} {Lancaster University} (\bibinfo {year} {2019})\BibitemShut {NoStop}%
\bibitem [{\citenamefont {Bhattacherjee}\ \emph {et~al.}(2017)\citenamefont {Bhattacherjee}, \citenamefont {Mukhopadhyay}, \citenamefont {Nojiri} \emph {et~al.}}]{Bhattacherjee_2017}%
  \BibitemOpen
  \bibfield  {author} {\bibinfo {author} {\bibfnamefont {B.}~\bibnamefont {Bhattacherjee}}, \bibinfo {author} {\bibfnamefont {S.}~\bibnamefont {Mukhopadhyay}}, \bibinfo {author} {\bibfnamefont {M.}~\bibnamefont {Nojiri}}, \emph {et~al.},\ }\bibfield  {title} {\bibinfo {title} {Quark-gluon discrimination in the search for gluino pair production at the {LHC}},\ }\href {https://doi.org/10.1007/JHEP01(2017)044} {\bibfield  {journal} {\bibinfo  {journal} {Journal of High Energy Physics}\ }\textbf {\bibinfo {volume} {2017}},\ \bibinfo {pages} {044} (\bibinfo {year} {2017})}\BibitemShut {NoStop}%
\bibitem [{\citenamefont {Chakraborty}\ \emph {et~al.}(2018)\citenamefont {Chakraborty}, \citenamefont {Iyer},\ and\ \citenamefont {Roy}}]{Chakraborty2018-wi}%
  \BibitemOpen
  \bibfield  {author} {\bibinfo {author} {\bibfnamefont {A.}~\bibnamefont {Chakraborty}}, \bibinfo {author} {\bibfnamefont {A.~M.}\ \bibnamefont {Iyer}},\ and\ \bibinfo {author} {\bibfnamefont {T.~S.}\ \bibnamefont {Roy}},\ }\bibfield  {title} {\bibinfo {title} {A framework for finding anomalous objects at the {LHC}},\ }\href {https://doi.org/10.1016/j.nuclphysb.2018.05.019} {\bibfield  {journal} {\bibinfo  {journal} {Nucl. Phys. B.}\ }\textbf {\bibinfo {volume} {932}},\ \bibinfo {pages} {439} (\bibinfo {year} {2018})}\BibitemShut {NoStop}%
\bibitem [{\citenamefont {Ferreira~de Lima}\ \emph {et~al.}(2017)\citenamefont {Ferreira~de Lima}, \citenamefont {Petrov}, \citenamefont {Soper},\ and\ \citenamefont {Spannowsky}}]{PhysRevD.95.034001}%
  \BibitemOpen
  \bibfield  {author} {\bibinfo {author} {\bibfnamefont {D.}~\bibnamefont {Ferreira~de Lima}}, \bibinfo {author} {\bibfnamefont {P.}~\bibnamefont {Petrov}}, \bibinfo {author} {\bibfnamefont {D.}~\bibnamefont {Soper}},\ and\ \bibinfo {author} {\bibfnamefont {M.}~\bibnamefont {Spannowsky}},\ }\bibfield  {title} {\bibinfo {title} {Quark-gluon tagging with shower deconstruction: Unearthing dark matter and higgs couplings},\ }\href {https://doi.org/10.1103/PhysRevD.95.034001} {\bibfield  {journal} {\bibinfo  {journal} {Phys. Rev. D}\ }\textbf {\bibinfo {volume} {95}},\ \bibinfo {pages} {034001} (\bibinfo {year} {2017})}\BibitemShut {NoStop}%
\bibitem [{\citenamefont {Sakaki}(2019)}]{PhysRevD.99.114012}%
  \BibitemOpen
  \bibfield  {author} {\bibinfo {author} {\bibfnamefont {Y.}~\bibnamefont {Sakaki}},\ }\bibfield  {title} {\bibinfo {title} {Quark jet rates and quark-gluon discrimination in multijet final states},\ }\href {https://doi.org/10.1103/PhysRevD.99.114012} {\bibfield  {journal} {\bibinfo  {journal} {Phys. Rev. D}\ }\textbf {\bibinfo {volume} {99}},\ \bibinfo {pages} {114012} (\bibinfo {year} {2019})}\BibitemShut {NoStop}%
\bibitem [{\citenamefont {Komiske}\ \emph {et~al.}(2022)\citenamefont {Komiske}, \citenamefont {Kryhin},\ and\ \citenamefont {Thaler}}]{Komiske_2022}%
  \BibitemOpen
  \bibfield  {author} {\bibinfo {author} {\bibfnamefont {P.~T.}\ \bibnamefont {Komiske}}, \bibinfo {author} {\bibfnamefont {S.}~\bibnamefont {Kryhin}},\ and\ \bibinfo {author} {\bibfnamefont {J.}~\bibnamefont {Thaler}},\ }\bibfield  {title} {\bibinfo {title} {Exploring the space of jets with {CMS} open data},\ }\href {https://doi.org/10.48550/arXiv.1908.08542} {\bibfield  {journal} {\bibinfo  {journal} {Physical Review D}\ }\textbf {\bibinfo {volume} {106}},\ \bibinfo {pages} {014009} (\bibinfo {year} {2022})}\BibitemShut {NoStop}%
\bibitem [{\citenamefont {Komiske}\ \emph {et~al.}(2018)\citenamefont {Komiske}, \citenamefont {Metodiev},\ and\ \citenamefont {Thaler}}]{Komiske_2018}%
  \BibitemOpen
  \bibfield  {author} {\bibinfo {author} {\bibfnamefont {P.~T.}\ \bibnamefont {Komiske}}, \bibinfo {author} {\bibfnamefont {E.~M.}\ \bibnamefont {Metodiev}},\ and\ \bibinfo {author} {\bibfnamefont {J.}~\bibnamefont {Thaler}},\ }\bibfield  {title} {\bibinfo {title} {An operational definition of quark and gluon jets},\ }\href {https://doi.org/10.1007/JHEP11(2018)059} {\bibfield  {journal} {\bibinfo  {journal} {Journal of High Energy Physics}\ }\textbf {\bibinfo {volume} {2018}},\ \bibinfo {pages} {059} (\bibinfo {year} {2018})}\BibitemShut {NoStop}%
\bibitem [{\citenamefont {Bhattacherjee}\ \emph {et~al.}(2015)\citenamefont {Bhattacherjee}, \citenamefont {Mukhopadhyay}, \citenamefont {Nojiri}, \citenamefont {Sakaki},\ and\ \citenamefont {Webber}}]{Bhattacherjee_2015}%
  \BibitemOpen
  \bibfield  {author} {\bibinfo {author} {\bibfnamefont {B.}~\bibnamefont {Bhattacherjee}}, \bibinfo {author} {\bibfnamefont {S.}~\bibnamefont {Mukhopadhyay}}, \bibinfo {author} {\bibfnamefont {M.~M.}\ \bibnamefont {Nojiri}}, \bibinfo {author} {\bibfnamefont {Y.}~\bibnamefont {Sakaki}},\ and\ \bibinfo {author} {\bibfnamefont {B.~R.}\ \bibnamefont {Webber}},\ }\bibfield  {title} {\bibinfo {title} {Associated jet and subjet rates in light-quark and gluon jet discrimination},\ }\bibfield  {journal} {\bibinfo  {journal} {Journal of High Energy Physics}\ }\textbf {\bibinfo {volume} {2015}},\ \href {https://doi.org/https://doi.org/10.1007/JHEP04(2015)131} {https://doi.org/10.1007/JHEP04(2015)131} (\bibinfo {year} {2015})\BibitemShut {NoStop}%
\bibitem [{\citenamefont {Jankowiak}\ and\ \citenamefont {Larkoski}(2011)}]{Jankowiak2011-ht}%
  \BibitemOpen
  \bibfield  {author} {\bibinfo {author} {\bibfnamefont {M.}~\bibnamefont {Jankowiak}}\ and\ \bibinfo {author} {\bibfnamefont {A.~J.}\ \bibnamefont {Larkoski}},\ }\bibfield  {title} {\bibinfo {title} {Jet substructure without trees},\ }\href {https://doi.org/10.1007/JHEP06(2011)057} {\bibfield  {journal} {\bibinfo  {journal} {J. High Energy Phys.}\ }\textbf {\bibinfo {volume} {2011}}\bibinfo  {number} { (6)}}\BibitemShut {NoStop}%
\bibitem [{\citenamefont {Kasieczka}\ \emph {et~al.}(2017)\citenamefont {Kasieczka}, \citenamefont {Plehn}, \citenamefont {Russell},\ and\ \citenamefont {Schell}}]{Kasieczka2017-bu}%
  \BibitemOpen
\bibfield  {number} {  }\bibfield  {author} {\bibinfo {author} {\bibfnamefont {G.}~\bibnamefont {Kasieczka}}, \bibinfo {author} {\bibfnamefont {T.}~\bibnamefont {Plehn}}, \bibinfo {author} {\bibfnamefont {M.}~\bibnamefont {Russell}},\ and\ \bibinfo {author} {\bibfnamefont {T.}~\bibnamefont {Schell}},\ }\bibfield  {title} {\bibinfo {title} {Deep-learning top taggers or the end of {QCD}?},\ }\href {https://doi.org/10.1007/JHEP05(2017)006} {\bibfield  {journal} {\bibinfo  {journal} {J. High Energy Phys.}\ }\textbf {\bibinfo {volume} {2017}}\bibinfo  {number} { (5)}}\BibitemShut {NoStop}%
\bibitem [{\citenamefont {Cheng}(2018)}]{Cheng2018-fg}%
  \BibitemOpen
\bibfield  {number} {  }\bibfield  {author} {\bibinfo {author} {\bibfnamefont {T.}~\bibnamefont {Cheng}},\ }\bibfield  {title} {\bibinfo {title} {Recursive neural networks in quark/gluon tagging},\ }\bibfield  {journal} {\bibinfo  {journal} {Comput. Softw. Big Sci.}\ }\textbf {\bibinfo {volume} {2}},\ \href {https://doi.org/10.1007/s41781-018-0007-y} {10.1007/s41781-018-0007-y} (\bibinfo {year} {2018})\BibitemShut {NoStop}%
\bibitem [{\citenamefont {Gras}\ \emph {et~al.}(2017{\natexlab{b}})\citenamefont {Gras}, \citenamefont {H{\"o}che}, \citenamefont {Kar}, \citenamefont {Larkoski}, \citenamefont {L{\"o}nnblad}, \citenamefont {Pl{\"a}tzer}, \citenamefont {Si{\'o}dmok}, \citenamefont {Skands}, \citenamefont {Soyez},\ and\ \citenamefont {Thaler}}]{Gras2017-me}%
  \BibitemOpen
  \bibfield  {author} {\bibinfo {author} {\bibfnamefont {P.}~\bibnamefont {Gras}}, \bibinfo {author} {\bibfnamefont {S.}~\bibnamefont {H{\"o}che}}, \bibinfo {author} {\bibfnamefont {D.}~\bibnamefont {Kar}}, \bibinfo {author} {\bibfnamefont {A.}~\bibnamefont {Larkoski}}, \bibinfo {author} {\bibfnamefont {L.}~\bibnamefont {L{\"o}nnblad}}, \bibinfo {author} {\bibfnamefont {S.}~\bibnamefont {Pl{\"a}tzer}}, \bibinfo {author} {\bibfnamefont {A.}~\bibnamefont {Si{\'o}dmok}}, \bibinfo {author} {\bibfnamefont {P.}~\bibnamefont {Skands}}, \bibinfo {author} {\bibfnamefont {G.}~\bibnamefont {Soyez}},\ and\ \bibinfo {author} {\bibfnamefont {J.}~\bibnamefont {Thaler}},\ }\bibfield  {title} {\bibinfo {title} {Systematics of quark/gluon tagging},\ }\href {https://doi.org/10.1007/JHEP07(2017)091} {\bibfield  {journal} {\bibinfo  {journal} {J. High Energy Phys.}\ }\textbf {\bibinfo {volume} {2017}}\bibinfo  {number} { (7)}}\BibitemShut {NoStop}%
\bibitem [{\citenamefont {Davighi}\ and\ \citenamefont {Harris}(2018)}]{Davighi_2018}%
  \BibitemOpen
\bibfield  {number} {  }\bibfield  {author} {\bibinfo {author} {\bibfnamefont {J.}~\bibnamefont {Davighi}}\ and\ \bibinfo {author} {\bibfnamefont {P.}~\bibnamefont {Harris}},\ }\bibfield  {title} {\bibinfo {title} {Fractal-based observables to probe jet substructure of quarks and gluons},\ }\href {https://doi.org/10.1140/epjc/s10052-018-5819-8} {\bibfield  {journal} {\bibinfo  {journal} {The European Physical Journal C}\ }\textbf {\bibinfo {volume} {78}},\ \bibinfo {pages} {334} (\bibinfo {year} {2018})}\BibitemShut {NoStop}%
\bibitem [{\citenamefont {Metodiev}\ and\ \citenamefont {Thaler}(2018)}]{Metodiev_2018}%
  \BibitemOpen
  \bibfield  {author} {\bibinfo {author} {\bibfnamefont {E.~M.}\ \bibnamefont {Metodiev}}\ and\ \bibinfo {author} {\bibfnamefont {J.}~\bibnamefont {Thaler}},\ }\bibfield  {title} {\bibinfo {title} {Jet topics: Disentangling quarks and gluons at colliders},\ }\href {https://doi.org/10.1103/PhysRevLett.120.241602} {\bibfield  {journal} {\bibinfo  {journal} {Phys. Rev. Lett.}\ }\textbf {\bibinfo {volume} {120}},\ \bibinfo {pages} {241602} (\bibinfo {year} {2018})}\BibitemShut {NoStop}%
\bibitem [{\citenamefont {Fodor}(1990)}]{PhysRevD.41.1726}%
  \BibitemOpen
  \bibfield  {author} {\bibinfo {author} {\bibfnamefont {Z.}~\bibnamefont {Fodor}},\ }\bibfield  {title} {\bibinfo {title} {How to see the differences between quark and gluon jets},\ }\href {https://doi.org/10.1103/PhysRevD.41.1726} {\bibfield  {journal} {\bibinfo  {journal} {Phys. Rev. D}\ }\textbf {\bibinfo {volume} {41}},\ \bibinfo {pages} {1726} (\bibinfo {year} {1990})}\BibitemShut {NoStop}%
\bibitem [{\citenamefont {Larkoski}\ \emph {et~al.}(2013)\citenamefont {Larkoski}, \citenamefont {Salam},\ and\ \citenamefont {Thaler}}]{Larkoski:2014gra}%
  \BibitemOpen
  \bibfield  {author} {\bibinfo {author} {\bibfnamefont {A.~J.}\ \bibnamefont {Larkoski}}, \bibinfo {author} {\bibfnamefont {G.~P.}\ \bibnamefont {Salam}},\ and\ \bibinfo {author} {\bibfnamefont {J.}~\bibnamefont {Thaler}},\ }\bibfield  {title} {\bibinfo {title} {Energy correlation functions for jet substructure},\ }\href {https://doi.org/10.1007/JHEP06(2013)108} {\bibfield  {journal} {\bibinfo  {journal} {Journal of High Energy Physics}\ }\textbf {\bibinfo {volume} {2013}},\ \bibinfo {pages} {108} (\bibinfo {year} {2013})}\BibitemShut {NoStop}%
\bibitem [{\citenamefont {Larkoski}\ \emph {et~al.}(2020)\citenamefont {Larkoski}, \citenamefont {Moult},\ and\ \citenamefont {Nachman}}]{larkoski2020jet}%
  \BibitemOpen
  \bibfield  {author} {\bibinfo {author} {\bibfnamefont {A.~J.}\ \bibnamefont {Larkoski}}, \bibinfo {author} {\bibfnamefont {I.}~\bibnamefont {Moult}},\ and\ \bibinfo {author} {\bibfnamefont {B.}~\bibnamefont {Nachman}},\ }\bibfield  {title} {\bibinfo {title} {Jet substructure at the large hadron collider: A review of recent advances in theory and machine learning},\ }\href {https://doi.org/https://doi.org/10.1016/j.physrep.2019.11.001} {\bibfield  {journal} {\bibinfo  {journal} {Physics Reports}\ }\textbf {\bibinfo {volume} {841}},\ \bibinfo {pages} {1} (\bibinfo {year} {2020})}\BibitemShut {NoStop}%
\bibitem [{\citenamefont {Mulligan}\ and\ \citenamefont {P{\l}osko{\'n}}(2020)}]{mulligan2020identifying}%
  \BibitemOpen
  \bibfield  {author} {\bibinfo {author} {\bibfnamefont {J.}~\bibnamefont {Mulligan}}\ and\ \bibinfo {author} {\bibfnamefont {M.}~\bibnamefont {P{\l}osko{\'n}}},\ }\bibfield  {title} {\bibinfo {title} {Identifying groomed jet splittings in heavy-ion collisions},\ }\href@noop {} {\bibfield  {journal} {\bibinfo  {journal} {Physical Review C}\ }\textbf {\bibinfo {volume} {102}},\ \bibinfo {pages} {044913} (\bibinfo {year} {2020})}\BibitemShut {NoStop}%
\bibitem [{\citenamefont {Salam}(2010)}]{salam2010towards}%
  \BibitemOpen
  \bibfield  {author} {\bibinfo {author} {\bibfnamefont {G.~P.}\ \bibnamefont {Salam}},\ }\bibfield  {title} {\bibinfo {title} {Towards jetography},\ }\href {https://doi.org/10.1140/epjc/s10052-010-1314-6} {\bibfield  {journal} {\bibinfo  {journal} {The European Physical Journal C}\ }\textbf {\bibinfo {volume} {67}},\ \bibinfo {pages} {637} (\bibinfo {year} {2010})}\BibitemShut {NoStop}%
\bibitem [{\citenamefont {Rabbertz}(2017)}]{klaus2018jet}%
  \BibitemOpen
  \bibfield  {author} {\bibinfo {author} {\bibfnamefont {K.}~\bibnamefont {Rabbertz}},\ }\href {https://doi.org/10.1007/978-3-319-42115-5} {\emph {\bibinfo {title} {{Jet Physics at the LHC}: {The Strong Force beyond the TeV Scale}}}},\ \bibinfo {series} {Springer Tracts in Modern Physics}, Vol.\ \bibinfo {volume} {268}\ (\bibinfo  {publisher} {Springer},\ \bibinfo {address} {Berlin},\ \bibinfo {year} {2017})\BibitemShut {NoStop}%
\bibitem [{\citenamefont {Marzani}\ \emph {et~al.}(2019)\citenamefont {Marzani}, \citenamefont {Soyez},\ and\ \citenamefont {Spannowsky}}]{Marzani_2019}%
  \BibitemOpen
  \bibfield  {author} {\bibinfo {author} {\bibfnamefont {S.}~\bibnamefont {Marzani}}, \bibinfo {author} {\bibfnamefont {G.}~\bibnamefont {Soyez}},\ and\ \bibinfo {author} {\bibfnamefont {M.}~\bibnamefont {Spannowsky}},\ }\bibfield  {title} {\bibinfo {title} {Looking inside jets},\ }\href {https://api.semanticscholar.org/CorpusID:119369716} {\bibfield  {journal} {\bibinfo  {journal} {Lecture Notes in Physics}\ } (\bibinfo {year} {2019})}\BibitemShut {NoStop}%
\bibitem [{\citenamefont {Bonilla}\ \emph {et~al.}(2022)\citenamefont {Bonilla} \emph {et~al.}}]{jet_subst}%
  \BibitemOpen
  \bibfield  {author} {\bibinfo {author} {\bibfnamefont {J.}~\bibnamefont {Bonilla}} \emph {et~al.},\ }\bibfield  {title} {\bibinfo {title} {Jets and jet substructure at future colliders},\ }\bibfield  {journal} {\bibinfo  {journal} {Frontiers in Physics}\ }\textbf {\bibinfo {volume} {10-2022}},\ \href {https://doi.org/https://doi.org/10.3389/fphy.2022.897719} {https://doi.org/10.3389/fphy.2022.897719} (\bibinfo {year} {2022})\BibitemShut {NoStop}%
\bibitem [{\citenamefont {Frixione}\ and\ \citenamefont {Ridolfi}(1997)}]{Frixione1997jet}%
  \BibitemOpen
  \bibfield  {author} {\bibinfo {author} {\bibfnamefont {S.}~\bibnamefont {Frixione}}\ and\ \bibinfo {author} {\bibfnamefont {G.}~\bibnamefont {Ridolfi}},\ }\bibfield  {title} {\bibinfo {title} {Jet photoproduction at hera},\ }\href {https://doi.org/https://doi.org/10.1016/S0550-3213(97)00575-0} {\bibfield  {journal} {\bibinfo  {journal} {Nuclear Physics B}\ }\textbf {\bibinfo {volume} {507}},\ \bibinfo {pages} {315} (\bibinfo {year} {1997})}\BibitemShut {NoStop}%
\bibitem [{\citenamefont {{ZEUS Collaboration}}\ \emph {et~al.}(2007)\citenamefont {{ZEUS Collaboration}}, \citenamefont {Chekanov}, \citenamefont {Derrick} \emph {et~al.}}]{chekanov2008three}%
  \BibitemOpen
  \bibfield  {author} {\bibinfo {author} {\bibnamefont {{ZEUS Collaboration}}}, \bibinfo {author} {\bibfnamefont {B.}~\bibnamefont {Chekanov}}, \bibinfo {author} {\bibfnamefont {M.}~\bibnamefont {Derrick}}, \emph {et~al.},\ }\bibfield  {title} {\bibinfo {title} {Jet-radius dependence of inclusive-jet cross sections in deep inelastic scattering at hera},\ }\href {https://doi.org/https://doi.org/10.1016/j.physletb.2007.03.039} {\bibfield  {journal} {\bibinfo  {journal} {Physics Letters B}\ }\textbf {\bibinfo {volume} {649}},\ \bibinfo {pages} {12} (\bibinfo {year} {2007})}\BibitemShut {NoStop}%
\bibitem [{\citenamefont {Jain}\ \emph {et~al.}(2023)\citenamefont {Jain}, \citenamefont {Aggarwal},\ and\ \citenamefont {Kaur}}]{PhysRevD.107.116002}%
  \BibitemOpen
  \bibfield  {author} {\bibinfo {author} {\bibfnamefont {S.}~\bibnamefont {Jain}}, \bibinfo {author} {\bibfnamefont {R.}~\bibnamefont {Aggarwal}},\ and\ \bibinfo {author} {\bibfnamefont {M.}~\bibnamefont {Kaur}},\ }\bibfield  {title} {\bibinfo {title} {Jet substructure in neutral current deep inelastic ${e}^{+}p$ scattering at the upcoming electron-ion collider},\ }\href {https://doi.org/10.1103/PhysRevD.107.116002} {\bibfield  {journal} {\bibinfo  {journal} {Phys. Rev. D}\ }\textbf {\bibinfo {volume} {107}},\ \bibinfo {pages} {116002} (\bibinfo {year} {2023})}\BibitemShut {NoStop}%
\bibitem [{\citenamefont {{ZEUS Collaboration}}\ \emph {et~al.}(1998)\citenamefont {{ZEUS Collaboration}}, \citenamefont {Breitweg} \emph {et~al.}}]{zeus1998measurement}%
  \BibitemOpen
  \bibfield  {author} {\bibinfo {author} {\bibnamefont {{ZEUS Collaboration}}}, \bibinfo {author} {\bibfnamefont {J.}~\bibnamefont {Breitweg}}, \emph {et~al.},\ }\bibfield  {title} {\bibinfo {title} {Measurement of jet shapes in photoproduction at {HERA}},\ }\href {https://doi.org/10.1007/PL00021562} {\bibfield  {journal} {\bibinfo  {journal} {The European Physical Journal C}\ }\textbf {\bibinfo {volume} {2}},\ \bibinfo {pages} {61} (\bibinfo {year} {1998})}\BibitemShut {NoStop}%
\bibitem [{\citenamefont {Arratia}\ \emph {et~al.}(2020)\citenamefont {Arratia}, \citenamefont {Song}, \citenamefont {Ringer},\ and\ \citenamefont {Jacak}}]{Arratia_2020}%
  \BibitemOpen
  \bibfield  {author} {\bibinfo {author} {\bibfnamefont {M.}~\bibnamefont {Arratia}}, \bibinfo {author} {\bibfnamefont {Y.}~\bibnamefont {Song}}, \bibinfo {author} {\bibfnamefont {F.}~\bibnamefont {Ringer}},\ and\ \bibinfo {author} {\bibfnamefont {B.~V.}\ \bibnamefont {Jacak}},\ }\bibfield  {title} {\bibinfo {title} {Jets as precision probes in electron-nucleus collisions at the future electron-ion collider},\ }\href {https://doi.org/10.1103/PhysRevC.101.065204} {\bibfield  {journal} {\bibinfo  {journal} {Phys. Rev. C}\ }\textbf {\bibinfo {volume} {101}},\ \bibinfo {pages} {065204} (\bibinfo {year} {2020})}\BibitemShut {NoStop}%
\bibitem [{\citenamefont {Skands}\ \emph {et~al.}(2014)\citenamefont {Skands}, \citenamefont {Carrazza},\ and\ \citenamefont {Rojo}}]{skands2014tuning}%
  \BibitemOpen
  \bibfield  {author} {\bibinfo {author} {\bibfnamefont {P.}~\bibnamefont {Skands}}, \bibinfo {author} {\bibfnamefont {S.}~\bibnamefont {Carrazza}},\ and\ \bibinfo {author} {\bibfnamefont {J.}~\bibnamefont {Rojo}},\ }\bibfield  {title} {\bibinfo {title} {Tuning pythia 8.1: the monash 2013 tune},\ }\href {https://doi.org/10.1140/epjc/s10052-014-3024-y} {\bibfield  {journal} {\bibinfo  {journal} {The European Physical Journal C}\ }\textbf {\bibinfo {volume} {74}},\ \bibinfo {pages} {3024} (\bibinfo {year} {2014})}\BibitemShut {NoStop}%
\bibitem [{\citenamefont {Deans}(2013)}]{deans2013progress}%
  \BibitemOpen
  \bibfield  {author} {\bibinfo {author} {\bibfnamefont {C.~S.}\ \bibnamefont {Deans}},\ }\bibfield  {title} {\bibinfo {title} {Progress in the nnpdf global analysis},\ }\href {https://doi.org/10.48550/arXiv.1304.2781} {\bibfield  {journal} {\bibinfo  {journal} {arXiv preprint arXiv:1304.2781}\ } (\bibinfo {year} {2013})}\BibitemShut {NoStop}%
\bibitem [{\citenamefont {Cornet}\ \emph {et~al.}(2003)\citenamefont {Cornet}, \citenamefont {Jankowski}, \citenamefont {Krawczyk},\ and\ \citenamefont {Lorca}}]{cornet2003new}%
  \BibitemOpen
  \bibfield  {author} {\bibinfo {author} {\bibfnamefont {F.}~\bibnamefont {Cornet}}, \bibinfo {author} {\bibfnamefont {P.}~\bibnamefont {Jankowski}}, \bibinfo {author} {\bibfnamefont {M.}~\bibnamefont {Krawczyk}},\ and\ \bibinfo {author} {\bibfnamefont {A.}~\bibnamefont {Lorca}},\ }\bibfield  {title} {\bibinfo {title} {New 5-flavor lo analysis and parametrization of parton distributions in the real photon},\ }\href {https://doi.org/https://doi.org/10.1103/PhysRevD.68.014010} {\bibfield  {journal} {\bibinfo  {journal} {Physical Review D}\ }\textbf {\bibinfo {volume} {68}},\ \bibinfo {pages} {014010} (\bibinfo {year} {2003})}\BibitemShut {NoStop}%
\bibitem [{\citenamefont {{ZEUS Collaboration}}\ \emph {et~al.}(2004)\citenamefont {{ZEUS Collaboration}}, \citenamefont {Chekanov}, \citenamefont {Derrick} \emph {et~al.}}]{Chekanov2004-te}%
  \BibitemOpen
  \bibfield  {author} {\bibinfo {author} {\bibnamefont {{ZEUS Collaboration}}}, \bibinfo {author} {\bibfnamefont {B.}~\bibnamefont {Chekanov}}, \bibinfo {author} {\bibfnamefont {M.}~\bibnamefont {Derrick}}, \emph {et~al.},\ }\bibfield  {title} {\bibinfo {title} {Substructure dependence of jet cross sections at {HERA} and determination of $\alpha_{s}$},\ }\href {https://doi.org/10.1016/j.nuclphysb.2004.08.049} {\bibfield  {journal} {\bibinfo  {journal} {Nucl. Phys. B.}\ }\textbf {\bibinfo {volume} {700}},\ \bibinfo {pages} {3} (\bibinfo {year} {2004})}\BibitemShut {NoStop}%
\bibitem [{\citenamefont {Chekanov}(2002)}]{chekanov2002jet}%
  \BibitemOpen
  \bibfield  {author} {\bibinfo {author} {\bibfnamefont {S.~V.}\ \bibnamefont {Chekanov}},\ }\href@noop {} {\bibinfo {title} {Jet algorithms: a minireview}} (\bibinfo {year} {2002}),\ \Eprint {https://arxiv.org/abs/hep-ph/0211298} {arXiv:hep-ph/0211298} \BibitemShut {NoStop}%
\bibitem [{\citenamefont {Catani}\ \emph {et~al.}(1993)\citenamefont {Catani}, \citenamefont {Dokshitzer}, \citenamefont {Seymour},\ and\ \citenamefont {Webber}}]{Catani:1993hr}%
  \BibitemOpen
  \bibfield  {author} {\bibinfo {author} {\bibfnamefont {S.}~\bibnamefont {Catani}}, \bibinfo {author} {\bibfnamefont {Y.~L.}\ \bibnamefont {Dokshitzer}}, \bibinfo {author} {\bibfnamefont {M.~H.}\ \bibnamefont {Seymour}},\ and\ \bibinfo {author} {\bibfnamefont {B.~R.}\ \bibnamefont {Webber}},\ }\bibfield  {title} {\bibinfo {title} {Longitudinally-invariant $k_t$-clustering algorithms for hadron-hadron collisions},\ }\href {https://doi.org/https://doi.org/10.1016/0550-3213(93)90166-M} {\bibfield  {journal} {\bibinfo  {journal} {Nuclear Physics B}\ }\textbf {\bibinfo {volume} {406}},\ \bibinfo {pages} {187} (\bibinfo {year} {1993})}\BibitemShut {NoStop}%
\bibitem [{\citenamefont {Cacciari}\ \emph {et~al.}(2012)\citenamefont {Cacciari}, \citenamefont {Salam},\ and\ \citenamefont {Soyez}}]{Cacciari_2012}%
  \BibitemOpen
  \bibfield  {author} {\bibinfo {author} {\bibfnamefont {M.}~\bibnamefont {Cacciari}}, \bibinfo {author} {\bibfnamefont {G.~P.}\ \bibnamefont {Salam}},\ and\ \bibinfo {author} {\bibfnamefont {G.}~\bibnamefont {Soyez}},\ }\bibfield  {title} {\bibinfo {title} {{FastJet} user manual},\ }\href {https://doi.org/10.1140/epjc/s10052-012-1896-2} {\bibfield  {journal} {\bibinfo  {journal} {The European Physical Journal C}\ }\textbf {\bibinfo {volume} {72}},\ \bibinfo {pages} {1896} (\bibinfo {year} {2012})}\BibitemShut {NoStop}%
\bibitem [{\citenamefont {Accardi}\ \emph {et~al.}(2016)\citenamefont {Accardi} \emph {et~al.}}]{Accardi:2012qut}%
  \BibitemOpen
  \bibfield  {author} {\bibinfo {author} {\bibfnamefont {A.}~\bibnamefont {Accardi}} \emph {et~al.},\ }\bibfield  {title} {\bibinfo {title} {Electron-ion collider: The next qcd frontier},\ }\href {https://doi.org/10.1140/epja/i2016-16268-9} {\bibfield  {journal} {\bibinfo  {journal} {The European Physical Journal A}\ }\textbf {\bibinfo {volume} {52}},\ \bibinfo {pages} {268} (\bibinfo {year} {2016})}\BibitemShut {NoStop}%
\bibitem [{\citenamefont {Bellm}\ \emph {et~al.}(2016)\citenamefont {Bellm}, \citenamefont {Gieseke}, \citenamefont {Grellscheid}, \citenamefont {Pl{\"a}tzer}, \citenamefont {Rauch}, \citenamefont {Reuschle}, \citenamefont {Richardson}, \citenamefont {Schichtel}, \citenamefont {Seymour}, \citenamefont {Si{\'o}dmok} \emph {et~al.}}]{bellm2016herwig}%
  \BibitemOpen
  \bibfield  {author} {\bibinfo {author} {\bibfnamefont {J.}~\bibnamefont {Bellm}}, \bibinfo {author} {\bibfnamefont {S.}~\bibnamefont {Gieseke}}, \bibinfo {author} {\bibfnamefont {D.}~\bibnamefont {Grellscheid}}, \bibinfo {author} {\bibfnamefont {S.}~\bibnamefont {Pl{\"a}tzer}}, \bibinfo {author} {\bibfnamefont {M.}~\bibnamefont {Rauch}}, \bibinfo {author} {\bibfnamefont {C.}~\bibnamefont {Reuschle}}, \bibinfo {author} {\bibfnamefont {P.}~\bibnamefont {Richardson}}, \bibinfo {author} {\bibfnamefont {P.}~\bibnamefont {Schichtel}}, \bibinfo {author} {\bibfnamefont {M.~H.}\ \bibnamefont {Seymour}}, \bibinfo {author} {\bibfnamefont {A.}~\bibnamefont {Si{\'o}dmok}}, \emph {et~al.},\ }\bibfield  {title} {\bibinfo {title} {Herwig 7.0/herwig++ 3.0 release note},\ }\href {https://doi.org/10.1140/epjc/s10052-016-4018-8} {\bibfield  {journal} {\bibinfo  {journal} {The European Physical Journal C}\ }\textbf {\bibinfo {volume} {76}},\ \bibinfo {pages} {196} (\bibinfo {year} {2016})}\BibitemShut {NoStop}%
\bibitem [{\citenamefont {Khalek}\ \emph {et~al.}(2022)\citenamefont {Khalek}, \citenamefont {Accardi}, \citenamefont {Adam}, \citenamefont {Adamiak}, \citenamefont {Akers}, \citenamefont {Albaladejo}, \citenamefont {Al-Bataineh}, \citenamefont {Alexeev}, \citenamefont {Ameli}, \citenamefont {Antonioli} \emph {et~al.}}]{khalek2022science}%
  \BibitemOpen
  \bibfield  {author} {\bibinfo {author} {\bibfnamefont {R.~A.}\ \bibnamefont {Khalek}}, \bibinfo {author} {\bibfnamefont {A.}~\bibnamefont {Accardi}}, \bibinfo {author} {\bibfnamefont {J.}~\bibnamefont {Adam}}, \bibinfo {author} {\bibfnamefont {D.}~\bibnamefont {Adamiak}}, \bibinfo {author} {\bibfnamefont {W.}~\bibnamefont {Akers}}, \bibinfo {author} {\bibfnamefont {M.}~\bibnamefont {Albaladejo}}, \bibinfo {author} {\bibfnamefont {A.}~\bibnamefont {Al-Bataineh}}, \bibinfo {author} {\bibfnamefont {M.}~\bibnamefont {Alexeev}}, \bibinfo {author} {\bibfnamefont {F.}~\bibnamefont {Ameli}}, \bibinfo {author} {\bibfnamefont {P.}~\bibnamefont {Antonioli}}, \emph {et~al.},\ }\bibfield  {title} {\bibinfo {title} {Science requirements and detector concepts for the electron-ion collider: Eic yellow report},\ }\href {https://doi.org/https://doi.org/10.1016/j.nuclphysa.2022.122447} {\bibfield  {journal} {\bibinfo  {journal} {Nuclear Physics A}\ }\textbf {\bibinfo {volume} {1026}},\ \bibinfo {pages} {122447} (\bibinfo
  {year} {2022})}\BibitemShut {NoStop}%
\bibitem [{\citenamefont {Page}\ \emph {et~al.}(2020{\natexlab{b}})\citenamefont {Page}, \citenamefont {Chu},\ and\ \citenamefont {Aschenauer}}]{PhysRevD_101}%
  \BibitemOpen
  \bibfield  {author} {\bibinfo {author} {\bibfnamefont {B.~S.}\ \bibnamefont {Page}}, \bibinfo {author} {\bibfnamefont {X.}~\bibnamefont {Chu}},\ and\ \bibinfo {author} {\bibfnamefont {E.~C.}\ \bibnamefont {Aschenauer}},\ }\bibfield  {title} {\bibinfo {title} {Experimental aspects of jet physics at a future eic},\ }\href {https://doi.org/10.1103/PhysRevD.101.072003} {\bibfield  {journal} {\bibinfo  {journal} {Phys. Rev. D}\ }\textbf {\bibinfo {volume} {101}},\ \bibinfo {pages} {072003} (\bibinfo {year} {2020}{\natexlab{b}})}\BibitemShut {NoStop}%
\bibitem [{\citenamefont {Kang}\ \emph {et~al.}(2023)\citenamefont {Kang}, \citenamefont {Larkoski},\ and\ \citenamefont {Yang}}]{Kang_2023}%
  \BibitemOpen
  \bibfield  {author} {\bibinfo {author} {\bibfnamefont {Z.-B.}\ \bibnamefont {Kang}}, \bibinfo {author} {\bibfnamefont {A.~J.}\ \bibnamefont {Larkoski}},\ and\ \bibinfo {author} {\bibfnamefont {J.}~\bibnamefont {Yang}},\ }\bibfield  {title} {\bibinfo {title} {Towards a nonperturbative formulation of the jet charge},\ }\href {https://doi.org/10.1103/PhysRevLett.130.151901} {\bibfield  {journal} {\bibinfo  {journal} {Phys. Rev. Lett.}\ }\textbf {\bibinfo {volume} {130}},\ \bibinfo {pages} {151901} (\bibinfo {year} {2023})}\BibitemShut {NoStop}%
\bibitem [{\citenamefont {Kang}\ \emph {et~al.}(2020)\citenamefont {Kang}, \citenamefont {Liu}, \citenamefont {Mantry},\ and\ \citenamefont {Shao}}]{Kang_2020}%
  \BibitemOpen
  \bibfield  {author} {\bibinfo {author} {\bibfnamefont {Z.-B.}\ \bibnamefont {Kang}}, \bibinfo {author} {\bibfnamefont {X.}~\bibnamefont {Liu}}, \bibinfo {author} {\bibfnamefont {S.}~\bibnamefont {Mantry}},\ and\ \bibinfo {author} {\bibfnamefont {D.~Y.}\ \bibnamefont {Shao}},\ }\bibfield  {title} {\bibinfo {title} {Jet charge: A flavor prism for spin asymmetries at the electron-ion collider},\ }\href {https://doi.org/10.1103/PhysRevLett.125.242003} {\bibfield  {journal} {\bibinfo  {journal} {Phys. Rev. Lett.}\ }\textbf {\bibinfo {volume} {125}},\ \bibinfo {pages} {242003} (\bibinfo {year} {2020})}\BibitemShut {NoStop}%
\end{thebibliography}%
\end{document}